\documentclass{aa}  

\usepackage{graphicx}
\usepackage{txfonts}
\usepackage{newtxtext,newtxmath}
\usepackage[T1]{fontenc}
\usepackage[export]{adjustbox}
\usepackage{graphicx}	
\usepackage{amsmath}	
\usepackage{amssymb}	

\usepackage{xcolor, ulem}

\usepackage{subcaption}

\newcommand{\nospix}{The first row shows from left to the right the 144-MHz maps at 6 and 20 arcsec resolution, respectively. Contours start at 3$\sigma$ with an increment of 2. The black circle shows the optical $D_{25}$ diameter and the blue ellipses show the 3-$\sigma$ contour extent used to integrate the flux density. The second row shows again the 6 and 20-arcsec images without contours. The third row shows the 6- and 20-arcsec contours overlayed on a \textit{rgb} SDSS image. The filled circle in the bottom-left corner shows the synthesised beam. A scale bar shows the projected size at the distance of the galaxy.}
\newcommand{\spix}[2]{The first row shows from left to the right the 144-MHz map at 6 and 20 arcsec resolution, respectively, and the radio spectral index between 144 and #1~MHz at #2~arcsec resolution. Contours start at 3$\sigma$ with an increment of 2. The black circle shows the optical $D_{25}$ diameter and the blue ellipses show the 3-$\sigma$ contour extent used to integrate the flux density. The second row shows again the 6- and 20-arcsec images without contours and the radio spectral index error. The third row shows the 6- and 20-arcsec contours overlayed on a \textit{rgb} SDSS image as well as contours of the radio spectral index. The filled circle shows the synthesised beam. A scale bar shows the projected size at the distance of the galaxy.}

\usepackage[colorlinks,linkcolor=blue,anchorcolor=blue,citecolor=blue]{hyperref}
\begin{document}

   \title{Nearby galaxies in LoTSS-DR2: insights into the non-linearity of the radio--SFR relation}


   \author{V. Heesen
          \inst{1}
          \and
          M.~Staffehl\inst{1}
          \and
          A.~Basu\inst{2}
          \and
          R.~Beck\inst{3}
          \and
          M.~Stein\inst{4}
          \and
          F.~S.~Tabatabaei\inst{5}
          \and
          M.~J.~Hardcastle\inst{6}
          \and
          K.~T.~Chy\.zy\inst{7}
          \and
          T.~W.~Shimwell\inst{8,9}
          \and 
          B.~Adebahr\inst{4}
          \and 
          R.~Beswick\inst{10}
          \and
          D.~J.~Bomans\inst{4}
          \and
          A.~Botteon\inst{9}
          \and
          E.~Brinks\inst{6}
          \and
          M.~Br\"uggen\inst{1}
          \and
          R.-J.~Dettmar\inst{4}
          \and
          A.~Drabent\inst{2}
          \and
          F.~de~Gasperin\inst{1}
          \and
          G.~G\"urkan\inst{2}
          \and
          G.~H.~Heald\inst{11}
          \and
          C.~Horellou\inst{12}
          \and
          B.~Nikiel-Wroczynski\inst{7}
          \and
          R.~Paladino\inst{13}
          \and
          J.~Piotrowska\inst{7}
          \and
          H.~J.~A.~R\"ottgering\inst{9}
          \and
          D.~J.~B.~Smith\inst{6}
          \and
          C.~Tasse\inst{14}
          }

   \institute{Hamburger Sternwarte, University of Hamburg, Gojenbergsweg 112, 21029 Hamburg, Germany\\
              \email{volker.heesen@hs.uni-hamburg.de}
         \and
             Th\"uringer Landessternwarte, Sternwarte 5, 07778 Tautenburg, Germany 
         \and     
             Max-Planck Institut f\"ur Radioastronomie, Auf dem H\"ugel 69, 53121 Bonn, Germany
         \and     
             Ruhr University Bochum, Faculty of Physics and Astronomy, Astronomical Institute, 44780 Bochum, Germany
         \and
             School of Astronomy, Institute for Research in Fundamental Sciences, 19395-5531 Tehran, Iran
         \and
             Centre for Astrophysics Research, University of Hertfordshire, College Lane, Hatfield AL10 9AB, UK
         \and
             Astronomical Observatory, Jagiellonian University, ul. Orla 171, 30- 244 Krak\'ow, Poland
         \and
             ASTRON, The Netherlands Institute for Radio Astronomy, Postbus 2, 7990 AA Dwingeloo, The Netherlands
         \and
             Leiden Observatory, Leiden University, PO Box 9513, 2300 RA Leiden, The Netherlands
         \and
             Jodrell Bank Centre for Astrophysics, Department of Physics and Astronomy, The University of Manchester, M13 9PL, UK
         \and
             CSIRO Astronomy and Space Science, PO Box 1130, Bentley, WA 6102, Australia
         \and
             Department of Space, Earth and Environment, Chalmers University of Technology, Onsala Space Observatory, 439 92 Onsala, Sweden
         \and
             INAF–Istituto di Radioastronomia, Via Gobetti 101, 40129 Bologna, Italy
         \and
             GEPI, Observatoire de Paris, Universit\'e PSL, CNRS, 5 place Jules Janssen, 92190 Meudon, France
             }

   \date{Accepted XXX. Received YYY; in original form ZZZ}
   
 
  \abstract
   {Cosmic rays and magnetic fields are key ingredients in galaxy evolution, regulating both stellar feedback and star formation. Their properties can be studied with low-frequency radio continuum observations, free from thermal contamination.}
   {We define a sample of 76 nearby ($<30~\rm Mpc$) galaxies, with rich ancillary data in the radio continuum and infrared from the CHANG-ES and KINGFISH surveys, which will be observed with the LOFAR Two-metre Sky Survey (LoTSS) at 144~MHz.}
   {We present maps for 45 of them as part of the LoTSS data release 2 (LoTSS-DR2), where we measure integrated flux densities and study integrated and spatially resolved radio spectral indices. We investigate the radio--SFR relation, using star-formation rates (SFR) from total infrared and H\,$\alpha$ + 24-$\mu\rm m$ emission.}
   {The radio--SFR relation at 144~MHz is clearly super-linear with $L_{144}\propto SFR^{1.4-1.5}$. The mean integrated radio spectral index between 144 and $\approx$1400 MHz is $\langle \alpha\rangle  = -0.56\pm 0.14$, in agreement with the injection spectral index for cosmic ray electrons (CRE). However, the radio spectral index maps show a variation of spectral indices with flatter spectra associated with star-forming regions and steeper spectra in galaxy outskirts and, in particular, in extra-planar regions. We found that galaxies with high star-formation rates (SFR) have steeper radio spectra; we find similar correlations with galaxy size, mass, and rotation speed.}
   {Galaxies that are larger and more massive are better electron calorimeters, meaning that the CRE lose a higher fraction of their energy within the galaxies. This explains the super-linear radio--SFR relation, with more massive, star-forming galaxies being radio bright. We propose a semi-calorimetric radio--SFR relation, which employs the galaxy mass as a proxy for the calorimetric efficiency.}


   \maketitle
%

\section{Introduction}

In the study of galaxy evolution, both cosmic rays and galactic magnetic fields are of special interest. Both have energy densities comparable to that of the gaseous interstellar medium \citep{cox_05a} and constitute the non-thermal components of the interstellar medium (ISM). With their high energy density, they have a major influence on the dynamics of galaxy development and are a key factor in the understanding of galaxy evolution. Cosmic rays for example are believed to be a driving factor in the creation of galactic winds, transporting material from the galaxy outwards \citep[e.g.][]{uhlig_12a,socrates_08a,everett_08a,salem_14a,girichidis_18a}. Cosmic rays are transported from the star-forming gaseous disc into the halo, where they can through their pressure gradient slowly accelerate the gas. This process is particularly important for cool galactic winds that contain mostly warm and neutral gas \citep{veilleux_20a}. Since galaxies require a steady influx of gas from the circum-galactic medium in order to compensate for gas depletion due to star formation and transport losses, the mechanisms of the entire disc-halo cycle including the impact of these galaxy-wide winds are an important factor in understanding the overall dynamics and evolution of galaxies \citep{tumlinson_17a}. Without magnetic fields, cosmic rays would quickly escape their host galaxy with relativistic velocities. Instead, they are spiralling along the magnetic field lines and are scattered in irregularities in the magnetic fields, making magnetic fields a key influence on the transport of cosmic rays \citep{zweibel_13a}. 

Over the past years, simulations of galaxies including outflows and winds, taking into account magnetic fields and cosmic rays, have advanced rapidly \citep{vogelsberger_20a}. However, many fundamental questions, e.g. regarding the transport processes of cosmic rays are still not understood. Much of our knowledge about cosmic-ray dynamics stems from observations of the Milky Way, but recently also other galaxies have come into focus since the view `from the outside' also includes the radio halo from these galaxies, which is harder to observe in the case of our Milky Way \citep[e.g.][]{heesen_18a,miskolczi_19a,schmidt_19a,mora_19a,stein_19a,stein_19b}. This is one of the reasons why observations of cosmic ray transport processes are of special interest. These relativistic cosmic rays also include cosmic ray electrons (CRE), and at radio frequencies we are sensitive to the emission of CRE in the GeV-energy range. This energy range also corresponds to the peak of the cosmic ray spectrum in general. Hence, we can probe the cosmic ray transport parameters, where they are dynamically most important.

Radio continuum emission offers further advantages. Radio emission is not attenuated by dust and since it originates from cosmic rays accelerated in supernova remnants of massive young stars, it can therefore be used as an extinction-free tracer for star formation \citep{condon_92a,murphy_11a}. Current radio surveys can serve to calibrate the radio--star formation rate (SFR) relation in order to extrapolate it to more distant galaxies at higher redshifts. The radio--SFR relation has a spread of only $\approx 0.15$~dex \citep{tabatabaei_17a}, which is comparable to uncertainty of the widely used hybrid star formation tracers such as the combination of H\,$\alpha$ and 24-$\mu$m mid-infrared emission \citep{calzetti_07a}. The synchrotron emission that we see as radio continuum depends both on the CRE and magnetic field content of galaxies. There is a degeneracy between having a high energy density of the CRE and weak magnetic fields, or contrary strong magnetic fields and a low CRE energy density. For reasons of dynamical stability one usually assumes energy equipartition between the cosmic rays and the magnetic field, which is a good approximation for the Milky Way \citep{boulares_90a}. Given the many dependencies of the synchrotron emission, it is astonishing that the radio--SFR relation is so tight. In particular, less-luminous galaxies should be deficient in radio luminosity as the CRE are expected to escape rapidly from their host. The fact that this appears to not be the case, when the SFR is measured by infrared star-formation tracers, has been explained by an invocation of a `conspiracy' between the escape of the CRE and dust-heating UV photons \citep{bell_03a,lacki_10a}.  While more sensitive spatially resolved observations of dwarf galaxies have finally shown that the synchrotron emission is indeed suppressed \citep{hindson_18a}, the influence of cosmic ray transport on the radio--SFR relation in $L_{\star}$ galaxies is still largely unknown \citep{lisenfeld_96a,li_16a}.

Low-frequency radio observations are particularly useful to study synchrotron emission from galaxies since the fraction of thermal radiation (i.e. bremsstrahlung from non-relativistic electrons) is less than 10 per cent \citep{tabatabaei_17a}. At 150~MHz, the radio--SFR relation shows a possible excess of radio emission in low-mass galaxies \citep{Brown_2017a,gurkan_18a} which is not seen at higher frequencies \citep{Schmitt_2006a, li_16a}. Radio continuum studies of dwarf galaxies are influenced by unrelated background sources \citep{hindson_18a}, so that spatially resolved studies are important. The most suitable telescope for these kinds of studies is the \textit{LOw Frequency ARray} \citep[LOFAR;][]{vanHaarlem_13a}, which is composed of 51 stations spread throughout Europe with two more stations under construction. With the LOFAR Two-metre Sky Survey \citep[LoTSS; see][]{shimwell_17a} it is currently producing a survey of the northern sky which is more sensitive than any radio survey at a similar angular resolution. It had its first data release \citep[LoTSS-DR1][]{shimwell_19a} and enabled successful work on cosmic-ray transport
and the radio--SFR relation \citep[e.g.][]{heesen_14a} explored for four LOFAR galaxies \citep{heesen_19a}.
Now the data release 2 \citep[LoTSS-DR2][]{shimwell_22a} covers a larger sky area at 144~MHz with a bandwidth of 48~MHz and in both 6 and 20~arcsec resolution with a sensitivity of 50--100~$\rm \mu Jy\,beam^{-1}$.

This paper is a first exploration of these data by investigating a sample of nearby galaxies at distances of $<$ 30~Mpc, which corresponds to a redshift of less than $z \approx 0.01$. Since spatially resolved scaling relations are our main aim, the sample selection was largely driven by the availability of ancillary data in the radio continuum and infrared. We selected galaxies from the CHANG-ES radio continuum survey (\textit{Continuum Halos in Nearby Galaxies: an EVLA Survey}; \citet{irwin_12a}), consisting of 35 edge-on galaxies, and from the KINGFISH far-infrared survey \citep[Key Insights on Nearby Galaxies: A Far-Infrared Survey with Herschel;][]{kennicutt_12a}, which in turn is a descendent of the SINGS survey \citep[The SIRTF Nearby Galaxies Survey;][]{kennicutt_03a}, the sample of the former consisting of 65 galaxies. From these samples, all galaxies also present in LoTSS-DR2 were included in our sample. The galaxy sample is very diverse in optical classification, ranging from elliptical galaxies, spiral galaxies to irregular galaxies. The sample has also a wide range of SFRs between $0.002$ and $15~\rm M_{\sun}\,yr^{-1}$. The galaxies also cover a wide range of nuclear types, ranging from the more quiescent H\,{\sc ii} nuclei over LINER to more active types of Seyfert nuclei. Similarly, the sample covers a wide range of total masses (within the optical radius) between $10^8$ and $10^{12}~\rm M_{\sun}$ as well a large size range between 1 and 70~kpc. The sample is not complete, but rather aims to capture a representative set of galaxies with a range of primary parameters such as SFR, morphology, mass, and size. Due to the limited sample size, secondary parameters such as nuclear type and environmental properties are not evenly sampled.

The structure of this paper is as follows: Section~\ref{s:methodology} presents the methodology including a description of the sample, a summary of how the LoTSS-DR2 data were obtained and reduced, and how flux densities and radio spectral indices were measured. This is followed by Section~\ref{s:results}, where we present the radio maps for one selected galaxy and also analyse integrated spectral indices (Section~\ref{ss:radio_spectral_index}), spatially resolved spectral indices (Section~\ref{ss:spatially_resolved_spectral_index}), and the integrated radio--SFR relation (Section~\ref{ss:radio_sfr_relation}). We discuss our findings in more detail in Section~\ref{s:discussion}, focusing on the non-linearity of the radio--SFR relation. Section~\ref{s:conclusions} presents our conclusions and a brief outlook on possible follow-up work. Our atlas of maps is presented in Appendix~\ref{as:atlas}.

\section{Methodology}
\label{s:methodology}

\subsection{Galaxy sample}
\label{ss:sample}

Appendix~\ref{as:the_full_lotss_galaxy_sample} lists the full sample that was considered. We have selected galaxies from samples that have either high-frequency radio continuum data, so that we can calculate radio spectral indices, or they have infrared data so that we can calculate star formation rates. For the edge-on sample, we chose from the CHANG-ES sample \citep{irwin_12a}, which contains 35 galaxies, 21 of which are located in the northern sky and so accessible to LoTSS. For the infrared data, we chose from the SINGS \citep{kennicutt_03a} and KINGFISH \citep{kennicutt_11a} samples. SINGS contains 75 representative star-forming galaxies within 30~Mpc distance, 48 of which are in the northern sky. Three galaxies were added, which are only in the KINGFISH survey (NGC~2146, IC~342, NGC 5457). Finally, we have added another 4 galaxies, which are IC~10, NGC~598 (M~33), NGC 4214, and NGC 4449. Three of them are star-burst dwarf irregular galaxies and M~33 is the most nearby galaxy, which we can easily image being less than 1 degree in apparent size. This results in 76 galaxies, which will be eventually covered by LoTSS. The four sub-samples are referred to as CHANG-ES, SINGS, KINGFISH+, and `Extra'. 

\begin{figure*}
\centering
\begin{subfigure}[t]{0.02\textwidth}
    \textbf{(a)}
    \end{subfigure}
\begin{subfigure}[t]{0.97\linewidth}
\includegraphics[width=1.0\linewidth,valign=t]{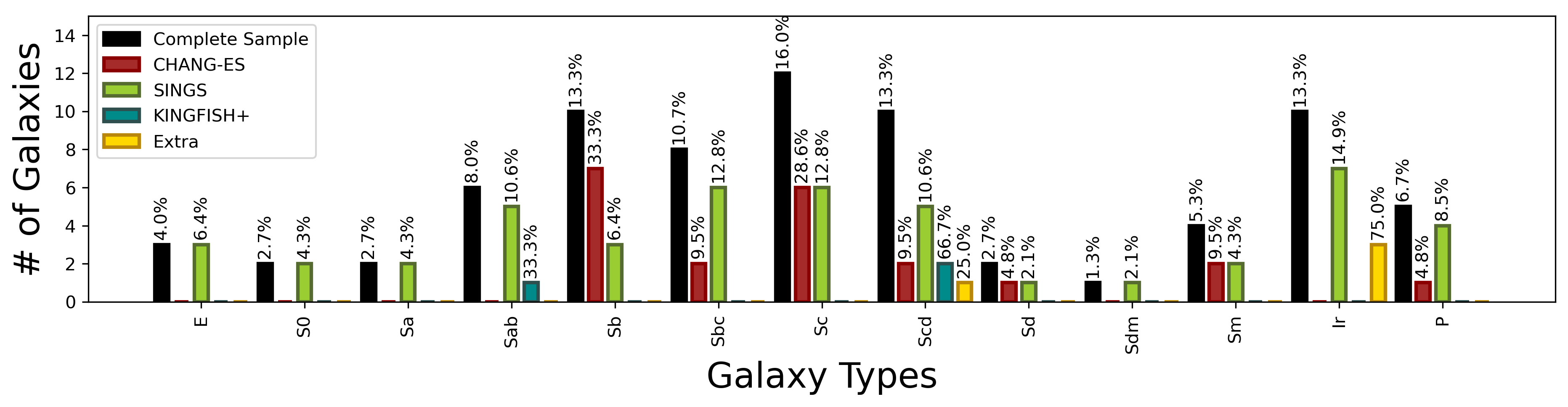}
\end{subfigure}
\\
\begin{subfigure}[t]{0.02\textwidth}
\textbf{(b)}
\end{subfigure}
\begin{subfigure}[t]{0.47\linewidth}
\includegraphics[width=1\linewidth,valign=t]{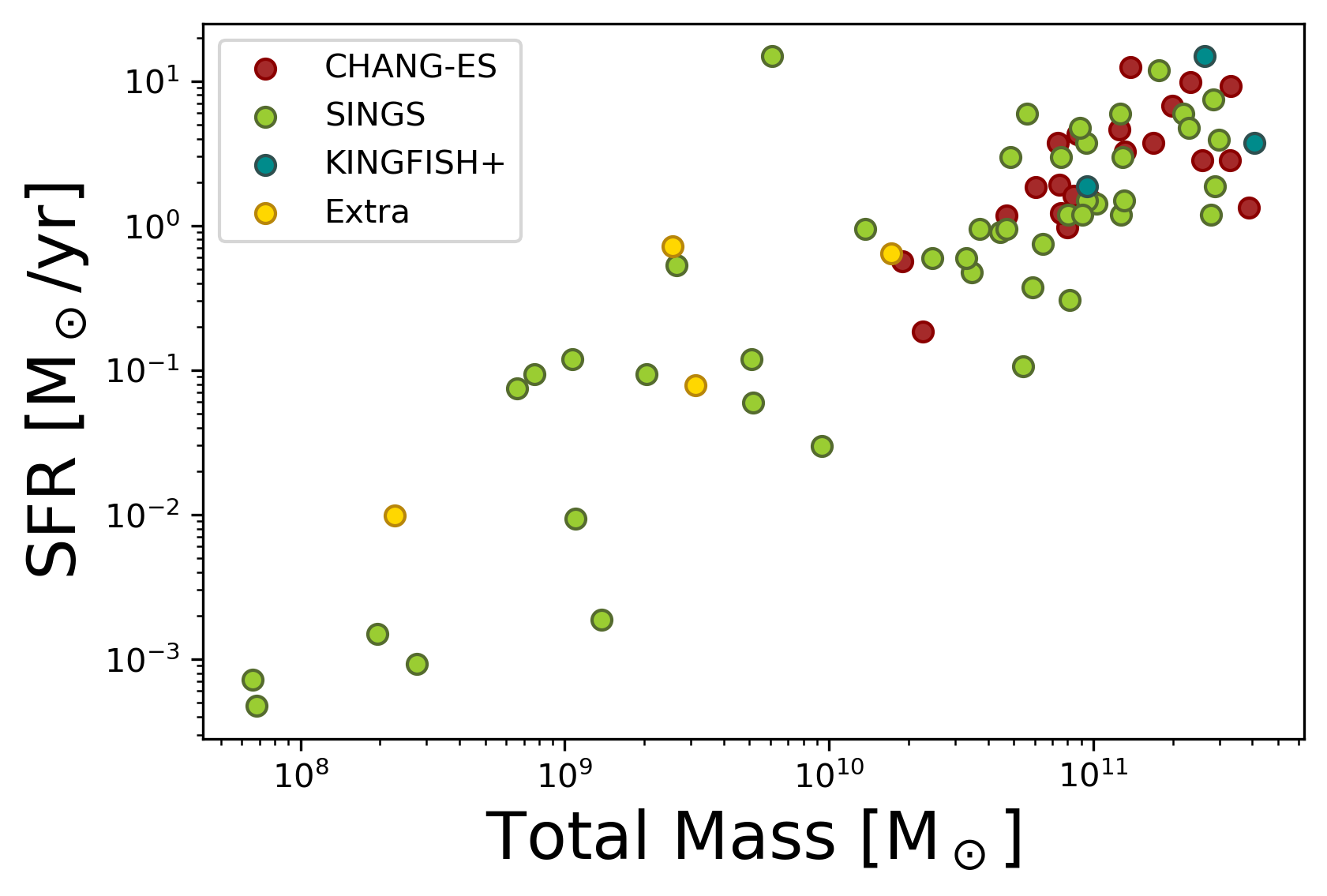}
\end{subfigure}
\begin{subfigure}[t]{0.02\textwidth}
\textbf{(c)}
\end{subfigure}
\begin{subfigure}[t]{0.47\linewidth}
\includegraphics[width=1\linewidth,valign=t]{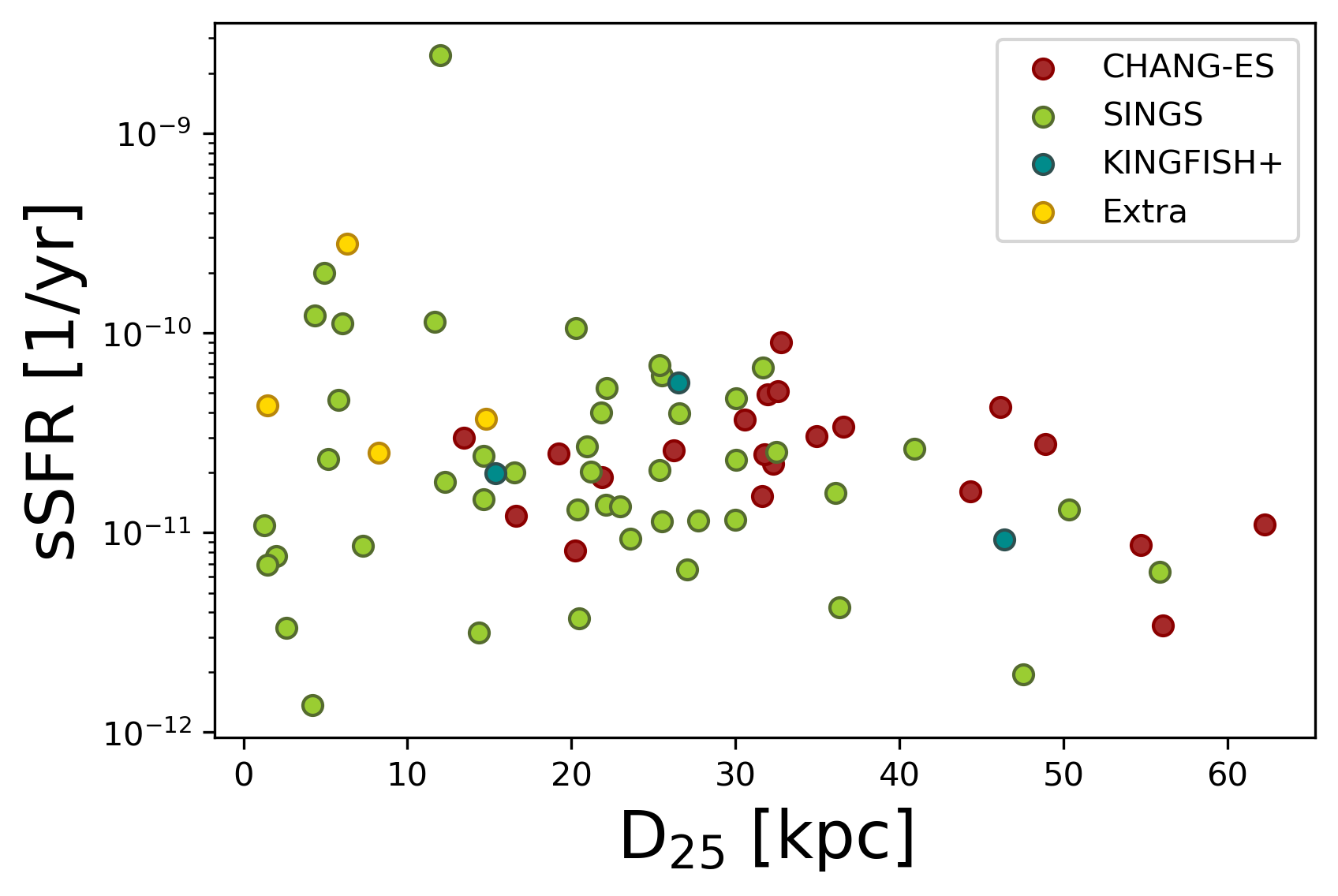}
\end{subfigure}
\caption{Properties of the full galaxy sample as presented in Appendix~\ref{as:the_full_lotss_galaxy_sample}. (a) Distribution of galaxy type; (b) star formation rate (SFR), as function of total mass enclosed within the radius of the star-forming disc; (c) specific star formation rate (sSFR; per total mass) as function of the optical galaxy diameter (projected apparent diameter $D_{25}$). The galaxies are grouped according to which respective sub-sample they belong (Section~\ref{ss:sample}).}
    \label{fig:sample}
\end{figure*}

In addition to the infrared data, we also have rich ancillary data in the radio continuum. First, all the CHANG-ES galaxies have 6-GHz data from \citet{wiegert_15a} at approximately 15 arcsec resolution, observed with the Karl G. Jansky Very Large Array (VLA) in D-configuration. There was also CHANG-ES data taken at 1.5 GHz in C-configuration with a similar resolution, which will be analysed in a forthcoming paper together with the LoTSS data (Stein et al. 2021, in preparation). The SINGS galaxies were observed at 1.4-GHz by the WSRT--SINGS survey \citep{braun_07a}, which includes 34 SINGS galaxies north of declination $12\fdg 5$ with an optical diameter of $>$5~arcmin. Approximately a third of the full sample galaxies have nuclei that can be classified as AGN \citep[LINER, Seyfert or transition object][]{ho_97a}, with more galaxies possibly containing hidden AGN that cannot be detected with optical spectroscopy \citep{satyapal_08a}. The AGN contribution to the radio flux was estimated from the work by \citet{baldi_18a} and \citet{irwin_19a}, who present high-resolution maps for the Palomar \citep{ho_95a} and CHANG-ES samples. We found that the the galaxies with a radio-detected AGN in LoTSS-DR2 (10 galaxies in our sample of 45) have mostly AGN contributions of less than 1 per cent at 1400~MHz. The exceptions are NGC~2683 (1.7 per cent), NGC~3079 (19 per cent) and NGC~5866 (35 per cent). Hence we conclude that our galaxies are mostly not AGN contaminated, but the mentioned galaxies require special attention.

One of the important characteristics of galaxies is their star formation rate (SFR). We use two approaches of calculating them. Our main method is to use total infrared (TIR) fluxes in the wavelength range 3--1100~$\mu$m. They do have the advantage that they are more reliable for our edge-on sample as the widely used mid-infrared (MIR) 24-$\mu$m emission becomes partially optically thick \citep{li_16a,vargas_18a}. Hence, in order to have consistent SFRs in our sample, this is the preferred method. Nevertheless, we also used SFRs from H\,$\alpha$ and MIR tracers, which should be particularly reliable for spatially resolved SFRs in moderately inclined galaxies \citep{murphy_11a,leroy_12a}. Since this paper is also a preparatory study for the spatially resolved radio--SFR relation, we compare the two methods. 
The TIR luminosities were calculated using the prescription of \citet{dale_02a}, or when available, we used the flux densities from \citet{dale_12a}. Since the CHANG-ES galaxies are mostly not included in SINGS or KINGFISH, we calculated their flux densities from IRAS data using the formulae by \citet{dale_02a}. The TIR luminosities were converted into SFRs using the calibration presented in \citet{kennicutt_12a}, which is based on work by \citet{hao_11a} and \citet{murphy_11a}. For SINGS and KINGFISH+, \citet{kennicutt_11a} provides SFRs from a combination of H\,$\alpha$ and 24-$\mu$m emission based on measurements by \citet{calzetti_10a}. The CHANG-ES galaxies have SFRs from 24-$\mu$m emission \citep{wiegert_15a} using the calibration of \citet{rieke_09a}.

 Our galaxy sample covers a wide range of galaxy parameters (see Fig.~\ref{fig:sample}). Nearly all types of optical galaxy classifications are included, and there is a range of almost $10^5$ in total infrared luminosity as well as a range of $10^{3}$ in physical size. SINGS was chosen specifically to have a fairly even distribution of optical galaxy types for the late-type spiral galaxies (Sa--Sd, and magellanic-type galaxies Sdm and Sm as well as irregular galaxies Ir). In addition, there are a few elliptical galaxies (E), lenticular galaxies (S0) and peculiar galaxies (P). With CHANG-ES this was not the case, where a radio flux density cut was used and so only galaxies with higher SFRs were included. This means the CHANG-ES galaxies tend to be heavier and larger. Hence, we are biased towards massive galaxies with total masses of $\gtrsim10^{10}~\rm M{\sun}$ and correspondingly larger SFRs. The 4 galaxies from the Extra sample add a few galaxies with lower masses to balance the distribution slightly (compare with Fig.~\ref{fig:sample}). Note that the total mass is calculated as $M_{\rm tot}=0.234(v_{\rm rot}/{100~\rm km\,s^{-1}})^2(r_\star/{\rm kpc})\times 10^{10}~{\rm M}_{\sun}$, where we have estimated $r_\star$, the size of the star-forming disc, by using the 3-$\sigma$ extent of the radio continuum emission at 144~MHz. Specifically, we projected the major axis of the $a_{20\arcsec}$ integration region (Section~\ref{ss:flux_estimation}) to obtain the physical size.

\begin{table*}[!th]
	\centering
	\caption{Integrated 144-MHz flux densities, map properties and star formation rates of the 45 galaxies in LoTSS-DR2.}
	\label{tab:flux}
	\begin{tabular}{lc ccc ccc c} 
		\hline
		Galaxy  & PA    & $\sigma_{6\arcsec}$        & $a_{6\arcsec}\times b_{6\arcsec}$         & $S_{6\arcsec}$  & $\sigma_{20\arcsec}$& $a_{20\arcsec}\times b_{20\arcsec}$  & $S_{20\arcsec}$ & $\rm SFR_{H\alpha/24 \mu m}$ \\
                        & ($\degr$) & ($\mu\rm Jy/b.$) & (arcmin$^2$) & (Jy)   & ($\mu\rm Jy/b.$) & (arcmin$^2$) & (Jy)    & ($\rm M_{\sun}\,yr^{-1}$)   \\
                (1)     & (2)   & (3)                       & (4)         & (5)    & (6)                       & (7)         & (8)      & (9)                       \\
		\hline
                N598   & 22.7 [LEDA]  & 170 & $14.0\times 10.0$  & $2.320\pm 0.234$ & 220 & $20.0\times 13.0$ & $3.627 \pm 0.363$       & 0.13  [H21] \\ 
                N855   & 67 [LEDA]    & 130 & $0.3 \times 0.2 $  & $0.007\pm 0.001$ & 150 & $0.7 \times 0.6 $ & $0.010 \pm 0.001$       & 0.043 [H21] \\
                N891   & 22 [LEDA]    & 160 & $4.6 \times 2.0 $  & $2.985\pm 0.299$ & 160 & $5.9 \times 4.6 $ & $3.294 \pm 0.329$       & 1.55  [W15] \\
                N925   & 287[dB08]    & 90  & $4.4 \times 2.3 $  & $0.197\pm 0.020$ & 130 & $5.6 \times 3.1 $ & $0.243 \pm 0.024$       & 0.54  [K11] \\
                N2683  & 43.6[LEDA]   & 70  & $2.6 \times 1.1 $  & $0.284\pm 0.028$ & 110 & $3.6 \times 2.4 $ & $0.362 \pm 0.036$       & 0.09  [W15] \\
                N2798  & 160 [LEDA]   & 80  & $0.6 \times 0.6 $  & $0.220\pm 0.022$ & 140 & $1.1 \times 0.8 $ & $0.222 \pm 0.022$       & 3.38  [K11] \\
                N2820  & 65 [K18]     & 60  & $2.9 \times 0.9 $  & $0.251\pm 0.025$ & 100 & $3.2 \times 1.2 $ & $0.261 \pm 0.026$       & 0.62  [W15] \\
                N2841  & 153 [dB08]   & 80  & $3.1 \times 2.3 $  & $0.463\pm 0.046$ & 110 & $3.9 \times 2.9 $ & $0.501 \pm 0.050$       & 2.45  [K11] \\
                N2976  & 143 [LEDA]   & 60  & $1.8 \times 1.0 $  & $0.109\pm 0.011$ & 140 & $3.0 \times 2.8 $ & $0.201 \pm 0.020$       & 0.082 [K11] \\
                N3003  & 78.8 [LEDA]  & 70  & $2.1 \times 0.9 $  & $0.132\pm 0.013$ & 100 & $3.1 \times 1.5 $ & $0.148 \pm 0.015$       & 0.67  [W15] \\
                N3031  & 330.2 [dB08] & 90  & $10.0\times 5.6 $  & $1.883\pm 0.189$ & 150 & $11.7\times 5.9 $ & $2.055 \pm 0.206$       & 0.39  [C10] \\
                N3079  & 167 [K18]    & 60  & $3.2 \times 2.0 $  & $3.708\pm 0.371$ & 110 & $3.9 \times 2.3 $ & $3.746 \pm 0.375$       & 3.46  [W15] \\
                N3077  & 49 [LEDA]    & 70  & $0.8 \times 0.6 $  & $0.059\pm 0.006$ & 110 & $1.4 \times 1.2 $ & $0.067 \pm 0.007$       & 0.094 [K11] \\
                N3184  & 0$^*$        & 60  & $3.2 \times 3.2 $  & $0.337\pm 0.034$ & 130 & $4.1 \times 4.1 $ & $0.377 \pm 0.038$       & 0.66  [K11] \\
                N3198  & 215 [dB08]   & 90  & $2.7 \times 1.0 $  & $0.099\pm 0.010$ & 170 & $3.4 \times 1.5 $ & $0.127 \pm 0.013$       & 1.01  [K11] \\
                N3265  & 73.4 [LEDA]  & 80  & $0.4 \times 0.5 $  & $0.023\pm 0.002$ & 190 & $0.6 \times 0.7 $ & $0.027 \pm 0.003$       & 0.38  [K11] \\
                Mrk33  & 129.3 [LEDA] & 60  & $0.5 \times 0.4 $  & $0.054\pm 0.005$ & 150 & $0.8 \times 0.7 $ & $0.055 \pm 0.006$       & 1.5   [K03] \\
                N3432  & 33 [LEDA]    & 70  & $3.7 \times 0.9 $  & $0.274\pm 0.028$ & 130 & $3.4 \times 1.9 $ & $0.304 \pm 0.030$       & 0.15  [W15] \\
                N3448  & 64.8 [LEDA]  & 80  & $1.1 \times 0.6 $  & $0.173\pm 0.017$ & 240 & $1.8 \times 1.2 $ & $0.195 \pm 0.019$       & 0.92  [W15] \\
                N3556  & 79 [LEDA]    & 70  & $3.4 \times 1.5 $  & $1.065\pm 0.107$ & 120 & $4.4 \times 3.2 $ & $1.215 \pm 0.122$       & 2.17  [W15] \\
                N3877  & 35 [K18]     & 60  & $1.8 \times 0.7 $  & $0.147\pm 0.015$ & 130 & $2.1 \times 1.0 $ & $0.169 \pm 0.017$       & 0.92  [W15] \\
                N3938  & 0$^*$        & 90  & $2.0 \times 2.0 $  & $0.334\pm 0.034$ & 300 & $3.0 \times 3.0 $ & $0.409 \pm 0.041$       & 1.77  [K11] \\
                N4013  & 65 [K18]     & 80  & $2.0 \times 0.6 $  & $0.144\pm 0.015$ & 170 & $2.1 \times 0.9 $ & $0.151 \pm 0.015$       & 0.48  [W15] \\
                N4096  & 20 [LEDA]    & 70  & $2.2 \times 0.9 $  & $0.169\pm 0.017$ & 120 & $3.2 \times 1.6 $ & $0.198 \pm 0.020$       & 0.27  [W15] \\
                N4125  & 95 [LEDA]    & 80  & $0.3 \times 0.3 $  & $0.012\pm 0.001$ & 110 & $0.3 \times 0.3 $ & $0.012 \pm 0.001$       & 0.58  [C10] \\
                N4157  & 66 [K18]     & 70  & $2.7 \times 1.4 $  & $0.921\pm 0.092$ & 170 & $4.8 \times 2.7 $ & $1.021 \pm 0.102$       & 1.25  [W15] \\
                N4214  & 0$^*$        & 60  & $1.5 \times 0.9 $  & $0.090\pm 0.009$ & 160 & $2.7 \times 1.4 $ & $0.126 \pm 0.013$       & 0.16  [H18] \\
                N4217  & 50 [K18]     & 70  & $2.0 \times 0.9 $  & $0.427\pm 0.043$ & 100 & $2.4 \times 1.8 $ & $0.446 \pm 0.045$       & 1.53  [W15] \\
                N4244  & 44.8 [LEDA]  & 60  & $5.9 \times 0.9 $  & $0.049\pm 0.005$ & 110 & $6.2 \times 1.1 $ & $0.050 \pm 0.005$       & 0.02  [W15] \\
                N4449  & 50.8 [LEDA]  & 70  & $3.2 \times 2.8 $  & $0.863\pm 0.086$ & 110 & $4.6 \times 4.6 $ & $1.032 \pm 0.103$       & 0.32  [C10] \\
                N4559  & 148.3 [LEDA] & 80  & $2.6 \times 1.4 $  & $0.164\pm 0.017$ & 120 & $4.7 \times 2.2 $ & $0.236 \pm 0.024$       & 0.37  [K11] \\
                N4625  & 0$^*$        & 70  & $0.8 \times 0.6 $  & $0.017\pm 0.002$ & 100 & $0.9 \times 0.7 $ & $0.017 \pm 0.002$       & 0.052 [K11] \\
                N4631  & 85.7 [LEDA]  & 100 & $6.5 \times 3.5 $  & $4.031\pm 0.403$ & 130 & $8.8 \times 7.5 $ & $4.640 \pm 0.464$       & 1.7   [K11] \\
                N4725  & 35.7 [LEDA]  & 70  & $3.4 \times 2.2 $  & $0.154\pm 0.016$ & 170 & $3.8 \times 2.4 $ & $0.220 \pm 0.022$       & 0.44  [K11] \\
                N4736  & 105 [LEDA]   & 80  & $2.3 \times 1.8 $  & $0.803\pm 0.080$ & 110 & $3.6 \times 2.9 $ & $0.897 \pm 0.090$       & 0.38  [K11] \\
                N5033  & 171.8 [LEDA] & 60  & $4.8 \times 3.1 $  & $1.133\pm 0.113$ & 100 & $5.0 \times 3.4 $ & $1.144 \pm 0.114$       & 1.06  [H21] \\
                N5055  & 102 [dB08]   & 60  & $5.2 \times 2.9 $  & $2.262\pm 0.226$ & 90  & $6.6 \times 4.9 $ & $2.286 \pm 0.229$       & 1.04  [K11] \\
                N5194  & 195 [H21]    & 50  & $6.2 \times 5.0 $  & $6.707\pm 0.671$ & 110 & $6.9 \times 6.7 $ & $6.363 \pm 0.636$       & 2.36  [C10] \\
                N5195  & 79 [LEDA]    & 50  & $2.0 \times 1.8 $  & $1.084\pm 0.108$ & 110 & $2.0 \times 1.8 $ & $1.004 \pm 0.100$       & 0.26  [H21] \\
                N5297  & 146.5 [LEDA] & 50  & $1.8 \times 0.8 $  & $0.107\pm 0.011$ & 90  & $2.5 \times 1.2 $ & $0.118 \pm 0.012$       & 1.27  [W15] \\
                N5457  & 0$^*$        & 80  & $8.6 \times 8.0 $  & $2.493\pm 0.250$ & 120 & $11.3\times 11.3$ & $3.226 \pm 0.323$       & 2.33  [K11] \\
                N5474  & 99.7 [LEDA]  & 60  & $0.5 \times 0.5 $  & $0.008\pm 0.001$ & 100 & $2.2 \times 1.6 $ & $0.021 \pm 0.002$       & 0.091 [K11] \\
                N5866  & 126.5 [LEDA] & 80  & $0.6 \times 0.4 $  & $0.059\pm 0.006$ & 220 & $1.0 \times 0.6 $ & $0.065 \pm 0.006$       & 0.26  [K11] \\
                N5907  & 155.6 [LEDA] & 100 & $3.8 \times 1.1 $  & $0.492\pm 0.049$ & 130 & $5.8 \times 2.0 $ & $0.594 \pm 0.059$       & 1.56  [W15] \\
                N7331  & 168 [dB08]   & 150 & $4.2 \times 2.0 $  & $2.077\pm 0.208$ & 250 & $5.4 \times 2.8 $ & $2.122 \pm 0.212$       & 2.74  [K11] \\
                \hline
	\end{tabular}
        \flushleft
            {\small {\bf Notes.}\\
              Column (1)  galaxy name; (2) position angle of the major axis, dB08 and O15 refer to the receding major axis, $^*$ indicates uncertain value; (3) rms noise of the 6-arcsec map; (4) semi-major and semi-minor axes of the integration ellipse for the 6-arcsec map; (5) integrated flux density from the 6-arcsec map; (6) rms noise of the 20-arcsec map; (7) semi-major and semi-minor axes of the integration ellipse for the 20-arcsec map; (8) integrated flux density from the 20-arcsec map; (9) alternative star formation rate from H\,$\alpha$ [H18], hybrid H\,$\alpha$ and $24~\mu\rm m$ [C10, K11, NGC~855, NGC~5033], and 24 $\mu\rm m$ [W15, NGC~5195].  \\
              {\bf References.}\\
              Data are from the from HyperLEDA (LED) and following references: dB08: \citet{de_blok_08a}; C10: \citet{calzetti_10a}; H18: \citet{hindson_18a}; H21: this paper, we used the H\,$\alpha$ fluxes of \citet{kennicutt_09a} and the 24-$\mu\rm m$ fluxes of \citet{dale_05a}; K02: \citet{kennicutt_03a}; K11: \citet{kennicutt_11a}; K18: \citet{krause_18a}; O15: \citet{oh_15a}; W08: \citet{walter_08a}; W15: \citet{wiegert_15a}}.
\end{table*}

\subsection{LoTSS-DR2 data}
\label{ss:lotss}
The LoTSS-DR2 data were observed and reduced as described in \citet{shimwell_17a,shimwell_19a}. Here, we briefly summarise the procedure. The LoTSS data are observed with the High Band Antenna (HBA) system of LOFAR. The data are taken with the \textsc{hba\_dual\_inner} configuration with 8~h dwell time and 120--168~MHz frequency coverage. The entire northern sky is covered with 3168 pointings. The publicly available LOFAR direction-independent calibration procedure was described in detail by \citet{van_weeren_16a} and \citet{williams_16a} and makes use of the LOFAR Default Preprocessing Pipeline \citep[\textsc{dppp};][]{van_diepen_18a} for averaging and calibration and BlackBoard Selfcal \citep[\textsc{bbs};][]{pandey_09a}. Because for LOFAR the direction-dependent effects (DDE) are important, a lot of work has gone into improving their treatment during calibration. LoTSS is using for its pipeline \textsc{KillMS} \cite[][]{smirnov_15a} to calculate DDE corrections and apply them to the data during imaging using {\sc ddfacet} \citep{tasse_18a}. Compared with LoTSS-DR1 \citep{shimwell_19a}, LoTSS-DR2 \citep{shimwell_22a} now provides a much improved imaging capability for diffuse extended emission \citep{tasse_21a}.

The LoTSS-DR2 data are provided in two ways: either as mosaics created from neighbouring pointings to improve sensitivity, or as re-calibrated ($u,v$) data sets, where the calibration was especially tailored to the target that we are interested in \citep{van_weeren_21a}. For most galaxies we used the re-calibrated (u,v) data and re-imaged and deconvolved the data with \textsc{wsclean v2.9} \citep{offringa_14a,offringa_17a}. For the imaging we produced maps with resolutions both at 6 and 20 arcsec resolution, hereafter referred to as high and low resolution maps, respectively. For the high-resolution maps, we used Brigg's robust weighting with $\tt robust=-0.5$, whereas for the low-resolution maps we used $\tt robust=-0.25$ together with a Gaussian taper of 10 arcsec. The maps were deconvolved with the multi-scale and auto masking options to remove any residuals comparable to the size of the galaxies. The maps were then restored with a Gaussian beam with a FWHM of 6 and 20 arcsec, respectively. We checked that the integrated flux densities of the 6- and 20-arcsec maps are identical for the same integration area (see Section~\ref{ss:flux_estimation}) and found and average ratio of $S_{6\arcsec}^{\star}/S_{20\arcsec}=1.01\pm 0.06$, where $S_{6\arcsec}^\star$ is the integrated flux density in the 20-arcsec integration area of the 6-arcsec map and $S_{20\arcsec}$ is the integrated 20-arcsec flux density. Hence, we are confident that we have sufficiently deconvolved both the high- and the low-resolution maps. In order to match the flux densities with the LoTSS-DR2 scale, we compared flux densities of point-like sources with the LoTSS-DR2 point source catalogue and applied a scaling factor to our maps.\footnote{We used 
the following script to perform the re-scaling:
\href{https://github.com/mhardcastle/ddf-pipeline/blob/master/scripts/align-extrationimage-fluxes.py}{https://github.com/mhardcastle/ddf-pipeline/blob/master/scripts/align-extrationimage-fluxes.py}} For a few galaxies (NGC 5055, NGC 5194/5, and NGC 5474) we found the mosaics to be superiour to the re-calibrated data sets, so we used them instead.

Galaxies included in this paper have been detected and are contained within usable LoTSS-DR2 mosaics. There were a few objects for which a mosaic was available, but no object was detected (M81DwA, IC~2574, DDO~154 and DDO~165). We did not attempt to improve detection rates by masking star-forming regions, reducing the area, and thus increasing the signal-to-noise ratio \citep{hindson_18a}. The analysis of dwarf irregular galaxies with low star-formation rates, which are particularly affected, will be deferred to future work (Sridhar et al. 2021, in preparation). For the 45 remaining galaxies, we made cutouts from the maps using the Common Astronomy Software Applications \citep[\textsc{casa};][]{mcmullin_07a}. These cutouts were square regions centred on the galaxy with an edge length equal to the optical diameter of the galaxy ($D_{25}$) plus $0\fdg 1$ rounded up to the nearest tenth of a degree. A minimum edge length of $0\fdg 3$ was chosen to ensure that the background can be measured reliably. The size of the cutouts was sufficient even to include the faint diffuse emission in the outer discs.

\subsection{Flux density estimation}
\label{ss:flux_estimation}

For all galaxies in the sample, we estimated the total radio flux density for both the high and low resolutions within $3\sigma$-contours. Around these contours, an elliptical region was created such that it encloses the $3\sigma$-line surrounding the entire galaxy. This approach is a compromise between choosing a large enough integration area to not underestimate the emission of the galaxy and minimising the contamination by background sources, which is $\approx 20~\mu\rm Jy\,beam^{-1}$ at 20~arcsec resolution \citep{williams_16a}. Another issue of extending the integration area below $3\sigma$ is that the flux is not properly deconvolved and the contribution from residual flux becomes an issue \citep{walter_08a}. Galaxies have fairly well-defined edges, the scale length of 2~kpc \citep{mulcahy_14a} is equivalent to 44~arcsec at the median distance of our sample. We estimate that the flux between the 1$\sigma$ and the 3$\sigma$ contour line is about 5 per cent of the integrated flux density. Our integration areas enclose the $3\sigma$ contours, so in many cases also include emission at the $2\sigma$ level, excluding some local radial extensions only. Hence, the 5 per cent estimate is an upper limit.

The position angle of the region was chosen to be the position angle of the galaxy as listed in Table~\ref{tab:flux}. Wherever other sources outside of the galaxy's disc were apparent or contours of the galaxy blended with those of other nearby sources, we exclude them from the region. In a few cases  this was not possible as the signal-to-noise ratio was not sufficient to discern the galaxy from the surroundings with only the $3\sigma$-lines. In these cases the region was drawn by eye and it was attempted to enclose the visibly discernible features of the galaxy as well as possible. 

The flux density $S_{\nu}$ was then measured by integrating the intensity $I_{\nu}$ in the integration regions. The error on the flux density was estimated by assuming a statistical error on the flux density caused by the background noise as well as a relative calibration error of $\epsilon =0.1$ \citep{shimwell_17a}, resulting in an error of the form:
\begin{equation}
    \label{eq:flux_density_error}
    \sigma_{S_{\nu}}  =  \sqrt{\left( \sigma \sqrt{N_{\text{beams}}} \right)^2 + \left( \epsilon S_{\nu} \right)^2},
\end{equation} 
where $N_{\rm beams}$ is the number of beams in the integration region. For most galaxies, except the faintest, the 10 per cent relative error is the dominating contribution. Table~\ref{tab:flux} shows the resulting flux densities, $S_{6\arcsec}$ and $S_{20\arcsec}$, for the galaxy sample for both high (6~arcsec) and low (20~arcsec) resolutions, respectively. The table also shows the rms intensities for the background noise for the respective cutout.

\subsection{Radio spectral indices}
\label{ss:radio_spectral_indices}

\subsubsection{Integrated spectral indices}
\label{ss:integrated_spectral_indices}

For further investigation we calculated integrated radio continuum spectral indices. For those galaxies, where $L$-band maps are available, we integrated the flux densities within the same ellipses as for the 20-arcsec LoTSS maps. These data were taken mostly from WSRT--SINGS \citep{braun_07a} at 1365 MHz and CHANG-ES \citep{wiegert_15a} at 1575 MHz. The CHANG-ES data are at a lower angular resolution of $\rm FWHM \gtrsim 30$~arcsec; while we did not use them for the resolved spectral index maps, they are sufficient to provide us with integrated spectral indices. For the remaining maps we used NVSS data at 1400~MHz, either from \citet{yun_01a}, or integrating the NVSS map \citep{condon_98a}. We computed integrated spectral indices using the low-resolution flux densities and the flux density $S_\text{ref}$ from the reference data at $\nu_\text{ref}\approx 1400~\rm MHz$ as:
\begin{equation}
    \alpha_{\rm low}\ = \frac{\log\left( \frac{S_{20\prime\prime}}{S_{\rm ref}} \right)}{\log\left( \frac{144~\rm MHz}{\nu_{\rm ref}} \right)}.
    \label{eq:integrated_spectral_index}
\end{equation}
The low-resolution data were chosen as they have a higher surface brightness sensitivity and so have more reliable integrated flux densities. Table~\ref{tab:spix} contains the resulting integrated radio spectral indices. These values are presented in Table~\ref{tab:spix}. We note that LOFAR data, and LoTSS data in particular, have been extensively compared with with random field sources in order to give confidence that overall the spectral indices between LoTSS and higher frequency surveys are reliable \citep{de_gasperin_18a,shimwell_19a}.

We also calculated high-frequency radio spectral indices using our $L$ band flux densities and $C$ band flux densities from the literature. These high-frequency spectral indices, which are calculated between $\approx$1400 and $\approx$5000~MHz, are referred to as $\alpha_\text{high}$ (Table~\ref{tab:spix}). These data are taken either from the 6000-MHz flux densities presented in \citet{wiegert_15a}, which are sensitive enough for the fairly small galaxies, so that missing spacings play no role for the interferometric data. The remaining flux densities were mostly single-dish data obtained at 4850~MHz with either the 100-m Effelsberg telescope \citep{tabatabaei_17a} or with the 110-m Green Bank Telescope \citep{gregory_91a}.

\subsubsection{Spatially resolved spectral index maps}
\label{sss:spatially_resolved_spectral_index_maps}

We also determined radio spectral index maps for a subset of galaxies. We used both $C$-band data from CHANG-ES survey \citep{wiegert_15a} at 6000~MHz and maps from the WSRT--SINGS survey \citep{braun_07a} at 1365~MHz. For two galaxies, instead of the 6000-MHz data from \citet{wiegert_15a}, we used the NGC~891 map from \citet{schmidt_19a} and the NGC~4631 map from \citet{mora_19a}. The reason is that these maps are merged with single-dish data and so have reliable flux densities. In total, we could compute spectral index maps for 30 galaxies.

 Prior to combination, at both frequencies, maps were masked below $3\sigma$. The error map of the spectral index was calculated assuming a systematic map error of $\epsilon_{\rm map}=0.05$ for both LoTSS and the reference data:
\begin{equation}
    \label{eq:results:spectral_indices:spatially_resolved_index_maps:error_map_formula}
    \sigma_{\alpha,\rm low} \ = \ \frac{\sqrt{\left( \frac{{\sigma}}{I_{\nu}} \right)^2 + \left( \frac{\sigma_{\rm ref}}{I_{\rm ref}} \right)^2 + 2\epsilon_{\rm map}^2}}{\log \left( \frac{\nu_{\rm ref}}{\nu} \right)}.
\end{equation}
This error underestimates the absolute error, but since in spectral index maps we are more interested in the spatial variation, this smaller error, dominated by the uncertainties in the deconvolution, is a good estimate. We present in Table~\ref{tab:spix} the mean radio spectral index in these maps within the flux integration areas and the corresponding standard deviation.

\subsection{Image atlas}
\label{ss:image_atlas}
\begin{figure*}
	\includegraphics[width=\textwidth]{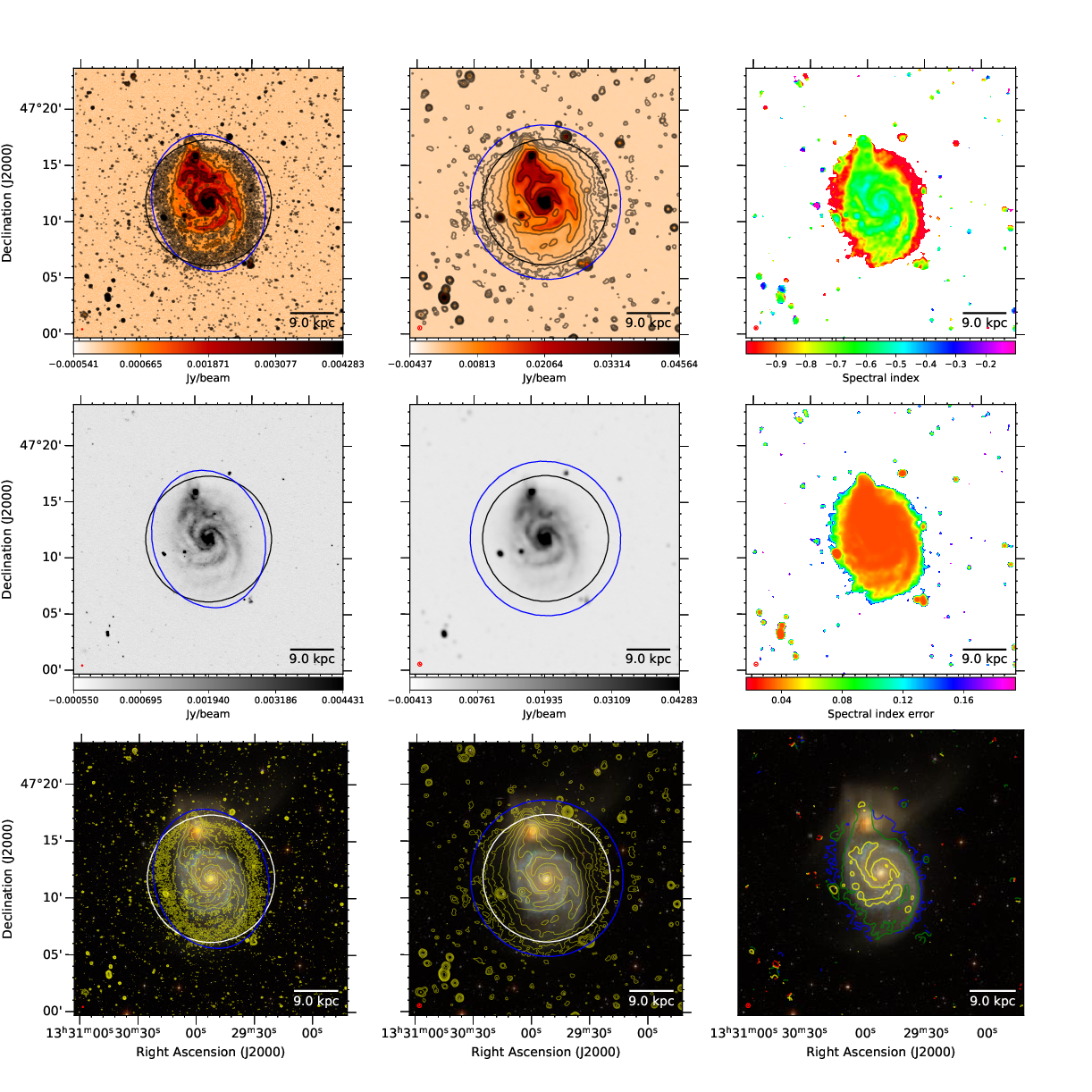}
    \caption{NGC~5194 (M~51). The first row shows from left to the right the 150-MHz map at 6 and 20~arcsec resolution, respectively, and the radio spectral index between 150 and 1400 MHz at 20 arcsec resolution. Contours start at $3\sigma$ with increments of powers of two. The black circle shows the optical $D_{25}$ diameter and the blue ellipses show the $3\sigma$ contour extent used to integrate the flux density. The second row shows again the 6 and 20~arcsec images without contours and the radio spectral index error. The third row shows the 6~ and 20~arcsec contours overlaid on a rgb SDSS image as well as contours of of the radio spectral index. Contours are at $-1.0$ (blue), $-0.8$ (green), $-0.6$ (yellow) and $-0.4$ (red). The filled circle shows the synthesised beam. The scale bar shows the projected size at the distance of the galaxy.}
    \label{fig:n5194}
\end{figure*}

Figure~\ref{fig:n5194} shows the maps of NGC~5194 (M~51). We present the LoTSS-DR2 intensity maps as contour and greyscale maps at both resolutions, and include the spectral index maps as well. The maps are also shown overlaid on the $rgb$ SDSS image. The full image atlas can be found in Appendix~\ref{as:atlas}.

\section{Results}
\label{s:results}

\begin{table*}
	\centering
	\caption{Radio spectral indices $\alpha_{\rm low}$ between 144~MHz and the reference frequency and $\alpha_{\rm high}$ between the two reference frequencies.}
	\label{tab:spix}
        \begin{tabular}{l cccc cccc} 
          \hline
          
          Galaxy & $\nu_{\rm ref}$ & $S_{\rm ref}$ & Ref. & $\alpha_{\rm low}$ & $\nu_{\rm ref,2}$ & $S_{\rm ref,2}$ & Ref. & $\alpha_{\rm high}$ \\
          & (MHz) & (Jy) & & & (MHz) & (Jy)\\\hline
          NGC 598     & 1400 & $1.99 \pm 0.10$     & T07  & $-0.34\pm 0.05$  &  4850  & $1.06   \pm 0.05   $ & T07 & $-0.51\pm 0.06$ \\
          NGC 855     & 1400 & $0.006\pm 0.001$    & C98  & $-0.22\pm 0.09$  &  4850  & $0.0032 \pm 0.0007 $ & T17 & $-0.51\pm 0.22$ \\
          NGC 891     & 1575 & $0.7432\pm 0.0105$  & W15  & $-0.62\pm 0.04$  &  6000  & $0.2520 \pm 0.027  $ & S19 & $-0.81\pm 0.08$ \\
          NGC 925     & 1365 & $0.078\pm 0.008$    & B07  & $-0.51\pm 0.06$  &  10700 & $0.038  \pm 0.006  $ & N95 & $-0.35\pm 0.09$ \\
          NGC 2683    & 1575 & $0.068\pm 0.007$    & W15  & $-0.70\pm 0.06$  &  6000  & $0.0203 \pm 0.0008 $ & W15 & $-0.90\pm 0.08$ \\
          NGC 2798    & 1400 & $0.086\pm 0.008$    & Y01  & $-0.42\pm 0.06$  &  4850  & $0.0338 \pm 0.0025 $ & T17 & $-0.75\pm 0.10$ \\
          NGC 2820    & 1575 & $0.0764\pm 0.0012$  & W15  & $-0.51\pm 0.04$  &  6000  & $0.0191 \pm 0.0004 $ & W15 & $-1.04\pm 0.02$ \\
          NGC 2841    & 1365 & $0.097\pm 0.007$    & B07  & $-0.73\pm 0.05$  &  4850  & $0.038  \pm 0.004  $ & T17 & $-0.74\pm 0.10$ \\
          NGC 2976    & 1365 & $0.072\pm 0.005$    & B07  & $-0.46\pm 0.05$  &  4850  & $0.039  \pm 0.003  $ & T17 & $-0.48\pm 0.08$ \\
          NGC 3003    & 1575 & $0.0349\pm 0.0041$  & W15  & $-0.60\pm 0.06$  &  6000  & $0.0122 \pm 0.0008 $ & W15 & $-0.79\pm 0.10$ \\
          NGC 3031    & 1400 & $0.624\pm 0.031$    & WB92 & $-0.52\pm 0.05$  &  4850  & $0.28931\pm 0.014  $ & B85 & $-0.62\pm 0.06$ \\
          NGC 3079    & 1575 & $0.808\pm 0.016$    & W15  & $-0.64\pm 0.04$  &  6000  & $0.3654 \pm 0.0073 $ & W15 & $-0.59\pm 0.02$ \\
          NGC 3077    & 1400 & $0.030\pm 0.003$    & Y01  & $-0.35\pm 0.06$  &  4850  & $0.023  \pm 0.001  $ & T17 & $-0.21\pm 0.09$ \\
          NGC 3184    & 1365 & $0.089\pm 0.005$    & B07  & $-0.64\pm 0.05$  &  4850  & $0.028  \pm 0.003  $ & T17 & $-0.91\pm 0.10$ \\
          NGC 3198    & 1365 & $0.039\pm 0.005$    & B07  & $-0.52\pm 0.07$  &  4850  & $0.012  \pm 0.001  $ & T17 & $-0.93\pm 0.12$ \\
          NGC 3265    & 1400 & $0.011\pm 0.001$    & Y01  & $-0.39\pm 0.06$  &  4850  & $0.0057 \pm 0.0006 $ & T17 & $-0.53\pm 0.11$ \\
          Mrk 33      & 1400 & $0.017\pm 0.002$    & C98  & $-0.52\pm 0.07$  &  N/A   & N/A                  & N/A & N/A             \\
          NGC 3432    & 1575 & $0.0833\pm 0.0019$  & W15  & $-0.54\pm 0.04$  &  6000  & $0.0263 \pm 0.0005 $ & W15 & $-0.86\pm 0.02$ \\
          NGC 3448    & 1575 & $0.0465\pm 0.0009$  & W15  & $-0.60\pm 0.04$  &  6000  & $0.0205 \pm 0.004  $ & W15 & $-0.61\pm 0.15$ \\
          NGC 3556    & 1575 & $0.2909\pm 0.0058$  & W15  & $-0.60\pm 0.04$  &  6000  & $0.0792 \pm 0.0047 $ & W15 & $-0.97\pm 0.05$ \\
          NGC 3877    & 1575 & $0.0423\pm 0.0009$  & W15  & $-0.58\pm 0.04$  &  6000  & $0.0129 \pm 0.0003 $ & W15 & $-0.89\pm 0.02$ \\
          NGC 3938    & 1365 & $0.082\pm 0.005$    & B07  & $-0.71\pm 0.05$  &  4850  & $0.0263 \pm 0.0015 $ & T17 & $-0.90\pm 0.07$ \\
          NGC 4013    & 1575 & $0.0378\pm 0.0008$  & W15  & $-0.58\pm 0.04$  &  6000  & $0.0126 \pm 0.0003 $ & W15 & $-0.82\pm 0.02$ \\
          NGC 4096    & 1575 & $0.0560\pm 0.0011$  & W15  & $-0.53\pm 0.04$  &  6000  & $0.0163 \pm 0.0003 $ & W15 & $-0.92\pm 0.02$ \\
          NGC 4125    & 1365 & $0.0018\pm 0.0002$  & B07  & $-0.84\pm 0.06$  &  N/A   & N/A                  & N/A & N/A             \\
          NGC 4157    & 1575 & $0.1845\pm 0.0037$  & W15  & $-0.72\pm 0.04$  &  6000  & $0.0551 \pm 0.0011 $ & W15 & $-0.90\pm 0.02$ \\
          NGC 4214    & 1400 & $0.062\pm 0.005$    & Y01  & $-0.31\pm 0.06$  &  6000  & $0.02316\pm 0.00009$ & H18 & $-0.68\pm 0.06$ \\
          NGC 4217    & 1575 & $0.1116\pm 0.0022$  & W15  & $-0.58\pm 0.04$  &  6000  & $0.0354 \pm 0.0007 $ & W15 & $-0.86\pm 0.02$ \\
          NGC 4244    & 1575 & $0.0165\pm 0.0006$  & W15  & $-0.46\pm 0.04$  &  6000  & $0.009  \pm 0.0006 $ & W15 & $-0.45\pm 0.06$ \\
          NGC 4449    & 1400 & $0.30\pm 0.03$      & C98  & $-0.54\pm 0.06$  &  4850  & $0.139  \pm 0.007  $ & C00 & $-0.62\pm 0.09$ \\
          NGC 4559    & 1365 & $0.081\pm 0.01$     & B07  & $-0.48\pm 0.07$  &  4850  & $0.038  \pm 0.003  $ & T17 & $-0.60\pm 0.12$ \\
          NGC 4625    & 1400 & $0.009\pm 0.003$    & Y01  & $-0.26\pm 0.15$  &  4850  & $0.0031 \pm 0.0003 $ & T17 & $-0.89\pm 0.27$ \\
          NGC 4631    & 1575 & $1.134\pm 0.037$    & W15  & $-0.59\pm 0.04$  &  4850  & $0.515  \pm 0.04   $ & M19 & $-0.70\pm 0.07$ \\
          NGC 4725    & 1365 & $0.054\pm 0.005$    & B07  & $-0.62\pm 0.06$  &  4850  & $0.03   \pm 0.002  $ & T17 & $-0.46\pm 0.09$ \\
          NGC 4736    & 1365 & $0.31\pm 0.02$      & B07  & $-0.47\pm 0.05$  &  4850  & $0.125  \pm 0.01   $ & T17 & $-0.72\pm 0.08$ \\
          NGC 5033    & 1365 & $0.22 \pm 0.01$     & B07  & $-0.73\pm 0.05$  &  4850  & $0.079  \pm 0.007  $ & G96 & $-0.81\pm 0.08$ \\
          NGC 5055    & 1365 & $0.42 \pm 0.01$     & B07  & $-0.75\pm 0.05$  &  4850  & $0.167  \pm 0.008  $ & T17 & $-0.73\pm 0.04$ \\
          NGC 5194    & 1365 & $1.41 \pm 0.01$     & B07  & $-0.67\pm 0.04$  &  4850  & $0.42   \pm 0.08   $ & T17 & $-0.96\pm 0.15$ \\
          NGC 5195    & 1365 & $0.20\pm 0.02$      & B07  & $-0.72\pm 0.06$  &  4850  & $0.054  \pm 0.005  $ & F11 & $-1.03\pm 0.11$ \\
          NGC 5297    & 1575 & $0.0238 \pm 0.0005$ & W15  & $-0.67\pm 0.04$  &  6000  & $0.0067 \pm 0.0004 $ & W15 & $-0.95\pm 0.05$ \\
          NGC 5457    & 1400 & $0.808\pm 0.08$     & WB92 & $-0.61\pm 0.06$  &  4850  & $0.31   \pm 0.002  $ & T17 & $-0.77\pm 0.08$ \\
          NGC 5474    & 1400 & $0.012\pm 0.01$     & C98  & $-0.25\pm 0.06$  &  4850  & $0.005  \pm 0.0006 $ & T17 & $-0.70\pm 0.12$ \\
          NGC 5866    & 1400 & $0.023\pm 0.002$    & Y01  & $-0.46\pm 0.06$  &  4850  & $0.0121 \pm 0.0008 $ & T17 & $-0.52\pm 0.09$ \\
          NGC 5907    & 1575 & $0.1187\pm 0.0036$  & W15  & $-0.67\pm 0.04$  &  6000  & $0.0515 \pm 0.001  $ & W15 & $-0.62\pm 0.03$ \\
          NGC 7331    & 1365 & $0.42\pm 0.02$      & B07  & $-0.72\pm 0.05$  &  4850  & $0.1738 \pm 0.0087 $ & T17 & $-0.70\pm 0.05$ \\
          \hline
        \end{tabular}
        \flushleft
              {\small {\bf References.}\\
            B07: \citet{braun_07a}; B85: \citet{beck_85a}; C00: \citet{chyzy_00a}; C98: \citet{condon_98a}; F11: \citet{fletcher_11a}; G96: \citet{gregory_96a}; H18: \citet{hindson_18a}; M19: \citet{mora_19a}; N95: \citet{niklas_95a}; S19: \citet{schmidt_19a}; T07: \citet{tabatabaei_07a}; T17: \citet{tabatabaei_17a}; W15: \citet{wiegert_15a}; WB92: \citet{white_92a}; Y01: \citet{yun_01a}.}
\end{table*}

\begin{figure}
    \includegraphics[width=1.0\columnwidth]{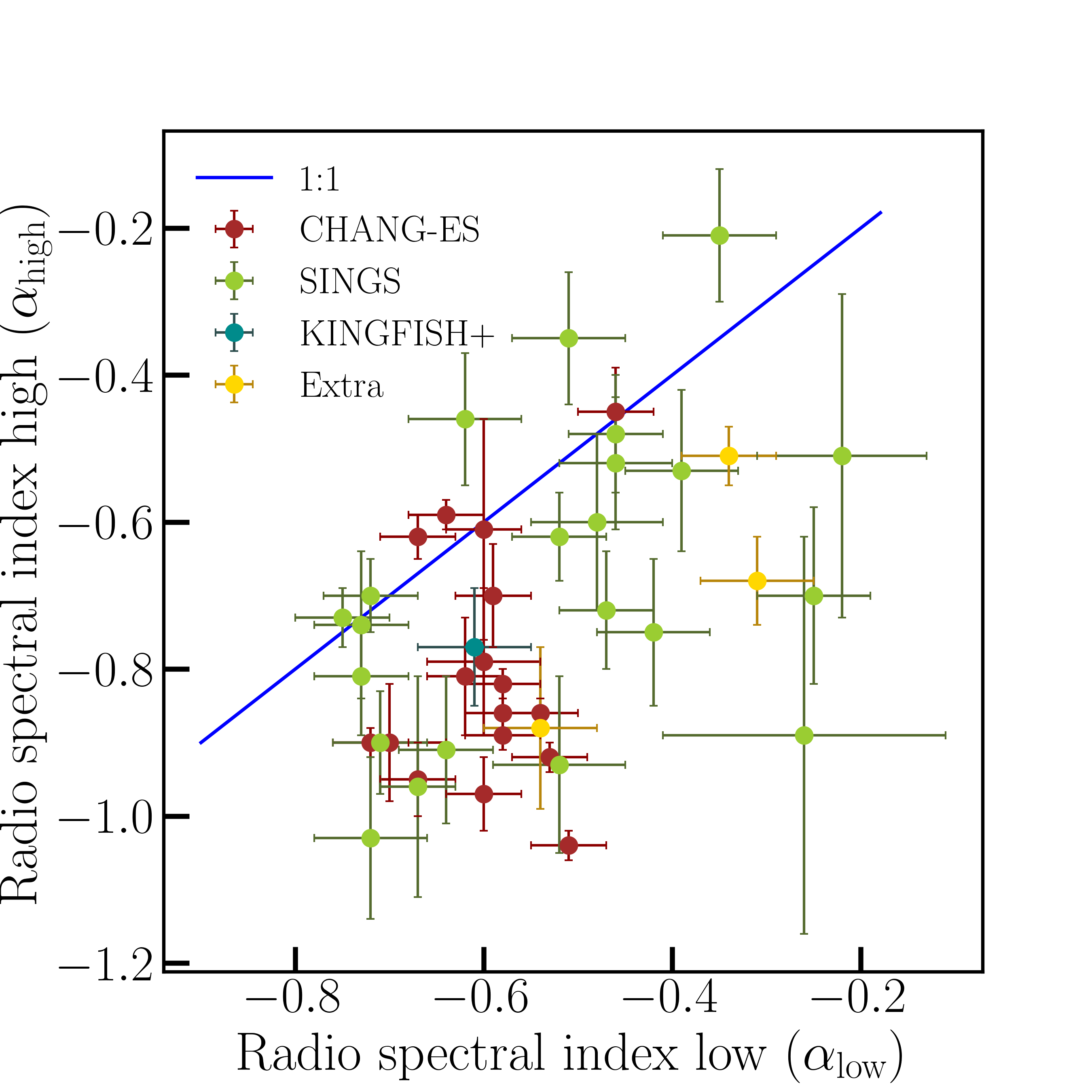}
    \caption{Radio colour--colour diagram with a comparison of the high frequency radio spectral index $\alpha_\text{high}$ (1400--5000 MHz) with the low-frequency radio spectral index $\alpha_\text{low}$ (144--1400~MHz) taken from Table~\ref{tab:spix}. The blue solid line indicates points of equal spectral index, indicating a power-law radio continuum spectrum. Most of the data points lie below the blue line indicating spectra with a convex curvature. Data points above the blue line indicate concave spectra, as may be explained by a higher thermal fraction at high frequencies.}
    \label{fig:spix_low_high}
\end{figure}

\begin{figure*}
    \centering
    \begin{subfigure}[t]{0.02\textwidth}
    \textbf{(a)}
  \end{subfigure}
      \begin{subfigure}[t]{0.47\linewidth}
    \includegraphics[width=1.0\linewidth,valign=t]{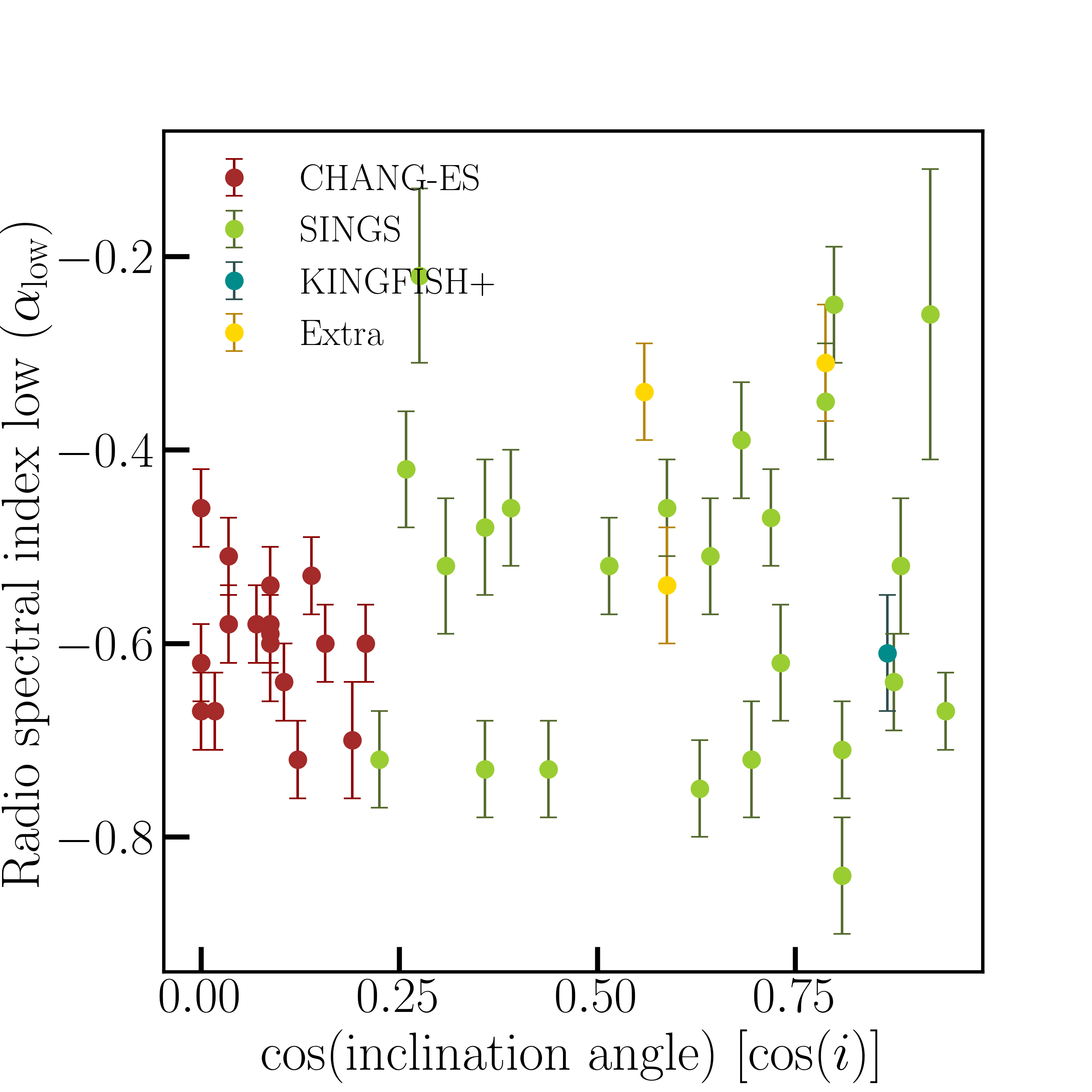}
    \end{subfigure}
    \begin{subfigure}[t]{0.02\textwidth}
    \textbf{(b)}
    \end{subfigure}
    \begin{subfigure}[t]{0.47\linewidth}
    \includegraphics[width=1.0\linewidth,valign=t]{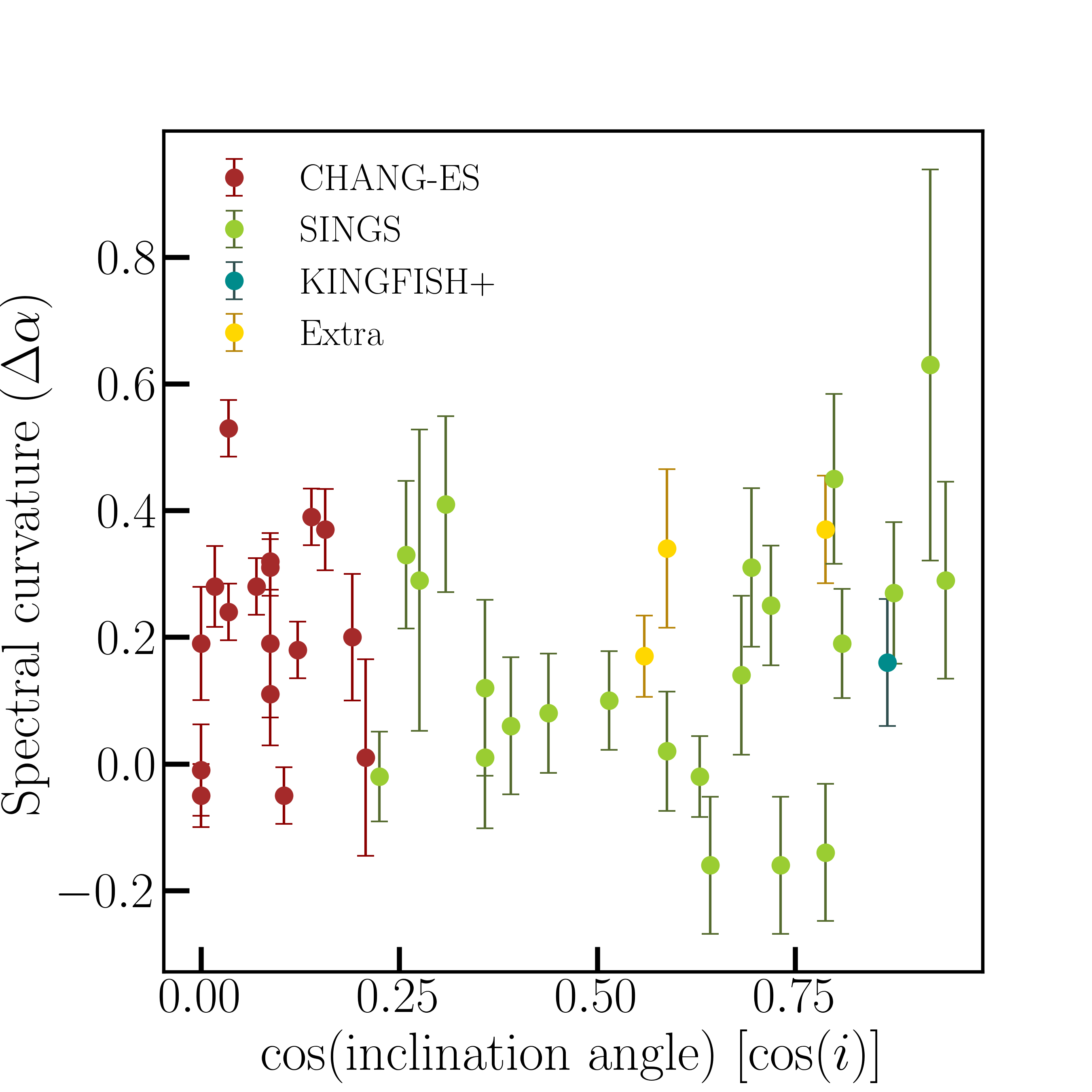}
    \end{subfigure}
    \caption{Radio spectral index as function of the inclination angle. Panel (a) shows the integrated radio spectral index between 144 and 1400~MHz as a function of $\cos(i)$. Panel (b) shows the spectral curvature $\Delta\alpha = \alpha_{\rm low} - \alpha_\text{high}$ as function of $\cos(i)$. The high frequency radio spectral index $\alpha_\text{high}$ is defined between 1400 and 5000~MHz. Both spectral index measurements show at most only a weak correlation with the inclination angle.}
    \label{fig:spix_incl}
\end{figure*}

\subsection{The low-frequency radio spectral index}
\label{ss:radio_spectral_index}

At the low frequencies of LoTSS, we expect the radio continuum emission to be dominated by non-thermal (synchrotron) emission. For synchrotron emission from CRE with a distribution $N\propto E^{-p}$, the observed radio spectral index is $\alpha=(1-p)/2$. When spectral ageing is important, the CRE spectrum steepens to $N\propto E^{-p-1}$ resulting in a spectral index of $\alpha=p/2$ \citep{longair_11a}. Hence, for CRE with an injection spectrum, we expect $\alpha\approx -0.5$, which is corroborated by both observations and theory. For example, the average injection spectral index of cosmic-ray energy spectra is measured from $\gamma$-ray emission of young supernova remnants to be $p=-2.3 \pm 0.3$ \citep[see fig.~4 in][]{mandel_15}, resulting in a non-thermal radio spectral index of $\alpha_{\rm inj}\approx -0.65$, consistent with models of diffusive shock acceleration \citep{bell_78a}. The CRE spectrum is also constrained by radio spectral index measurements of supernova remnants, which provide a value of $\alpha_{\rm inj}= -0.5\pm 0.2$ \citep{reynolds_12a}. Some remnants show a low-frequency ($<100$~MHz) spectral turnovers, but some do not. This is usually explained with an inhomogeneous ionized medium in the Galactic foreground \citep{Kassim_1989a}.

The non-thermal radio spectral index is altered by several processes which broadly fall into two categories. First, the observed radio continuum emission of a galaxy is affected by radiation transfer. At low frequencies, both free--free absorption \citep[e.g][]{carilli_1996,adebahr_13a} and synchrotron self-absorption \citep[e.g.][]{carilli_1991,irwin_15a,callingham_2015,Kapinska_2017} can lower the radio continuum emission, thus flatten the radio spectrum. The latter is in particular associated with AGN, although the spectrum may be better explained by a low-energy CRE cut-off \citep{lazio_2006,mckean_2016}. Second, loss processes of the CRE, which have various dependencies on energy, will change radio spectrum. Both ionization losses and bremsstrahlung flatten the radio continuum spectrum to $\alpha\approx -0.2$ in dense star-formation regions \citep[e.g.][]{basu_15a}. In contrast, both inverse Compton (IC) and synchrotron radiation depends on the energy of CRE, so that the radio spectrum is steepened to $\alpha\approx -0.7$ if the synchrotron loss time-scale of $\sim 100~\rm Myr$ is similar to the CRE escape time-scale. For the case where a galaxy is an \textit{electron calorimeter} with no CRE escape, $\alpha=\alpha_{\rm inj}-0.50\approx -1.1$ \citep[e.g.][]{lisenfeld_96a}. Finally, adiabatic losses and escape of the CRE will result in an unchanged injection spectrum \citep[e.g.][]{krause_18a}. These various processes we have to disentangle in order to understand the low-frequency radio continuum spectrum.

The mean integrated radio spectral index and standard deviation in our sample are $\langle \alpha_{\rm low}\rangle \ = \ -0.55 \pm 0.14$. Several galaxies in our sample have spectral indices that are lower than the injection spectral index. Most noticeable among them are NGC 855 and 5474, which have almost flat spectra with spectral indices of $-0.22\pm 0.09$ and $-0.25\pm 0.06$, respectively. Both galaxies are ellipticals and so have low star formation rates and may host AGN, which are likely affected by absorption, which could explain their flat spectra. Late-type spiral and irregular galaxies also have spectral indices $\alpha_{\rm low} > -0.5$, which are NGC 598, NGC 2798, NGC~3077, NGC~3265, NGC 4214 and NGC~4625.  Only a very large fraction of thermal radiation could, in theory at least, flatten the spectrum. While the thermal contribution is hard to estimate for our galaxies without radio data beyond 10\,GHz, the thermal fraction is expected to be only at a few per cent level at the LoTSS frequencies \citep[e.g.,][]{basu12a, tabatabaei_17a, klein18}. Hence, thermal emission is an insufficient explanation for the galaxies with significant deviations from the injection spectrum, which leaves absorption effects and CRE energy losses as the most likely cause.

\citet{basu_15a} modelled the relative importance of the various energy CRE loss mechanisms in galaxy discs and showed that, at 100~MHz, bremsstrahlung and ionization losses are of similar importance and dominate over synchrotron losses. An integrated radio spectral index in agreement with the injection spectral index may be the result of a superposition of both spectral flattening and steepening, such as by a flat radio spectrum in the gaseous disc and a steep radio spectrum in the halo \citep{adebahr_13a}. Of course, the spectrum could be explained also by a fast escape of the CRE, where energy losses due to escape and adiabatic expansion dominates over synchrotron and IC radiation losses \citep{krause_18a}. However, this would mean small radiation losses due to synchrotron emission and so the galaxies would be underluminous in the radio continuum, as it is the case in dwarf irregular galaxies \citep{hindson_18a}. This is not observed, however, as the galaxies are in good agreement with the radio--SFR relation even at the low frequencies of LoTSS \citep{gurkan_18a, smith_21a}. We conclude that synchrotron losses of the CRE are still important. We will return to this point in more detail in Section~\ref{s:discussion}.

\subsection{Spectral curvature}
\label{ss:spectral_curvature}

If both spectral flattering at low frequencies and spectral steepening at high frequencies occurs (Section~\ref{ss:radio_spectral_index}), the resulting spectrum is {\it convex} and has spectral curvature. Figure~\ref{fig:spix_low_high} shows the high-frequency spectral indices as function of our previously calculated low-frequency spectral indices between 144 and 1400~MHz. It can be seen that the $\alpha_\text{low}$-values are mostly smaller than the $\alpha_{\rm high}$-values. If we define the spectral curvature as $\Delta \alpha = \alpha_{\rm low}-\alpha_{\rm high}$, we find a mean value of $\langle \Delta \alpha \rangle = 0.20 \pm 0.18$. This indicates a convex spectral curvature, in agreement with previous studies \citep{marvil_15a,chyzy_18a}.

In order to distinguish free--free absorption from CRE bremsstrahlung and ionisation losses, we now investigate the effect of the inclination angle and thus the length of the line-of-sight on the spectral index \citep{israel_90a}. We show in Figure~\ref{fig:spix_incl}(a) the radio spectral index as a function of $\cos(i)$, where $i$ is the inclination angle. If free--free absorption was important, we would expect to see a flattening of the spectrum for high inclination angles (see also \citet{chyzy_18a}). Clearly, the expected increase of $\alpha$ at low values of $\cos (i)$ is not observed; at best a weak trend is visible (Spearman's rank correlation coefficient of $\rho_{\rm s} = 0.29$), where the spectral index seems to be increasing with $\cos(i)$, opposite to the expected trend. However, there is a bias for massive galaxies to have high inclination angles caused by the CHANG-ES sample. They have intrinsically steeper spectra since they are more massive (we will return to this in Section~\ref{ss:semi_calorimetric_radio_sfr_relation}). 

Hence, in order to take out such possible dependence of the spectral index, we also study the dependence of spectral curvature on the inclination angle. Since free--free absorption depends strongly on frequency, the low frequency radio spectral index should be affected, whereas the high-frequency one should be unchanged. As Fig.~\ref{fig:spix_incl}(b) shows, the spectral curvature has no correlation with $\cos(i)$ either ($\rho_{\rm s}=0.08$). \citet{chyzy_18a} modelled the low-frequency radio continuum emission of galaxies and showed that spectral flattening towards lower frequencies cannot be solely due to free--free absorption. We note that inclination dependence alone is not a sufficient argument to rule out free--free absorption since in highly inclined galaxies, a mix of absorption in the inner disc, and steep spectrum in the outer disc and halo (due to older CRE) could compensate. This claim requires more rigorous demonstration, for instance by observing at 57~MHz \citep{de_gasperin_21a}.

\begin{figure*}
    \centering
    \begin{subfigure}[t]{0.02\textwidth}
    \textbf{(a)}
  \end{subfigure}   
    \begin{subfigure}[t]{0.47\linewidth}
    \includegraphics[width=1\linewidth,valign=t]{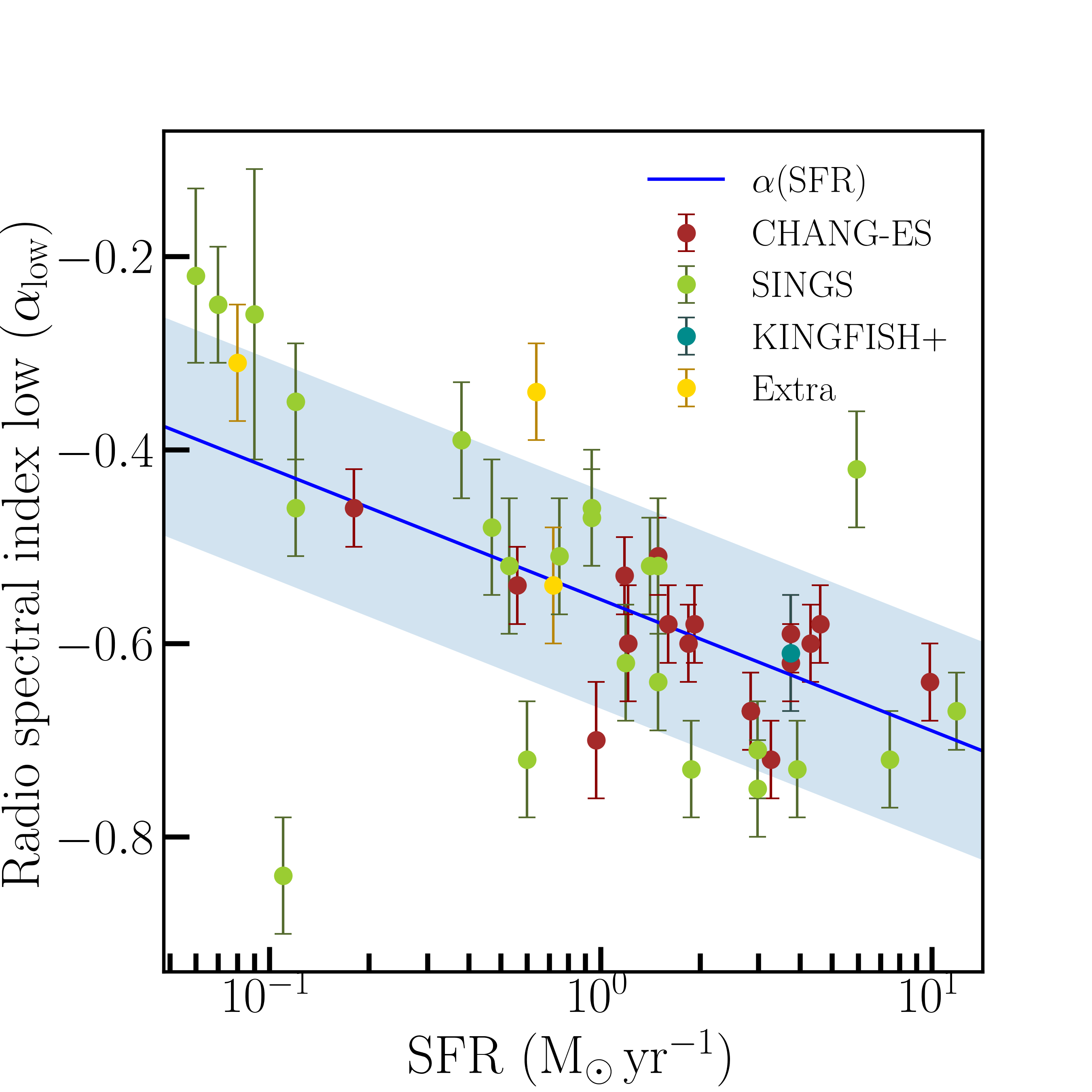}
    \end{subfigure}
    \begin{subfigure}[t]{0.02\textwidth}
    \textbf{(b)}
  \end{subfigure}   
    \begin{subfigure}[t]{0.47\linewidth}
    \includegraphics[width=1\linewidth,valign=t]{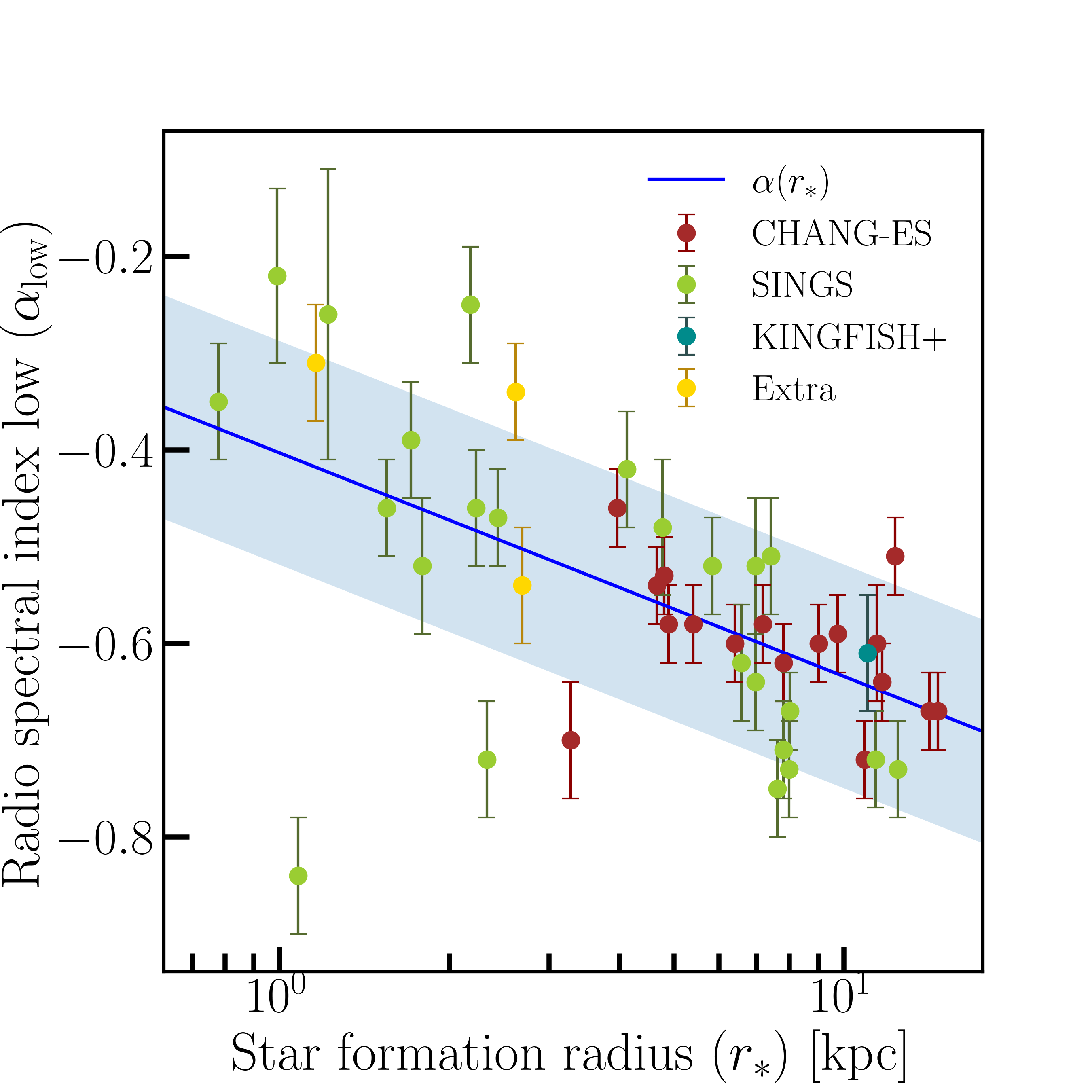}
    \end{subfigure}
    \\
    \begin{subfigure}[t]{0.02\textwidth}
    \textbf{(c)}
  \end{subfigure}   
    \begin{subfigure}[t]{0.47\linewidth}
    \includegraphics[width=1\linewidth,valign=t]{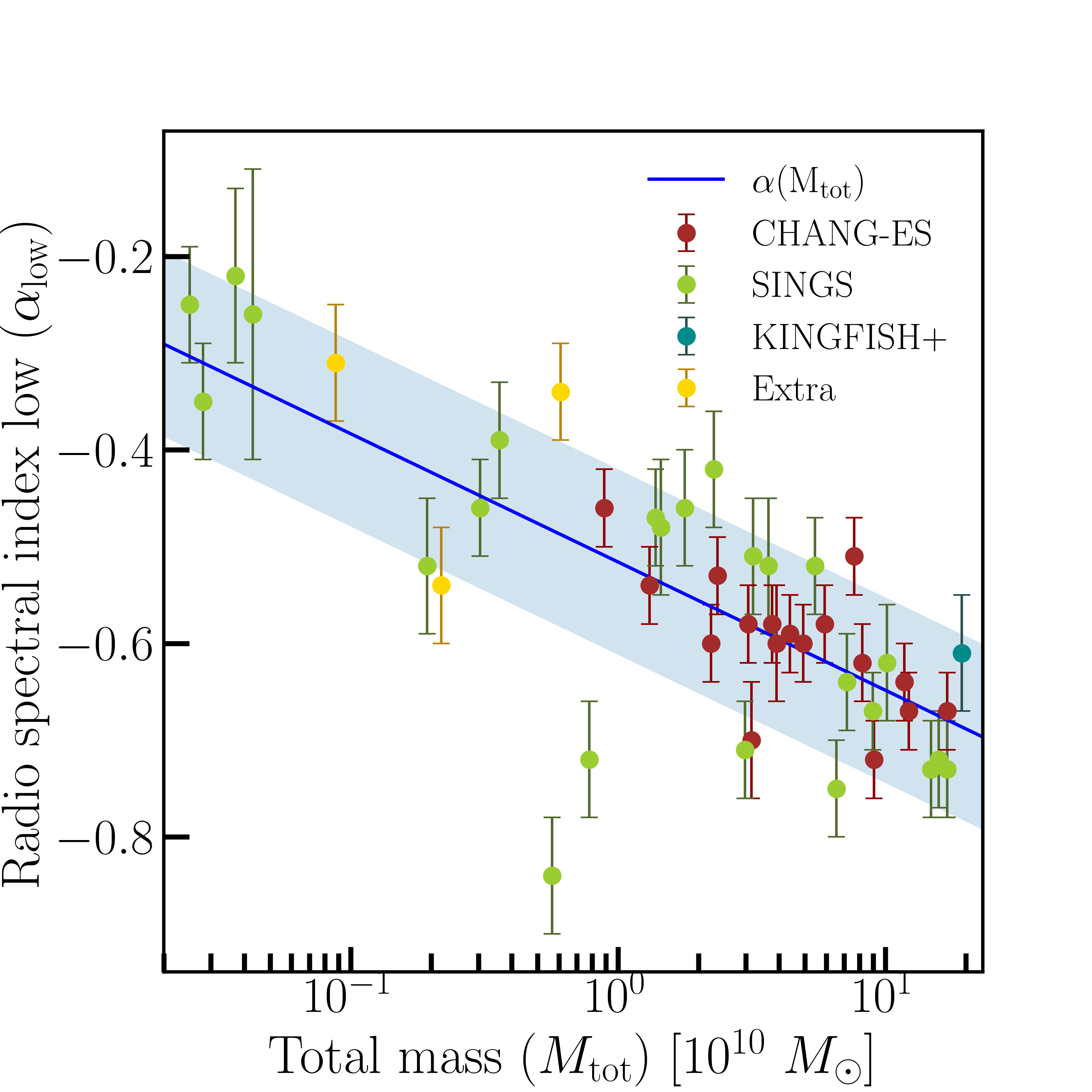}
    \end{subfigure}
    \begin{subfigure}[t]{0.02\textwidth}
    \textbf{(d)}
  \end{subfigure}   
    \begin{subfigure}[t]{0.47\linewidth}
    \includegraphics[width=1\linewidth,valign=t]{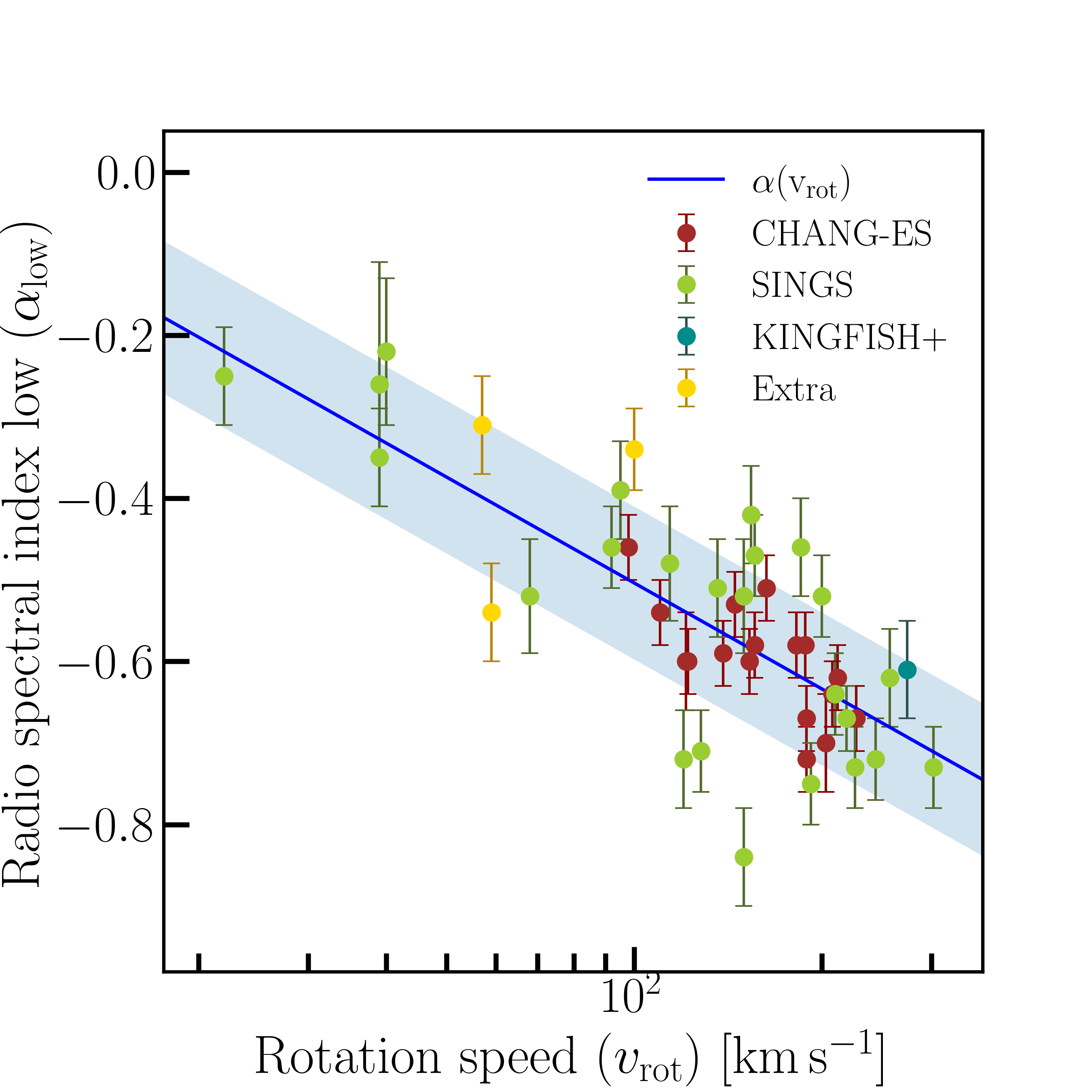}
    \end{subfigure}
    \caption{The radio spectral index as function of several fundamental galaxy parameters. Panel (a) shows the dependence on SFR as calculated from the TIR luminosity; (b) shows the dependence of the star formation radius, which is derived from the extent of the 3-$\sigma$ contours in the 20~arcsec LoTSS images; (c) shows the dependence on the total mass calculated; (d) shows the dependence on the rotation speed. In all panels, the blue line is the best-fitting relation and the shaded areas show $1\sigma$ confidence intervals. The colour of the data points in panel (b) indicates the rotation velocity, in the other panels the different sub-samples.}
    \label{fig:alpha_sfr}
\end{figure*}

\subsection{Radio spectral index and galaxy properties}
\label{ss:radio_spectral_index_properties}

In this section, we compare the radio spectral index with several fundamental galaxy parameters. Our galaxies span a wide range in mass, size and star-formation activity, the influence of which we now explore. In Fig.~\ref{fig:alpha_sfr}(a), we show the radio spectral index as a function of the SFR. A clear trend is visible ($\rho_{\rm s}=-0.61$), with the best-fitting relation:
\begin{equation}
\alpha_\text{low} = (-0.136\pm 0.026)\log_{10}({\rm SFR/M_{\sun} yr^{-1} }) -(0.555\pm 0.015).
\label{eq:alpha_sfr}
\end{equation}

For higher SFRs, the radio continuum spectra become steeper. Similar results were obtained by \citet{marvil_15a}, who studied 250 bright galaxies and measured mean spectral indices of $-0.45$ between 74\,MHz and 325\,MHz, $-0.55$ between 325\,MHz and 1.4\,GHz, and $-0.69$ between 1.4\,GHz and 4.85\,GHz. The spectra between 325\,MHz and 1.4\,GHz were found to become steeper with increasing radio luminosity and hence with increasing SFR.

Naively, one may expect galaxies with lower SFRs to exhibit steeper spectra due to the ceased acceleration and subsequent ageing of CREs. However, Fig.~\ref{fig:alpha_sfr}(a) shows the opposite trend where steeper spectra are found for galaxies with higher SFRs. Explanations for this effect could be the combination of CRE escape, free--free absorption, the fraction of thermal emission, and CRE energy losses. While the escaping CRE leave mostly the flat injection spectrum visible, absorption would further flatten this spectrum which may explain the observed spectral indices $\alpha>-0.5$. A high fraction of thermal emission could have a similar effect which is especially likely in dwarf galaxies \citep[e.g.,][]{basu_17a}, but also in other galaxies with very low surface SFR (below about $10^{-4}\,\rm M_{\sun}\,kpc^{-2}\,yr^{-1}$), as these are deficient in injecting CREs to produce synchrotron radiation, leaving thermal radiation as a noticeable contribution to the spectrum \citep{schleicher_16a}. However, all galaxies of our sample have larger values of surface SFR, so that dominant thermal emission cannot explain the trend seen in Fig.~\ref{fig:alpha_sfr}(a). This leaves the competition between escape and synchrotron energy loss of CRE as the probable reason for the dependence of spectral index on SFR.

In Fig~\ref{fig:alpha_sfr}(b), we show the radio spectral index as function of the size of star-forming disc. This correlation is significant ($\rho_{\rm s}=-0.62$):
\begin{equation}
\alpha_\text{low} = (-0.231\pm 0.047)\log_{10}(r_{\star}/{\rm kpc}) -(0.403\pm 0.039).
\label{eq:alpha_radius}
\end{equation}
The radio spectral index depends also clearly on total mass enclosed within the star-forming disc (Fig.~\ref{fig:alpha_sfr}(c)) with ($\rho_{\rm s}=-0.70$):
\begin{equation}
\alpha_\text{low} = (-0.133\pm 0.019)\log_{10}(M_{\rm tot}/{10^{10}~\rm M_{\sun}}) -(0.516\pm 0.015).
\label{eq:alpha_mass}
\end{equation}
Similarly, we find a strong dependence on rotation speed ($\rho_{\rm s}=-0.70$)
as shown in Fig.~\ref{fig:alpha_sfr}(d): 
\begin{equation}
\alpha_\text{low} = (-0.432\pm 0.062)\log_{10}(\rm{v_{\rm rot}}/{\rm km\,s^{-1}}) + (0.359\pm 0.134).
\label{eq:alpha_vrot}
\end{equation}

It is unclear whether there is a primary relation that drives the remaining correlations. For instance, larger galaxies with large values of $r_{\star}$ tend to be also to be more massive, have larger values of SFR as well, and rotate faster. The correlation between mass and SFR is well known, and is usually referred to as the `main sequence of galaxies' \citep[][compare with Fig.~\ref{fig:sample}(b)]{brinchmann_04a}. Similarly, there is a mass-size relation for star-forming galaxies \citep{van_der_wel_14a}, where larger galaxies have also a larger stellar mass. Also, there is a size-velocity relation, where galaxies with larger sizes have increased rotation speeds \citep{meurer_18a}. Thus if size were the primary driver of the spectral index variation, the other relations would follow.

Ordered magnetic fields could be an additional factor why the radio spectral index depends on rotation speed. 
As \citet{tabatabaei_16a} have shown, the strength of the ordered magnetic field in galaxy discs correlates with the rotation speed. Diffusion and streaming of cosmic rays is enhanced along the field lines, so that CREs may be stored for longer in fast-rotating galaxies.

\subsection{Spatially resolved radio spectral index and galaxy morphology}
\label{ss:spatially_resolved_spectral_index}

 In the radio spectral index maps, we found in many late-type spiral galaxies regions with $\alpha_\text{low} \gtrsim -0.4$, which are clearly flatter than the injection spectrum. Such regions are nearly always compact and cover only a small fraction of the galaxies, with the exception of NGC~2976 and NGC~4713 where they are more widely distributed. In the moderately inclined galaxies, these areas appear to be mostly co-spatial with star-formation regions, whereas for most of the edge-on galaxies, these regions are co-spatial with the nucleus. Also, in the edge-on galaxies, the spectral index is relatively flat in the galactic mid-plane with $\alpha_\text{low}\gtrsim -0.7$, but then decreases at larger distances from the mid-plane, reaching values of $-0.9\lesssim \alpha_\text{low}\lesssim -0.7$, with even steeper spectra ($\alpha_\text{low} \approx -1.1$) found far away from the disc.

The galaxies which are not classified as late-type spirals, with classifications as Sa, S0, E, P, or Ir, have visibly different radio continuum emission. They have integrated spectral indices of $\alpha\approx -0.3$. Since their radio continuum emission is compact, they may be more affected by free--free absorption and CRE bremsstrahlung and ionization losses and CRE escape. The only notable exception is NGC~4125, which has the steepest spectrum in our sample ($\alpha_\text{low}=-0.84$). There appears to be little or no star formation, corroborated by the relatively red appearance of the optical image. Possibly, star formation has ceased and with it the injection of young CRE, causing strong spectral ageing. Most likely though, the emission is caused by the weak AGN, which is classified as a transition object between H\,{\sc ii} and LINER object \citep{ho_97a}.

\begin{figure*}
    \centering
     \begin{subfigure}[t]{0.02\textwidth}
    \textbf{(a)}
  \end{subfigure}   
     \begin{subfigure}[t]{0.47\linewidth}
    \includegraphics[width=1.0\linewidth,valign=t]{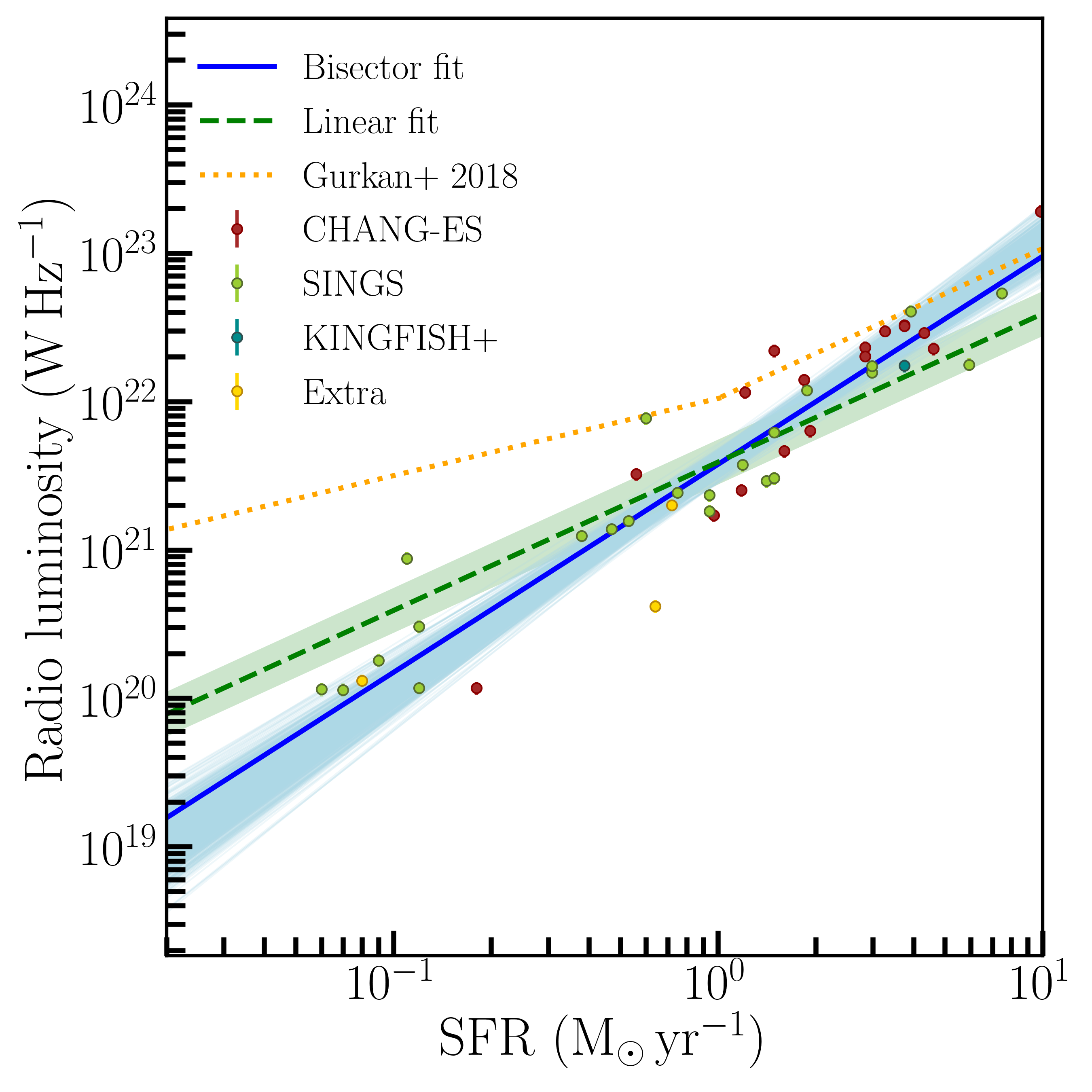}
    \end{subfigure}
    \begin{subfigure}[t]{0.02\textwidth}
    \textbf{(b)}
  \end{subfigure}   
    \begin{subfigure}[t]{0.47\linewidth}
     \includegraphics[width=1.0\linewidth,valign=t]{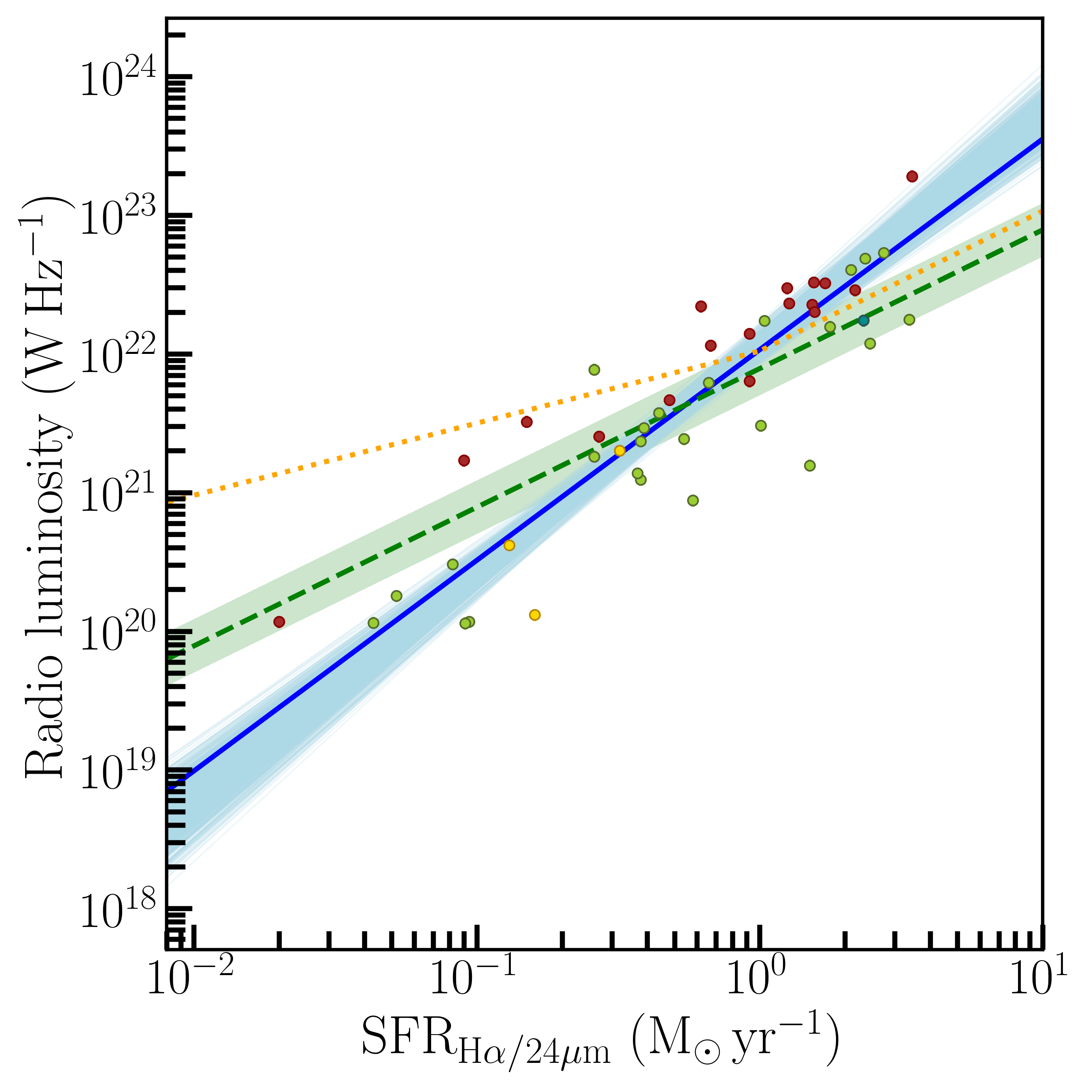}
     \end{subfigure}
    \caption{Integrated radio--SFR relation at 144~MHz. Blue lines indicate the best-fitting relations using a bisector fitting method as fitted in the log--log space. The dashed green line shows the best-fitting linear relation as fitted in linear space. Shaded bands represent the 
    $3\sigma$ confidence intervals from the 1000 Monte-Carlo
    realizations. In the left panel, the SFR is calculated from the TIR luminosity; in the right panel, the SFR is calculated either using the hybrid H\,$\alpha+24\,\mu$m SFR calibration or using a $24~\mu\rm m$ only calibration. For both SFR calibrations, the radio--SFR relation is clearly non-linear.}
    \label{fig:radio_sfr}
\end{figure*}

\subsection{Radio--SFR relation}
\label{ss:radio_sfr_relation}

In this section, we study the integrated radio--SFR relation at 144~MHz. We use the low-resolution flux densities ($S_{20\arcsec}$) measured from the 20-arcsec maps to calculate the spectral radio luminosity ($L_{144\,\rm MHz} = 4\pi d^2 S_{20\arcsec} $) as:
\begin{equation}
        \left(\dfrac{L_{144\,\rm MHz}}{\rm W\,Hz^{-1}}\right) \ = 
        1.197 \times 10^{20} \left( \dfrac{d}{\rm Mpc} \right )^2 \left( \dfrac{S_{20\arcsec}}{\rm Jy} \right),
\end{equation}
where we have neglected the $K$-correction since our redshifts are low with $z\lesssim 0.01$. These spectral luminosities are to be compared to the SFR as listed in Table~\ref{tab:sample}, which were calculated from the total infrared luminosities (Section~\ref{ss:sample}). We also used the SFRs calculated from H\,$\alpha$ and 24-$\mu\rm m$ emission (Section~\ref{ss:sample}) as listed in Table~\ref{tab:flux}.

Figure~\ref{fig:radio_sfr} shows both resulting plots for the integrated radio--SFR relation. We expect a scatter of at least $0.18$~dex since that is the accuracy of the integrated SFR value from either method \citep{leroy_12a}. The scatter of the data points is slightly larger than that, as expected, and a clear general non-linear trend is found.  We applied a least-square bisector fit, where it is best practice to fit the relation log--log space, which is be equivalent to fitting $\log(L_{\rm 144\,MHz}) = \log\,a + b\log({\rm SFR})$. Below are the results of bisector fit \citep{isobe90} for radio--SFR, with SFR determined using TIR: 
\begin{equation}
    \label{eq:radio_sfr}
 \left( \frac{L_{\rm 144\,MHz}}{\rm W\,Hz^{-1}} \right)  = 10^{(21.576 \pm 0.044)} \left( \frac{\rm SFR}{\rm M_{\sun}\, yr^{-1}} \right)^{(1.402 \pm 0.072)}.
\end{equation}
For this best-fitting relation, we find $\chi_{\nu}^2 = 0.9$ and a rms scatter of $\sigma = 0.27~\rm dex$. We assess the confidence intervals of derived model parameters using the Markov chain Monte Carlo (MCMC) method. If we force a linear fit, we find the following relation:
\begin{equation}
    \left( \frac{L_{\rm 144\,MHz}}{\rm W\,Hz^{-1}} \right)  = 10^{(21.61 \pm 0.05)} \left( \frac{\rm SFR}{\rm M_{\sun}\, yr^{-1}} \right),
\end{equation}
this time with $\chi_{\nu}^2 = 2.1$. Similarly, below are the results of bisector fit for radio--SFR, with SFR determined using H$\alpha + 24\,\mu$m:
\begin{equation}
  \left( \frac{L_{\rm 144\,MHz}}{\rm W\,Hz^{-1}} \right)  = 10^{(22.030 \pm 0.060)} \left( \frac{\rm SFR_{H\alpha + 24\mu m}}{\rm M_{\sun}\, yr^{-1}} \right)^{(1.519 \pm 0.110)},
\end{equation}
with $\chi_{\nu}^2 = 1.8$ and $\sigma = 0.38~\rm dex$. If we force a linear fit, we find:
\begin{equation}
    \left( \frac{L_{\rm 144\,MHz}}{\rm W\,Hz^{-1}} \right)  = 10^{(21.91 \pm 0.06)} \left( \frac{\rm SFR_{H\alpha + 24\mu m}}{\rm M_{\sun}\, yr^{-1}} \right),
\end{equation}
with $\chi_{\nu}^2 = 17.4$. These best-fitting non-linear and linear relations are presented in Fig.~\ref{fig:radio_sfr} with their respective $\pm 1\sigma$ confidence intervals. For the H\,$\alpha$ + 24-$\mu\rm m$ derived SFRs, the scatter is slightly larger with $0.38~\rm dex$.  As Fig.~\ref{fig:radio_sfr}(b) shows, the CHANG-ES data points tend to mostly lie above the best-fitting non-linear relation. This means their radio luminosity is too large for a given SFR, or, as would argue, their SFR measurement is too low. This has been for the case of the CHANG-ES galaxies investigated by \citet{vargas_18a} who conclude that the 24-$\mu$m MIR emission becomes partially optically thick in edge-on galaxies \citep[see also][]{li_16a}. Hence, in what follows we only study the relation further using the TIR SFR measurements.

Now, we compare our results with previous studies of the low-frequency radio--SFR relation using LOFAR. \citet{gurkan_18a} studied approximately 1000 star-forming galaxies at 150~MHz using photometric star formation rates in a red shift range of $0\lesssim z \lesssim 0.3$. They also found a super-linear relation with a $L_{150} / {\rm W\,Hz^{-1}}=10^{22.06\pm 0.01}({\rm SFR/M_{\sun}\,yr^{-1})}^{1.07\pm 0.01}$. They also posited that the stellar mass is a second parameter in the relation, with higher masses having a higher radio luminosity. These results were largely confirmed by an even larger sample of galaxies at red shifts of $z<1$ by \citet{smith_21a}, with an intrinsic scatter around the relation of $\approx 0.3$~dex, similar to our results.

The previous studies and our data show that it is possible to establish a robust radio--SFR relation for low radio frequencies. Of course, the scatter with $0.27$\,dex exceeds the margin of error of the data which is the flux density uncertainty of 10 per cent (0.041\,dex). Hence, our scatter represents the intrinsic scatter of the radio--SFR relation. The scatter of $0.27$\,dex is nearly on par with the best state-of-art star formation tracers which have an uncertainty of 50 per cent, equivalent to 0.18\,dex \citep{leroy_12a}. What is clear now is that a significantly super-linear slope of the integrated radio--SFR relation. This can also be seen in Fig.~\ref{fig:radio_sfr}, where the linear fit is shown as a green dashed line. It is dominated  by the data points at lower SFRs, with most points at higher SFRs lying above the curve, deviating from it much more than a few standard deviations.  We do not expect this super-linear slope to be the result of contribution from AGN. The mean AGN luminosity for spiral galaxies is $\nu L_{\rm AGN} = 10^{36}~\rm erg\,s^{-1}$, which is equivalent to a 144-MHz spectral luminosity of $\log_{10}(L_{\rm AGN}/{\rm W\,Hz^{-1}})=20.5$ \citep[assuming $\alpha=-0.7$;][]{baldi_21a}. Hence, the contribution from AGN to the radio continuum should be sub-dominant except possibly at the lowest SFRs.

\section{Discussion}
\label{s:discussion}


\subsection{The non-linear radio--SFR relation}
\label{ss:the_non-linear_radio-sfr_relation}

The low frequency radio--SFR relation is well represented by a slightly super-linear fit (Section~\ref{ss:radio_sfr_relation}), consistent with earlier studies of LOFAR data with significantly larger samples by both \citet{gurkan_18a} and \citet{smith_21a}. This implies that the radio emission is enhanced in comparison to injection of CRE linearly proportional to the star-formation rate,
if the model of a CRE calorimeter applies \citep[][see below]{voelk_89a}. \citet{gurkan_18a} suggested, based on the stellar mass dependence of the radio--SFR relation, that the escape of the CRE from galaxies plays a role as well. Galaxies with higher stellar masses tend to be larger and so take longer to escape from the galaxy, boosting radio continuum luminosity.

The slope of the radio--SFR relation of (in logarithmic scales) gives important information about the propagation of CREs and the relation between CRs and magnetic fields in galaxies. A linear slope indicates that CREs lose all their energy within the galaxy, so that the radio luminosity depends only on the density of CREs, not on field strength \citep[CRE calorimeter;][]{voelk_89a}. A super-linear relation can be interpreted as a signature of energy equipartition between CRs and magnetic fields \citep{niklas_97a}, where the slope $b$ depends on the synchrotron spectral index $\alpha$ and on the exponents of the relation between magnetic field and gas density ($B \propto\rho^\kappa$) and of the global Kennicutt--Schmidt law (SFR~$\propto\rho^n$). Using $\alpha\approx -0.7$ (Section~\ref{ss:radio_spectral_index}) and generic values of $\kappa\approx 0.5$ and $n=1.4\pm0.15$ \citep{kennicutt_98a} yields $b\approx 1.3$. A sub-linear radio--SFR relation is expected under non-equipartition conditions, e.g. if the CRs diffuse rapidly away from their places of origin in star-forming regions \citep{berkhuijsen_13a,heesen_19a}.
The slope of the radio--SFR correlation shown in Fig.~\ref{fig:radio_sfr} is in excellent agreement with the equipartition case.

The super-linear radio--SFR relation found here is consistent with the findings of \citet{smith_21a}, for a large sample of galaxies at redshifts below 1. These authors also found a super-linear relation between SFR and radio luminosity, though with a slope closer to one, between the pure equipartition case ($b\approx 1.3$) and the pure calorimeter case ($b=1.0$). The actual conditions in the galaxies of the sample seem to be between these two cases. Galaxies with higher SFRs, in particular LIRGs and ULIGRS, have stronger magnetic fields, which should bring them nearer to the calorimeter case, as argued by \citet{li_16a}. Indeed, the mean synchrotron spectral index of a sample of 19 starburst galaxies is $\alpha \approx -1.06$ \citep{galvin_18a}.

\citet{smith_21a} proposed, from the mass dependence of the radio/SFR ratio as seen in their data, that for lower mass galaxies CREs escape the galaxy fast, thus removing energy from the galaxy instead of radiating it as synchrotron radiation. This is in agreement with the findings of this work, as the previous results have also already advocated for the loss of cosmic rays by advection and winds \citep{gurkan_18a}. The mass dependence might also be an explanation for the high scatter in the data points: \citet{smith_21a} elaborate that higher-mass galaxies have a higher luminosity at 150~MHz for a fixed SFR. It is possible that the scatter between our points in Fig.~\ref{ss:radio_sfr_relation} is a result of large differences in mass at a given SFR.

It has been noted by \citet{li_16a} that the slope of the radio--SFR relation is frequency-dependent. \citet{li_16a} find a slope of $1.13\pm 0.07$ at $1.6$~GHz, but only a slope of $1.06\pm 0.08$ at 6 GHz. Similarly, \citet{hindson_18a} find only a slope of $0.93\pm 0.14$ in their sample of dwarf irregular galaxies at 6 GHz. Hence, both relations are in agreement with linearity at 6~GHz. In contrast, \citet{heesen_14a} found a slope of $1.24\pm 0.04$ at 1.4~GHz. In galaxies out to a redshift of $1.2$, \citet{basu_15a} found a slope of $1.11\pm 0.04$ at 1.4~GHz. However, \citet{davies_17a} find slopes of $1.3$ and $1.5$ at 1.4~GHz for TIR + ultra-violet and photometric SFRs, respectively. We can test this trend for our sample, exploiting the radio spectral index information. Using a standard least-square fit, we find slopes of $1.34\pm 0.06$ at 144~MHz, $1.16\pm 0.06$ at $1.4$~GHz, and $1.08\pm 0.06$ at 5~GHz. A flattening of the radio--SFR relation at 1.4~GHz in comparison to 144~MHz across one sample was also found by \citet{gurkan_18a}. We propose that the flattening of the relation is caused by CRE propagation (Section~\ref{ss:cosmic_ray_transport}).

\begin{figure}
    \centering
    \includegraphics[width=1.0\columnwidth]{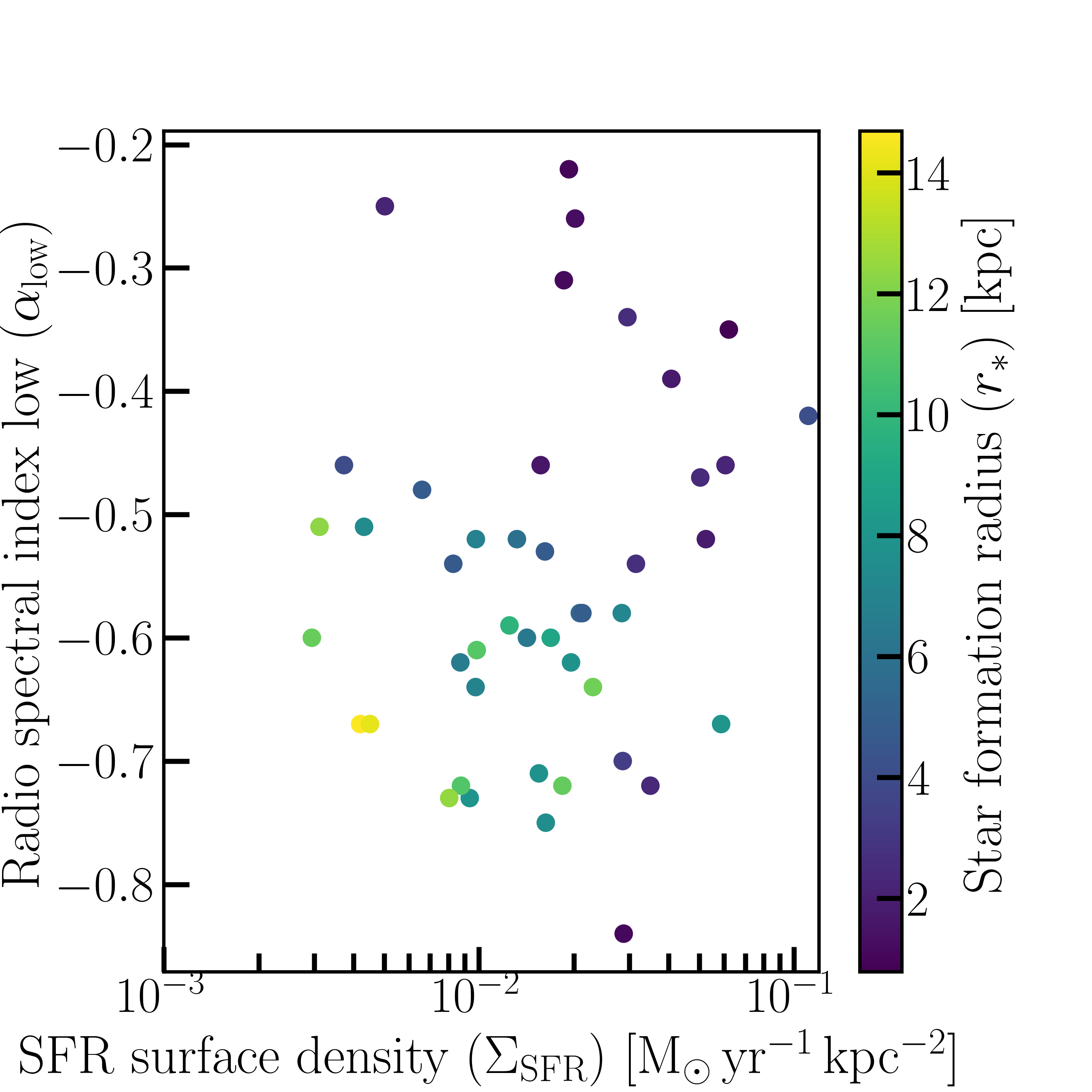}
    \caption{Radio spectral index as function of the SFR surface density $\Sigma_{\rm SFR}=SFR/(\pi r_{\star}^2)$. Data points are coloured which indicates the radius of the star-forming disc $r_{\star}$. The radio spectral index depends only weakly on $\Sigma_{\rm SFR}$ but significantly on $r_{\star}$ (compare with Fig.~\ref{fig:alpha_sfr}(b)). This is in agreement with escape of CRs by advection in a wind (see Eq.~\ref{eq:advective_escape}).}
    \label{fig:alpha_sfrd}
\end{figure}

\subsection{The influence of energy losses and cosmic ray transport}
\label{ss:cosmic_ray_transport}

The findings in Section~\ref{ss:the_non-linear_radio-sfr_relation} suggest that energy-dependent CRE escape may be behind the non-linearity of the radio--SFR relation. The best proxy for the importance of energy losses for the CRE is the radio spectral index.

We now review which conditions regulate the radio spectral index. \citet{krause_18a} have shown that the radio continuum scale height is related to the radius $r_{\star}$ of the radio disc, 
similar the radius of the star-forming disc $r_{\star}$, as $h=0.05\,r_{\star}$.  For advective transport the escape time is hence $t_{\rm esc} \propto h / {\rm v} \propto r_{\star}/{\rm v}$, with $\rm v$ the outflow speed. The synchrotron lifetime scales as $t_{\rm syn} \propto \nu^{-0.5} B^{-1.5} \propto \nu^{-0.5}{\rm \Sigma_{SFR}}^{-0.5}$. As the wind speed increases with ${\rm SFR}$ as ${\rm v} \propto \Sigma_{\rm SFR}^{0.4}$ \citep{heesen_18b,heesen_21a}, the ratio of escape time-scale to synchrotron time-scale is:
\begin{equation}
    \frac{t_{\rm esc}}{t_{\rm syn}} \propto \nu^{-0.5}\, r_\mathrm{\star} \, \Sigma_{\rm SFR}^{0.1} \propto \nu^{-0.5}\, {\rm SFR}^{0.1} r_{\star}^{0.8}.
    \label{eq:advective_escape}
\end{equation}
For galaxy which is a CRE calorimeter, $t_{\rm esc}\gg t_{\rm syn}$, all the CRE energy is radiated away via synchrotron emission and IC emission. Contrary, for a highly non-calorimetric galaxy, $t_{\rm esc}\ll t_{\rm syn}$, meaning almost free CRE escape. The ratio $t_{\rm esc}/t_{\rm syn}$ can be related to the (non-thermal) radio spectral index $\alpha$, since we expect  $\alpha=\alpha_{\rm inj}$ in the non-calorimetric case, meaning for a fast CRE escape the injection radio spectral index is observed. Contrary, in the calorimetric case when synchrotron radiation losses dominate, the radio spectral index is $\alpha = \alpha_{\rm inj}-0.5$ \citep{lisenfeld_00a}. 
Hence, Eq.~\eqref{eq:advective_escape} predicts strong dependence of the radio spectral index on the radius with steeper spectra at larger radii as indeed found in Fig.~\ref{fig:alpha_sfr}(b). On the other hand, there is no correlation between the radio spectral index and the SFR surface density ($\rho_{\rm s}=0.16$) as shown in Fig.~\ref{fig:alpha_sfrd}. Again, this is in good agreement with the prediction by Eq.~\eqref{eq:advective_escape}.

\subsection{Semi-calorimetric radio--SFR relation}
\label{ss:semi_calorimetric_radio_sfr_relation}

As shown in this paper, the radio--SFR relation has a slope (in logarithmic scales) slightly above one, whereas results from previous works reveal some variation. The radio--SFR relation is has a slope in agreement with unity in several cases, including spatially resolved observations \citep{dumas_11a}. While this could be explained with a calorimetric radio--SFR relation, our spectral indices in agreement with the injection spectrum argue against such a picture, although this might be accurate in starburst galaxies \citep{galvin_18a}.
The radio spectral index variation suggests that more massive galaxies with steeper spectral indices are better electron calorimeters than less massive galaxies.
Radio continuum luminosity is not only driven by SFR, the dependence on radio spectral index indicates that another physical parameter is required. One way to is to include the calorimetric synchrotron fraction $\eta$ of the CRE, so that the semi-calorimetric radio--SFR relation, which takes CRE escape into account, becomes \citep[e.g.][]{pfrommer_21a}\footnote{Where we have neglected the small super-linearity that \citet{pfrommer_21a} have predicted due to the changing magnetic field strength of their model as function of the SFR}:
\begin{equation}
    L_{144} \propto \eta {\rm SFR}.
    \label{eq:semi_calorimetric_radio_sfr_model}
\end{equation}
The calorimetric fraction is closely related to the radio spectral index, which itself is a function of star-forming radius, rotation speed, and total mass (Section~\ref{ss:cosmic_ray_transport}). 

Hence, we may parametrise $\eta$ with the total mass,  as the second parameter of the `semi-calorimetric' radio--SFR relation, as this is a quantity is closely related to the stellar mass, which can be fairly easily measured from photometric measurements even for high-$z$ galaxies. The parametrisation of \citet{gurkan_18a} is then used where assume a linear dependence on the SFR:
\begin{equation}
    L_{144} = L_C \left (\frac{\rm SFR}{1~\rm M_{\sun}\, yr^{-1}} \right ) \left (\frac{M_{\rm tot}}{10^{10}~\rm M_\sun}\right )^{\gamma}
    \label{eq:semi_calorimetric_radio_SFR}
\end{equation}
results in $\gamma=0.25\pm 0.05$ and a normalisation of $L_C=10^{21.518\pm 0.044}~\rm W\,Hz^{-1}$. We have also attempted to fit a non-linear SFR dependence, as \citet{gurkan_18a} have done, however this does not result in a significantly improved fit. The parametrisation of Eq.~\eqref{eq:semi_calorimetric_radio_SFR} takes hence CRE calorimetric fraction into account in a simple way. The CRE calorimetric fraction is the fraction of CRE energy that is radiated via synchrotron emission before the CRE are able to escape from the galaxy.

We now try a heuristic derivation of the found relation. For this we assume the calorimetric fraction the CRE as $\eta=t_{\rm CRE}/t_{\rm syn}$, where $t_{\rm CRE}$ is the effective CRE lifetime, considering both escape and synchrotron losses, and $t_{\rm syn}$ is the synchrotron lifetime. In our model, the effective CRE lifetime is then:
\begin{equation}
    \frac{1}{t_{\rm CRE}}=\frac{1}{t_{\rm syn}}+\frac{1}{t_{\rm esc}}.
    \label{eq:effective_lifetime}
\end{equation}
It can be shown that for $t_{\rm esc}\sim t_{\rm syn}$, which is a reasonable assumption for galaxies as they fall in between the calorimetric and the non-calorimetric case \citep{lacki_10a}, we have:
\begin{equation}
    \eta = \frac{1}{\frac{t_{\rm syn}}{t_{\rm esc}}+1} \approx \frac{1}{2}\sqrt{\frac{t_{\rm esc}}{t_{\rm syn}}}.
    \label{eq:calorimetric_fraction}
\end{equation}
For $\eta\approx 0.5$, the approximation in Eq.~\eqref{eq:calorimetric_fraction} is  valid.\footnote{Eq.~\eqref{eq:calorimetric_fraction} can be expressed as $\eta=x/(1+x)$ with $x=t_{\rm esc}/t_{\rm syn}$. The first Taylor expansion around $x\approx 1$ leads to $\eta\approx 1/2(1+x/2)$. The second Taylor expansions then gives $\eta\approx \sqrt{x}/2$.} With $M_{\rm tot}\propto r_{\star}^{1/3}$ \citep[e.g.][]{mo_98a}, we obtain using Eq.~\eqref{eq:advective_escape}:
\begin{equation}
    \eta \propto {\rm SFR}^{0.05} M_{\rm tot}^{0.27}.
\end{equation}
This is in good agreement with the found  mass dependency of the radio--SFR relation of Eq.~\eqref{eq:semi_calorimetric_radio_SFR}. Obviously, such a derivation can be only approximate. Numerical models by \citet{werhahn_21a} support the semi-calorimetric model, although their calorimetric synchrotron fractions of the CRE may be as small as $\eta \approx 10^{-2}$ due to low magnetic field strengths. We note that by combining Eqs.~\eqref{eq:semi_calorimetric_radio_sfr_model} and \eqref{eq:effective_lifetime} we can also explain a small super-linearity of the radio--SFR relation. Our found mass dependence is slightly smaller than that of \citet{gurkan_18a} and \citet{smith_21a}, who find $\gamma >0.3$.

\begin{figure}
    \centering
    \includegraphics[width=\columnwidth]{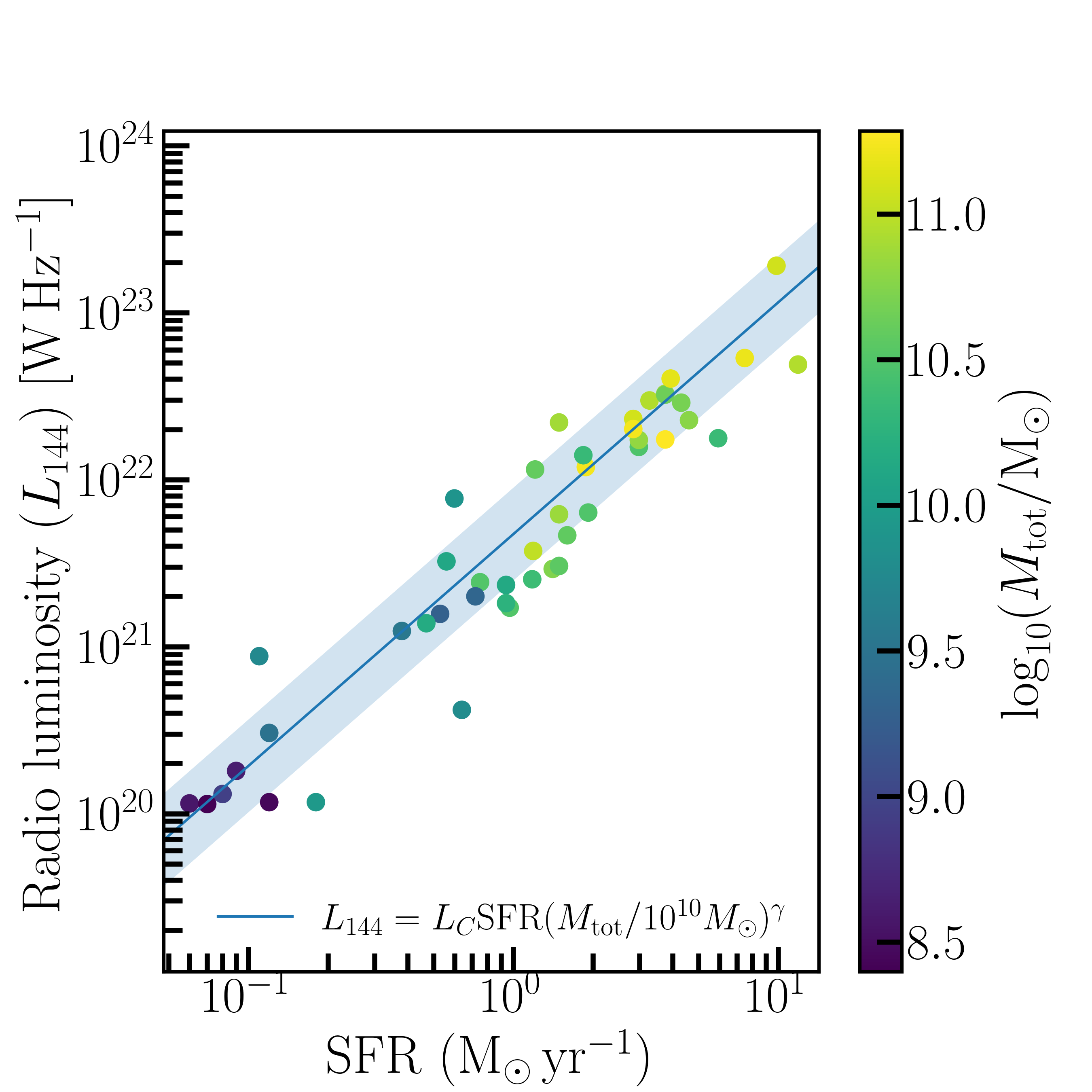}
    \caption{Semi-calorimetric model fitted to the integrated radio--SFR relation at 144~MHz. The blue line shows the best-fitting semi-calorimetric relation as as defined by Eq.~\eqref{eq:semi_calorimetric_radio_SFR}. The shaded area indicates $\pm 1\sigma$. Data points are coloured according to the total mass of the galaxies.}
    \label{fig:radio_sfr_mass}
\end{figure}

In Fig.~\ref{fig:radio_sfr_mass}, we show the radio--SFR relation when using the parametrisation of Eq.~\eqref{eq:semi_calorimetric_radio_SFR}. As can be seen, the super-linear relation is well described. The advantage of such a semi-calorimetric model is that it can both explain super-linear radio--SFR relations particularly at low frequencies and nearly linear relations at high frequencies (Section~\ref{ss:the_non-linear_radio-sfr_relation}). In essence, at low frequencies the radiation efficiency has a larger relative variation compared to higher frequencies, resulting in a super-linear relation, whereas the radio--SFR relation at higher frequencies is closer to a linear relation. In contrast, at high frequencies the radiation efficiency is so high that only a small variation is expected. This means the radio--SFR relation is closer to linear. Similarly, if the samples are restricted to galaxies with a fairly uniform distribution of spectral indices, then the relation will be close to be linear as well.

\subsection{Equipartition magnetic fields}
\label{ss:equipartition_magnetic_fields}

Even though in our non-calorimetric model the magnetic field strength does not enter the formula for the radio continuum luminosity directly, the magnetic field still plays an important role by determining the CRE lifetime. Also, the magnetic field might regulate the escape of cosmic rays by diffusion, streaming and advection in a wind. Since there are good reasons to believe that galaxies are generally in energy equipartition, we now try to establish which conditions do then generally have to hold. For energy equipartion to hold, the cosmic ray energy has to be equivalent to the magnetic energy. In a stationary state the cosmic-ray injection rate $\dot E_{\rm CR}$ and the escape time-scale $t_{\rm esc}$ determine the total cosmic-ray energy:
\begin{equation}
    \dot E_{\rm CR} t_{\rm esc} = u_{\rm B} V = \frac{B^2}{8\pi}\pi r^2 2h,
\end{equation}
where $u_{\rm B}=B^2/(8\pi)$ is the magnetic energy density, $V=2\pi r^2h$ is the volume of the galaxy approximated by a cylinder, and $h$ is the cosmic-ray scale height. The cosmic-ray lifetime is set by the escape time only since we neglect any cosmic ray losses other than escape. With $t_{\rm esc}=h/{\rm v}$, we obtain:
\begin{equation}
    \epsilon_{\rm CR} \, \Sigma_{\rm SFR} = u_{\rm B}\, \rm{v},
\end{equation}
where we have used the assumption that the cosmic-ray injection rate is proportional to the SFR $\dot E_{\rm CR}=\epsilon_{\rm CR} {\rm SFR}$ with $\epsilon_{\rm CR}=(1$--$3)\times 10^{40}~\rm erg\,s^{-1}\,M_{\sun}^{-1}\,yr$ \citep{socrates_08a}. If energy equipartition holds, then $u_{\rm B}=u_{\rm CR}$ and this equation states that in a steady state the cosmic ray injection is balanced by the escape of cosmic rays in a wind. We can now express the equipartition magnetic field strength as:
\begin{equation}
    B = \left (4\epsilon_{\rm CR} \frac{\Sigma_{\rm SFR}}{\rm{v}}\right )^{1/2}.
\end{equation}
Observations indicate that $\rm{v}\propto \Sigma_{\rm SFR}^{1/3}$, which is what is measured using galaxies viewed from an edge-on orientation with spectral ageing models in radio haloes \citep{heesen_18b}. So then we can derive $B\propto \Sigma_{\rm SFR}^{1/3}$, which is the equipartition scaling. However, we notice that in dwarf irregular galaxies the advection speed is generally smaller than what would be expected from the SFR surface density \citep{heesen_18c}. This means these galaxies have higher magnetic field strengths than what would be estimated from energy equipartition. This has been shown by \citet{hindson_18a} for an entire sample of dwarf irregular galaxies.

This result is consistent with \citet{tabatabaei_17a}, who found $B
\propto {\rm SFR}^{0.34 \pm 0.04}$. It should be noted that we have
obtained the $B\textrm{--}{\rm SFR}$ relation indirectly from the
radio continuum luminosity instead of directly using magnetic field
strengths. It can be refined further by actually calculating magnetic
field strengths and directly finding a $B$--SFR relationship and
comparing it with this relation \citep[see e.g.,][]{chyzy_11,
  basu_17a}. However, an estimate of magnetic fields from the data at
hand is beyond the scope of this work. The combination of $B\propto
\Sigma_{\rm SFR}^{1/3}$ with the global Schmidt--Kennicutt law yields
$B\propto\rho^{0.5}$, which is consistent with Zeeman measurements of
Galactic H\,{\sc i} clouds \citep{crutcher_10a} and with estimates in galaxies both for global averages \citep{niklas_97a} and spatially resolved measurements \citep{chyzy_11,basu12b, basu_17a}.

\section{Conclusions}
\label{s:conclusions}

The LOFAR Two-metre Sky Survey (LoTSS) maps the entire northern hemisphere at 144~MHz with an unsurpassed combination of sensitivity and angular resolution \citep{shimwell_17a,shimwell_19a,shimwell_22a}. In this work, we define the LoTSS nearby galaxies sample, consisting of 76 nearby galaxies within 30~Mpc ($z\lesssim 0.01$) distance. This sample is largely drawn from the CHANG-ES radio continuum survey of edge-on galaxies \citep{irwin_12a} and the KINGFISH infrared survey of moderately inclined galaxies \citep{kennicutt_11a}, providing us with rich ancillary data both in the radio continuum and infrared. Here, we present our first look at the LoTSS data release 2 (LoTSS-DR2), in which 45 of these galaxies have maps that are suitable for further analysis. We measure integrated flux densities at 144~MHz, calculate integrated and spatially resolved spectral indices between 144 and 1400~MHz, and investigate the integrated radio--SFR relation.

We found that radio spectral indices are flat in general with a mean spectral index and standard deviation of $\langle \alpha \rangle = -0.56\pm 0.14$ (Section~\ref{ss:radio_spectral_index}). Of the 43 galaxies with measured spectral indices, 6 galaxies have spectral indices $>-0.5$ even when considering the $1\sigma$ uncertainties. These flat spectral indices, not in agreement with the injection spectral index of $\alpha_{\rm inj}\approx -0.6$, cannot be explained by thermal free--free emission. Instead they require either a low-energy flattening of the CRE by ionization losses or are caused by absorption of radio waves such as free--free or synchrotron self-absorption. We also find these flat spectral indices $\alpha \gtrsim -0.4$ in the spatially resolved spectral index maps (Section~\ref{ss:spatially_resolved_spectral_index}). These are usually compact regions co-spatial with star formation regions (e.g.\ NGC 2976 and 4736), or they are located in the nucleus of galaxies (e.g.\ NGC 3556 and 4631). The nuclei of galaxies have plenty of ionized gas \citep{law_09a} and AGN activity \citep{satyapal_08a}, both of which can affect the radio continuum spectra. The morphology of the spectral index maps suggests that the flat spectral index is confined to the star formation regions; in contrast, the full emitting volume of the galaxy, as measured by the global, integrated radio spectral index, is instead dominated by steep spectrum regions, in particular in large galaxies. These areas of steep spectral indices ($\alpha\lesssim -0.8$) can be best seen in the haloes of edge-on galaxies, but do also show up in the outskirts of some spiral galaxies (e.g.\ NGC 5055 and 5194).

These findings are corroborated by the observed correlations of the radio spectral index with star formation rate, radius of the star-forming disc, rotation speed, and total mass. These relations are of course closely connected because of the correlations between mass, SFR, radius, and rotation speed (Fig.~\ref{fig:sample} and Section~\ref{ss:radio_spectral_index_properties}). It is not clear which of these parameters is the primary one, if there indeed is any, or whether it is a combination of them. The competing factors are advection speed, regulating the escape of the CRE, and magnetic fields, regulating their synchrotron lifetime. Simply speaking, more massive galaxies have stronger magnetic fields reducing the synchrotron lifetime of the CRE, but they have also higher winds speeds, reducing the escape time scale. The radio spectral index is determined by the ratio of escape to synchrotron time-scale, so these effects do partially compensate each other. A wind velocity scaling with the SFR surface density of $\rm{v}\propto \Sigma_{\rm SFR}^{1/3}$ would suggest that the radius of the star-forming disc is the main physical parameter, whereas the SFR surface density is of little importance (Eq.~\ref{eq:advective_escape}). This is in good agreement with our observational data (Fig.~\ref{fig:alpha_sfr}(b) and Fig.~\ref{fig:alpha_sfrd}). What makes such a velocity scaling so intriguing is also that it ensures energy equipartition between the cosmic rays and the magnetic field (Section~\ref{ss:equipartition_magnetic_fields}), something that appears to be the natural outcome of stability considerations in galaxies \citep{crocker_21b}.

The variation of the radio spectral index as a function of galaxy mass means that more massive galaxies are closer to an electron calorimeter and so the CRE are losing a larger fraction of their energy due to synchrotron losses. This in turn means that galaxies are brighter with respect to the SFR, assuming that the CRE injection luminosity is proportional to the SFR. Such a dependence of the radio--SFR relation on mass has been posited already by \citet{gurkan_18a} and \citet{smith_21a} and we can now confirm their suggestion that this is indeed related to the escape of CRE. In our sample this manifests itself as a super-linear radio--SFR relation (Section~\ref{ss:radio_sfr_relation}), with galaxies having a higher SFR also being more massive (Fig.~\ref{fig:sample}). In order to quantify the influence of CRE escape on the radio--SFR relation, we use the parametrisation of \citet{gurkan_18a}, replacing the stellar mass with the closely related total mass within the star-forming radius (Section~\ref{ss:semi_calorimetric_radio_sfr_relation}), which can account for the super-linearity of the radio--SFR relation (Fig.~\ref{fig:radio_sfr_mass}). In dwarf galaxies, we would expect the CRE escape to be fast and so they should be underluminous with respect to the SFR. \citet{hindson_18a} show this to be the case, where galaxies with $\rm SFR < 0.1~M_{\sun}\,yr^{-1}$ can be a factor of 10 below the radio--SFR relation extrapolated from massive spiral galaxies. On the other hand, \citet{gurkan_18a} suggest just the opposite trend where galaxies with $\rm SFR<1~M_{\sun}\,yr^{-1}$ are radio bright. Part of this difference may be attributed to the fact that the stellar mass in these two studies is very different with \citet{gurkan_18a} probing more massive galaxies ($>10^{9}~\rm M_{\sun}$) and \citet{hindson_18a} dwarfs with stellar masses of $\approx 10^{8}~\rm M_{\sun}$. This warrants some additional LOFAR observations of more dwarf irregular galaxies with LOFAR (Sridhar et al., in preparation).

This study presents an analysis of integrated relations of the data from LoTSS-DR2 and should be seen as a preparatory work for exploiting the spatially resolved observations. For instance, the radio--SFR relation can be studied at kpc-resolution and the influence of the spectral index can be tested. As many of our galaxies have H\,{\sc i} maps from the THINGS survey \citep{walter_08a}, we can compare equipartition magnetic fields with turbulent energy densities in the ISM and explore the B--$\rho$ relation. Edge-on galaxies will provide us with an excellent view on the vertical transport of cosmic rays and so scaling laws such as with SFR surface density can be explored which is important to understand whether energy equipartition might hold (Section~\ref{ss:equipartition_magnetic_fields}).

\section*{Data Availability}

The LoTSS-DR2 data have been made publicly available on the website of the LOFAR surveys project (\href{https://www.lofar-surveys.org}{https://www.lofar-surveys.org}).  In addition, our cutouts made from the re-calibrated LoTSS-DR2 data and spectral index maps will be made available on the website of the Centre de Donn\'ees astronomiques de Strasbourg (CDS) (\href{http://cds.u-strasbg.fr}{http://cds.u-strasbg.fr)}.

\begin{acknowledgements}

We thank the anonymous referee for a constructive report that helped to improve the paper. LOFAR is the Low Frequency Array designed and constructed by ASTRON. It has observing, data processing, and data storage facilities in several countries, which are owned by various parties (each with their own funding sources), and which are collectively operated by the ILT foundation under a joint scientific policy. The ILT resources have benefited from the following recent major funding sources: CNRS-INSU, Observatoire de Paris and Université d'Orléans, France; BMBF, MIWF-NRW, MPG, Germany; Science Foundation Ireland (SFI), Department of Business, Enterprise and Innovation (DBEI), Ireland; NWO, The Netherlands; The Science and Technology Facilities Council, UK; Ministry of Science and Higher Education, Poland; The Istituto Nazionale di Astrofisica (INAF), Italy.

This research made use of the Dutch national e-infrastructure with support of the SURF Cooperative (e-infra 180169) and the LOFAR e-infra group. The Jülich LOFAR Long Term Archive and the German LOFAR network are both coordinated and operated by the Jülich Supercomputing Centre (JSC), and computing resources on the supercomputer JUWELS at JSC were provided by the Gauss Centre for Supercomputing e.V. (grant CHTB00) through the John von Neumann Institute for Computing (NIC).

This research made use of the University of Hertfordshire high-performance computing facility and the LOFAR-UK computing facility located at the University of Hertfordshire and supported by STFC [ST/P000096/1], and of the Italian LOFAR IT computing infrastructure supported and operated by INAF, and by the Physics Department of Turin university (under an agreement with Consorzio Interuniversitario per la Fisica Spaziale) at the C3S Supercomputing Centre, Italy.

The J\"ulich LOFAR Long Term Archive and the German LOFAR network are both coordinated and operated by the J\"ulich Supercomputing Centre (JSC), and computing resources on the supercomputer JUWELS at JSC were provided by the Gauss Centre for Supercomputing e.V. (grant CHTB00) through the John von Neumann Institute for Computing (NIC).

MB acknowledges support from the Deutsche Forschungsgemeinschaft under Germany's Excellence Strategy - EXC 2121 "Quantum Universe" - 390833306.  AD acknowledges support by the BMBF Verbundforschung under the grant 05A20STA. MJH acknowledges support from the UK Science and Technology Facilities Council (ST/R000905/1). AB acknowledges support from the VIDI research programme with project number 639.042.729, which is financed by the Netherlands Organisation for Scientific Research (NWO). FdG acknowledges support from the Deutsche Forschungsgemeinschaft under Germany's Excellence Strategy - EXC 2121 “Quantum Universe” - 390833306.
\end{acknowledgements}


%
%

\bibliographystyle{aa}
\bibliography{dr2_data} 


\appendix


\section{The full LoTSS galaxy sample}
\label{as:the_full_lotss_galaxy_sample}

\begin{table*}
	\centering
	\caption{The full LoTSS sample of 76 galaxies, a subset of 45 galaxies are included in LoTSS-DR2.}
	\label{tab:sample}
	\begin{tabular}{lc cc cc ccc ccc cc} 
		\hline
		Galaxy      & RA            &   Dec              & $i$          & $D_{25}$ & $d$            & $M_B$     & Type & Nuc.    & $\rm v_{\rm rot}$ & SFR \\
                            & (h m s)&($\degr~\arcmin~\arcsec$)  & ($\degr$)    & ($\prime$) & (Mpc)       & (mag) & & & ($\rm km\,s^{-1}$) & ($\rm M_{\sun}\,yr^{-1}$) \\
                (1)         & (2)           &   (3)              & (4)          & (5)     & (6)            & (7)        & (8)   & (9)     & (10)      & (11) \\
		\hline
                IC 10       &  00 20 17.3   &   59	18  14   &  47  [O15]   &  7.2    &  0.7    [H12]  &  $-15.52$  & Ir    &  H       & 37  [O15] &  0.010  \\
                NGC 598     &  01 33 50.9   &   30	39  37   &  56  [T88]   &  56.5   &  0.9    [P11]  &  $-19.00$  & Scd   &  H       & 100 [LED] &  0.64   \\
                NGC 628     &  01 36 41.7   &   15	47  01   &  15  [D08]   &  7.0    &  7.2    [K11]  &  $-19.94$  & Sc    &  H*      & 217 [L08] &  1.19   \\
                NGC 855     &  02 14 30.5   &   27	52  38   &  74  [LED]   &  9.8    &  9.73   [K11]  &  $-16.98$  & E     &  N/A     & 40  [LED] &  0.059  \\
                NGC 891     &  02 22 33.4   &   42	20  57   &  90  [S19]   &  12.2   &  9.1    [W15]  &  $-20.10$  & Sb    &  H       & 212 [LED] &  3.75   \\
                NGC 925     &  02 27 16.9   &   33	34  45   &  50  [D08]   &  11.3   &  9.12   [K11]  &  $-20.03$  & Sd    &  H       & 136 [L08] &  0.75   \\
                IC 342      &  03 46 48.5   &   68	05  47   &  20  [T88]   &  16.1   &  3.28   [K11]  &  $-21.44$  & Scd   &  H       & 230 [LED] &  1.88   \\
                NGC 2146    &  06 18 37.7   &   78	21  25   &  36  [T88]   &  5.3    &  17.2   [K11]  &  $-21.19$  & Sab   &  H       & 292 [LED] &  14.93  \\
                NGC 2403    &  07 36 51.4   &   65	36  09   &  55  [D08]   &  23.8   &  3.06   [T08]  &  $-19.30$  & Scd   &  H       & 134 [L08] &  0.90   \\
                Ho II       &  08 19 05.0   &   70	43  12   &  31  [D08]   &  8.2    &  3.05   [K11]  &  $-16.68$  & Ir    &  N/A     & 36  [L08] &  0.009  \\
                M81 DwA     &  08 23 55.1   &   71	01  56   &  27  [D08]   &  1.3    &  3.4    [J09]  &  $-11.39$  & Ir    &  N/A     & 21  [LED] &  0.0007 \\
                DDO 53      &  08 34 07.2   &   66	10  54   &  33  [D08]   &  1.9    &  3.61   [K11]  &  $-13.49$  & Ir    &  N/A     & 29  [O15] &  0.0015 \\
                NGC 2683    &  08 52 41.3   &   33	25  18   &  79  [I12]   &  9.1    &  6.27   [W15]  &  $-19.89$  & Sb    &  L2/S2   & 203 [LED] &  0.97   \\
                NGC 2798    &  09 17 22.8   &   41	59  59   &  75  [T88]   &  2.7    &  25.8   [K11]  &  $-19.56$  & Sa    &  L2/S2   & 154 [LED] &  5.94   \\
                NGC 2820    &  09 21 45.6   &   64	15  29   &  88  [K18]   &  4.1    &  26.5   [W15]  &  $-20.76$  & Sc    &  N/A     & 163 [LED] &  1.49   \\
                NGC 2841    &  09 22 02.6   &   50	58  35   &  69  [D08]   &  6.6    &  14.1   [K11]  &  $-21.21$  & Sb    &  L2      & 302 [L08] &  1.88   \\
                NGC 2903    &  09 32 10.1   &   21	30  03   &  66  [D08]   &  11.6   &  8.9    [W08]  &  $-20.92$  & Sbc   &  H       & 190 [D08] &  5.96   \\
                Ho I        &  09 40 35.1   &   71	10  46   &  27  [D08]   &  3.7    &  3.9    [K11]  &  $-14.84$  & Ir    &  N/A     & 53  [L08] &  0.0019 \\
                NGC 2976    &  09 47 15.5   &   67	54  59   &  54  [D08]   &  5.0    &  3.55   [K11]  &  $-17.73$  & Sc    &  H       & 92  [L08] &  0.12   \\
                NGC 3003    &  09 48 36.1   &   32	25  17   &  85  [K18]   &  6.0    &  25.4   [K11]  &  $-20.86$  & Sbc   &  H       & 121 [LED] &  1.21   \\
                NGC 3031    &  09 55 33.2   &   69	03  55   &  59  [D08]   &  22.1   &  3.44   [J09]  &  $-20.71$  & Sab   &  S1.5    & 200 [D08] &  1.48   \\
                NGC 3034    &  09 55 52.7   &   69	40  46   &  80  [L15]   &  11.7   &  3.52   [J09]  &  $-20.07$  & P     &  H       & 66  [LED] &  14.92  \\
                Ho IX       &  09 57 32.0   &   69	02  45   &  41  [K03]   &  2.5    &  3.61   [D09]  &  $-13.76$  & Ir    &  N/A     & 30  [LED] &  0.0009 \\
                NGC 3079    &  10 01 57.8   &   55	40  47   &  84  [K18]   &  7.7    &  20.6   [W15]  &  $-21.60$  & Sd    &  S2      & 208 [LED] &  9.86   \\
                NGC 3077    &  10 03 19.1   &   68	44  02   &  38  [LED]   &  5.4    &  3.83   [K11]  &  $-17.72$  & P     &  H       & 39  [LED] &  0.12   \\
                M81 DwB     &  10 05 30.6   &   70	21  52   &  28  [D08]   &  1.4    &  3.6    [K11]  &  $-13.72$  & Ir    &  N/A     & 20  [LED] &  0.0005 \\
                NGC 3190    &  10 18 05.6   &   21	49  52   &  71  [T88]   &  4.2    &  19.3   [K11]  &  $-20.03$  & Sa    &  L2      & 215 [LED] &  1.19   \\
                NGC 3184    &  10 18 18.9   &   41	25  27   &  29  [D08]   &  7.5    &  11.7   [K11]  &  $-20.03$  & Scd   &  H       & 210 [L08] &  1.49   \\
                NGC 3198    &  10 19 55.0   &   45	32  59   &  72  [D08]   &  8.8    &  14.1   [K11]  &  $-20.81$  & Sc    &  H       & 150 [L08] &  1.49   \\
                IC 2574     &  10 28 23.5   &   68	24  44   &  51  [D08]   &  13.0   &  3.79   [K11]  &  $-17.90$  & Sm    &  H*      & 75  [D08] &  0.030  \\
                NGC 3265    &  10 31 06.8   &   28	47  48   &  47  [LED]   &  9.8    &  19.6   [K11]  &  $-17.53$  & E     &  H*      & 95  [LED] &  0.38   \\
                Mrk 33      &  10 32 32.0   &   54	24  02   &  28  [T88]   &  1.1    &  15.4   [T16]  &  $-18.07$  & Ir    &  H*      & 68  [LED] &  0.53   \\
                NGC 3351    &  10 43 57.7   &   11	42  14   &  56  [T88]   &  7.5    &  9.33   [K11]  &  $-19.71$  & Sb    &  H*      & 196 [L08] &  1.19   \\
                NGC 3432    &  10 52 31.1   &   36	37  08   &  85  [K18]   &  4.9    &  9.42   [W15]  &  $-19.28$  & Scd   &  H       & 110 [LED] &  0.56   \\
                NGC 3448    &  10 54 39.2   &   54	18  18   &  78  [I12]   &  4.9    &  24.5   [W15]  &  $-20.40$  & P     &  H       & 122 [LED] &  1.84   \\
                NGC 3556    &  11 11 31.0   &   55	40  27   &  81  [I12]   &  7.8    &  14.09  [W15]  &  $-20.78$  & Scd   &  H$^1$   & 153 [LED] &  4.31   \\
                NGC 3627    &  11 20 15.0   &   12	59  30   &  61  [D08]   &  8.0    &  9.38   [K11]  &  $-20.77$  & Sb    &  T2/S2   & 192 [L08] &  3.75   \\
                NGC 3628    &  11 20 17.0   &   13	35  23   &  87  [T88]   &  14.8   &  8.5    [W15]  &  $-20.50$  & Sb    &  T2      & 216 [LED] &  6.75   \\
                NGC 3735    &  11 35 57.3   &   70	32  08   &  84  [K18]   &  4.0    &  42     [W15]  &  $-21.94$  & Sc    &  S2?     & 241 [LED] &  9.21   \\
                NGC 3773    &  11 38 12.9   &   12	06  43   &  41  [T88]   &  1.6    &  12.4   [K11]  &  $-17.29$  & S0    &  H*      & 55  [LED] &  0.094  \\
                NGC 3877    &  11 46 07.7   &   47	29  40   &  85  [K18]   &  5.1    &  17.7   [W15]  &  $-20.60$  & Scd   &  H       & 156 [LED] &  1.92   \\
                NGC 3938    &  11 52 49.5   &   44	07  15   &  36  [T88]   &  4.9    &  17.9   [K11]  &  $-20.52$  & Sc    &  H??$^1$ & 128 [LED] &  2.98   \\
                NGC 4013    &  11 58 31.2   &   41	56  48   &  88  [K18]   &  4.7    &  16     [W15]  &  $-19.65$  & Sbc   &  T2      & 182 [LED] &  1.60   \\
                NGC 4096    &  12 06 01.1   &   47	28  42   &  82  [T88]   &  6.4    &  10.32  [W15]  &  $-20.06$  & Sc    &  H       & 145 [LED] &  1.18   \\
                NGC 4125    &  12 08 06.0   &   65	10  27   &  36  [T88]   &  6.6    &  24.77  [NED]  &  $-21.43$  & E     &  T2      & 150 [P10] &  0.11   \\
                NGC 4157    &  12 11 04.4   &   50	29  05   &  83  [K18]   &  7.0    &  15.6   [W15]  &  $-19.99$  & Sb    &  H       & 189 [LED] &  3.27   \\
                NGC 4214    &  12 15 39.2   &   36	19  37   &  38  [D08]   &  9.6    &  2.95   [J09]  &  $-17.44$  & Ir    &  H       & 57  [L08] &  0.078  \\
                NGC 4217    &  12 15 50.9   &   47	05  03   &  86  [K18]   &  5.1    &  20.6   [W15]  &  $-20.38$  & Sb    &  H       & 188 [LED] &  4.61   \\
                NGC 4236    &  12 16 42.1   &   69	27  45   &  73  [T88]   &  19.6   &  4.45   [K11]  &  $-19.22$  & Sdm   &  H       & 68  [LED] &  0.94   \\
                NGC 4244    &  12 17 29.7   &   37	48  27   &  90  [I12]   &  15.8   &  4.4    [W15]  &  $-18.49$  & Scd   &  H       & 98  [LED] &  0.18   \\
                NGC 4254    &  12 18 49.6   &   14	24  59   &  33  [T88]   &  5.0    &  14.4   [K11]  &  $-20.62$  & Sc    &  H       & 300 [LED] &  5.94   \\
                NGC 4321    &  12 22 54.8   &   15	49  19   &  37  [T88]   &  6.1    &  14.3   [K11]  &  $-20.95$  & Sbc   &  T2$^1$  & 279 [LED] &  4.72   \\
                NGC 4449    &  12 28 11.1   &   44	05  37   &  54  [LED]   &  5.4    &  4.02   [S18]  &  $-19.04$  & Ir    &  H       & 59  [LED] &  0.72   \\
                NGC 4450    &  12 28 29.6   &   17	05  06   &  50  [T88]   &  5.0    &  14.07  [NED]  &  $-20.28$  & Sab   &  L1.9    & 185 [LED] &  0.30   \\
                NGC 4536    &  12 34 27.1   &   02	11  17   &  69  [T88]   &  6.3    &  14.5   [K11]  &  $-20.54$  & Sbc   &  H$^1$   & 156 [LED] &  2.98   \\
                NGC 4559    &  12 35 57.6   &   27	57  36   &  69  [T88]   &  11.3   &  6.98   [K11]  &  $-19.62$  & Scd   &  H       & 114 [LED] &  0.47   \\
                NGC 4565    &  12 36 20.8   &   25	59  16   &  86  [S19]   &  16.2   &  11.9   [W15]  &  $-21.41$  & Sb    &  S1.9    & 244 [LED] &  1.33   \\
                NGC 4625    &  12 41 52.7   &   41	16  26   &  23  [T88]   &  1.6    &  9.3    [K11]  &  $-17.11$  & Sm    &  N/A     & 39  [LED] &  0.094  \\
                NGC 4631    &  12 42 08.0   &   32	32  29   &  85  [I12]   &  14.7   &  7.62   [K11]  &  $-21.42$  & Sd    &  H       & 139 [LED] &  3.75   \\
                NGC 4725    &  12 50 26.6   &   25	30  03   &  43  [T88]   &  10.5   &  11.9   [K11]  &  $-20.70$  & Sab   &  S2?     & 257 [LED] &  1.19   \\
               \hline                                                                                                            
	\end{tabular}                                                                                                            
\end{table*}

\setcounter{table}{0}
\begin{table*}
	\centering
	\caption{-- continued.}
	\label{tab:example_table}
        \begin{tabular}{lc cc cc ccc ccc cc} 
	        \hline
	        Galaxy      & RA    & Dec     & $i$          & $D_{25}$ & $d$            & $M_B$     & Type & Nuc.    & $\rm v_{\rm rot}$ & SFR \\
                            &(h m s)&($\degr~\arcmin~\arcsec$)&($\degr$)    & ($\prime$) & (Mpc)       & (mag) & & & ($\rm km\,s^{-1}$) & ($\rm M_{\sun}\,yr^{-1}$) \\
                 (1)        & (2)           & (3)           & (4)          & (5)     & (6)            & (7)        & (8)   & (9)     & (10)      & (11) \\
		\hline
                NGC 4736    &  12 50 53.1   &  41 07  14    &  44  [D08]   &  12.2   &  4.66   [K11]  &  $-19.80$  & Sab &  L2       & 156 [L08] &  0.94   \\
                DDO 154     &  12 54 05.3   &  27 08  59    &  31  [T88]   &  2.5    &  4.3    [K11]  &  $-14.44$  & Ir  &  N/A      & 50  [L08] &  N/A    \\
                NGC 4826    &  12 56 43.6   &  21 40  59    &  64  [D08]   &  8.0    &  5.27   [K11]  &  $-19.90$  & Sab &  T2       & 152 [LED] &  0.59   \\
                DDO 165     &  13 06 24.9   &  67 42 25     &  50  [T88]   &  3.5    &  4.57   [K11]  &  $-16.02$  & Ir  &  N/A      & 20  [LED] &  N/A    \\
                NGC 5033    &  13 13 27.5   &  36 35 38     &  64  [T88]   &  10.1   &  17.13  [NED]  &  $-21.09$  & Sc  &  S1.5     & 226 [LED] &  3.93   \\
                NGC 5055    &  13 15 49.3   &  42 01 45     &  51  [D08]   &  13.0   &  7.94   [K11]  &  $-20.61$  & Sbc &  T2$^1$   & 192 [L08] &  2.98   \\
                NGC 5194    &  13 29 52.7   &  47 11 43     &  20  [T74]   &  13.6   &  8.0    [W08]  &  $-21.18$  & Sbc &  S2       & 219 [L08] &  11.89  \\
                NGC 5195    &  13 29 59.6   &  47 15 58     &  46  [T88]   &  6.3    &  8.0    [W08]  &  $-19.26$  & P   &  L2??     & 120 [LED] &  0.60   \\
                NGC 5297    &  13 46 23.7   &  43 52 20     &  89  [I12]   &  5.3    &  40.4   [W15]  &  $-21.70$  & Sc  &  L2       & 189 [LED] &  2.84   \\
                NGC 5457    &  14 03 12.5   &  54 20 56     &  30  [D08]   &  23.8   &  6.7    [K11]  &  $-20.84$  & Scd &  H        & 274 [LED] &  3.75   \\
                NGC 5474    &  14 05 01.6   &  53 39 44     &  37  [T88]   &  5.9    &  6.8    [K11]  &  $-18.05$  & Scd &  H        & 22  [LED] &  0.075  \\
                NGC 5775    &  14 53 57.6   &  03 32 40     &  86  [K18]   &  3.9    &  28.9   [W15]  &  $-21.32$  & Sc  &  H        & 190 [LED] &  12.42  \\
                NGC 5866    &  15 06 29.5   &  55 45 48     &  67  [T88]   &  7.3    &  15.3   [K11]  &  $-20.25$  & S0  &  T2       & 185 [N99] &  0.94   \\
                NGC 5907    &  15 15 53.8   &  56 19 44     &  90  [I12]   &  11.2   &  16.8   [W15]  &  $-21.42$  & Sc  &  H?       & 227 [LED] &  2.84   \\
                NGC 6946    &  20 34 52.3   &  60 09 14     &  35  [D08]   &  11.2   &  6.8    [K11]  &  $-20.93$  & Scd &  H        & 186 [L08] &  4.72   \\
                NGC 7331    &  22 37 04.0   &  34 24 56     &  77  [D08]   &  9.7    &  14.5   [K11]  &  $-21.62$  & Sbc &  T2       & 244 [L08] &  7.48   \\
                \hline
	\end{tabular}
        \flushleft
            {\small {\bf Notes.}\\
              Column (1)  galaxy name; (2) and (3) right ascension and declination at epoch J2000.0 from NED; (4) inclination angle (0 is face-on, 90 is edge-on); (5) optical diameter to the 25 $\rm mag\,arcsec^{-2}$ isophote from T88; (6) distance; (7) absolute $B$-band magnitude corrected for foreground extinction (from T88, corrected to our distances); (8) galaxy type from the Third Reference Catalog of Bright Galaxies \citep{de_vaucouleurs_91a}; (9) spectral classification of the nucleus from \citet{ho_97a} or with `*' are from NED, $^1$ means classified as AGN by \citet{satyapal_08a}; (10) rotation velocity; (11) star formation rate from total infrared luminosity 3--1100~$\mu$m (see Section~\ref{ss:sample}) using ${\rm SFR}=0.39 L_{\rm TIR} / (10^{43}~{\rm erg\,s^{-1})}$ from \citet{kennicutt_12a} .\\
              {\bf References.}\\
              Data are from the NASA extragalactic data base (NED) or from HyperLEDA (LED) and following references; D08: \citet{de_blok_08a}; H12: \citet{hunter_12a}; I12: \citet{irwin_12a}; J09: \citet{jacobs_09a}; K11: \citet{kennicutt_11a}; K18: \citet{krause_18a}; N99: \citet{neistein_99a}; L08: \citet{leroy_08a}; L15: \citet{leroy_15a}; O15:\citet{oh_15a}; P10: \citet{pu_10a}; P11: \citet{pellerin_11a}; S19: \citet{schmidt_19a}; T88: \citet{tully_88a}; W15: \citet{wiegert_15a}}
\end{table*}

In this section, we present in Table~\ref{tab:sample} more information about the full sample of galaxies contained in LoTSS. Of these 76 galaxies, a subset of 45 galaxies are in LoTSS-DR2 and presented in this paper.

\section{Atlas of galaxies}
\label{as:atlas}

We present maps for 45 galaxies, which are included in LoTSS-DR2. There are two different sets of figures, with the first set containing 6 panels for galaxies where no spectral index maps were made (14 galaxies). The second set of figures includes 3 additional panels, showing the radio spectral index. 

In the first row, we show the 144-MHz radio continuum intensity in units of $\rm Jy\,beam^{-1}$ as heat maps with contour lines, at 6 and 20 arcsec resolution, respectively. We also show intensity contour lines, beginning at $3\sigma$ and incrementing by powers of 2 with each higher contour level. The same data are also presented as a grey-scale images without contour lines in the second row. In the third row, we present overlays of these contours on optical colour images either from SDSS or DSS.\footnote{DSS images only used for NGC~891 and NGC~925 as these are outside of the SDSS footprint. In these cases, optical observations from DSS in the blue band were used in grey scale.} In all panels, flux regions are shown in blue, $D_{25}$ optical diameters are shown as either black or white. Where available, we also present radio spectral maps in the third column. In the first row, we show the spectral index; in the second row the spectral index error map; the third row presents the radio spectral index as contours overlaid on the optical images from either SDSS or DSS. Contour lines are at values of $\alpha = -1$ (blue), $-0.8$ (green), $-0.6$ (yellow) and $-0.4$ (red), respectively. Note that due to the $3\sigma$-clipping prior to combination, the radio spectral index maps are blanked in parts of the maps, so that the contour lines may be interrupted in places. 

All of these maps are presented in Figs~\ref{fig:n855}--\ref{fig:n7331}.


\begin{figure*}
	\centering
	\includegraphics[width=0.8\textwidth]{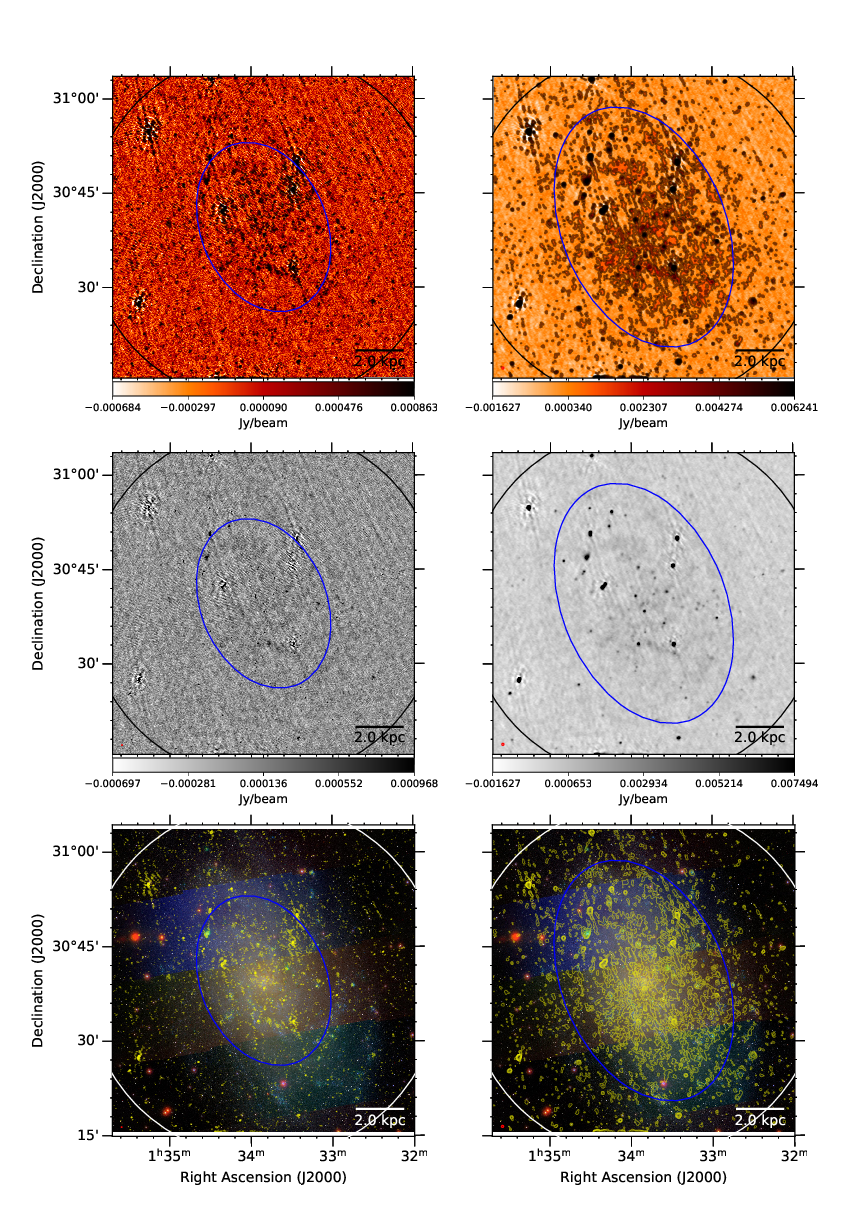}
    \caption{NGC~598 (M~33). \nospix}
    \label{fig:n598}
\end{figure*}
\addcontentsline{toc}{subsection}{NGC 598}

\begin{figure*}
	\centering
	\includegraphics[width=0.8\textwidth]{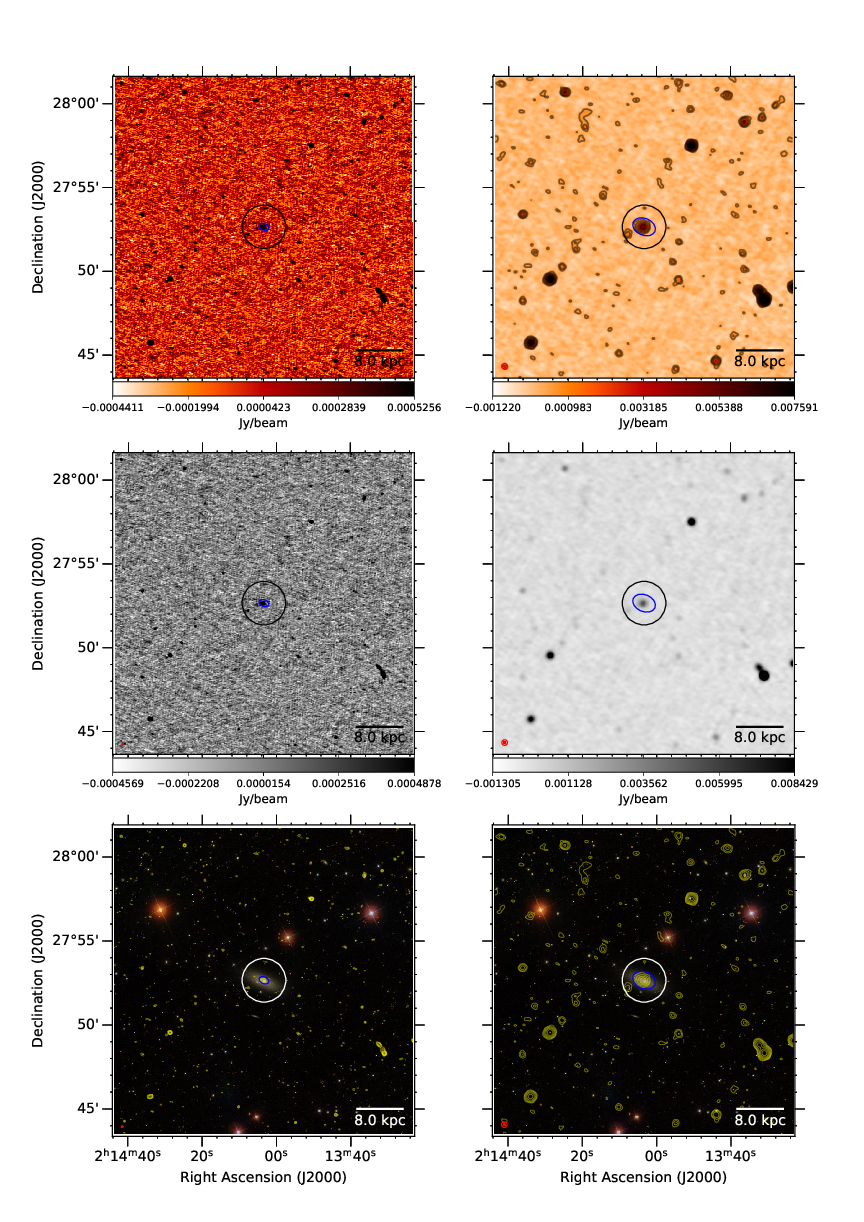}
    \caption{NGC~855. \nospix}
    \label{fig:n855}
\end{figure*}
\addcontentsline{toc}{subsection}{NGC 855}

\begin{figure*}
	\centering
	\includegraphics[width=\textwidth]{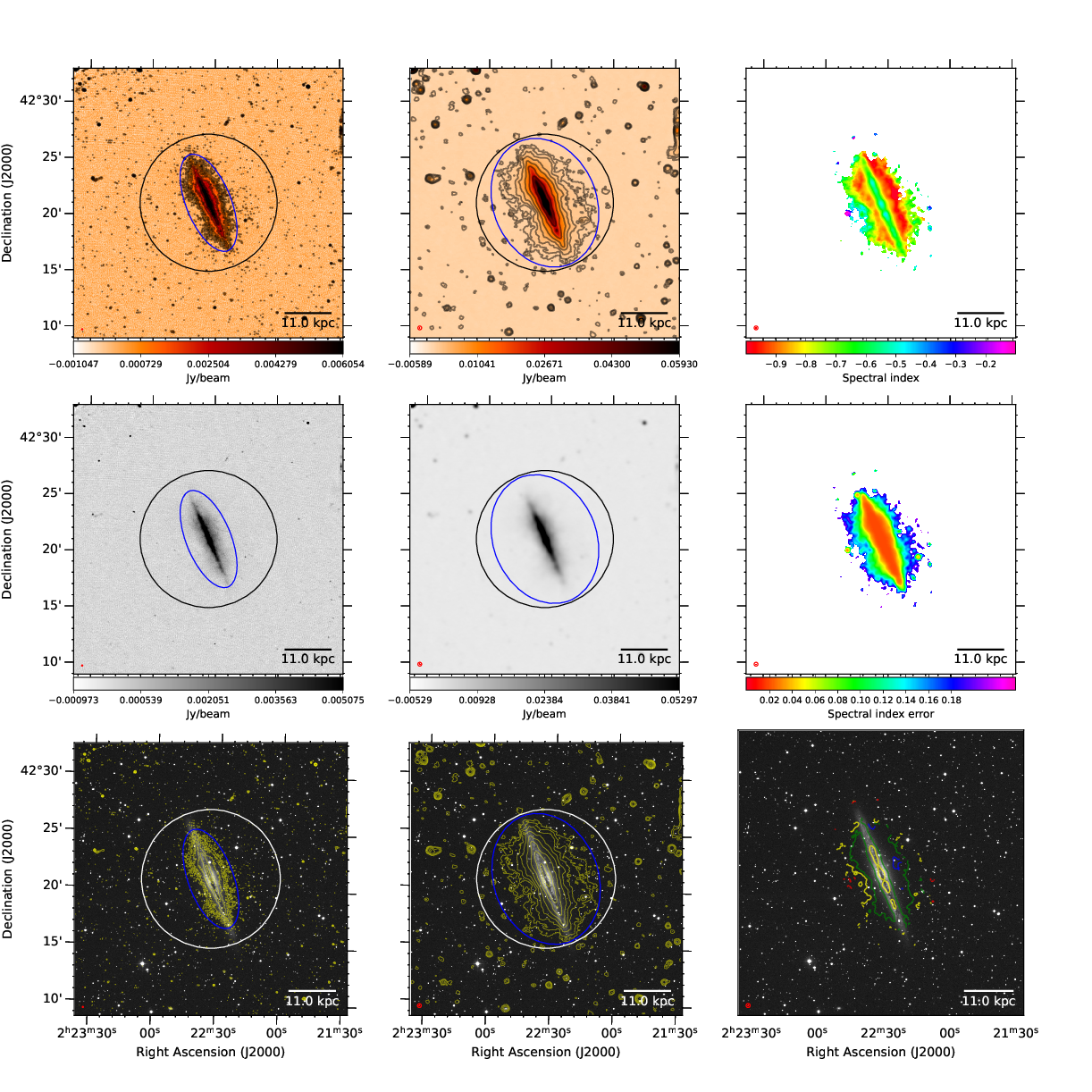}
    \caption{NGC~891. \spix{6000}{20}}
    \label{fig:n891}
\end{figure*}
\addcontentsline{toc}{subsection}{NGC 891}

\begin{figure*}
	\centering
	\includegraphics[width=\textwidth]{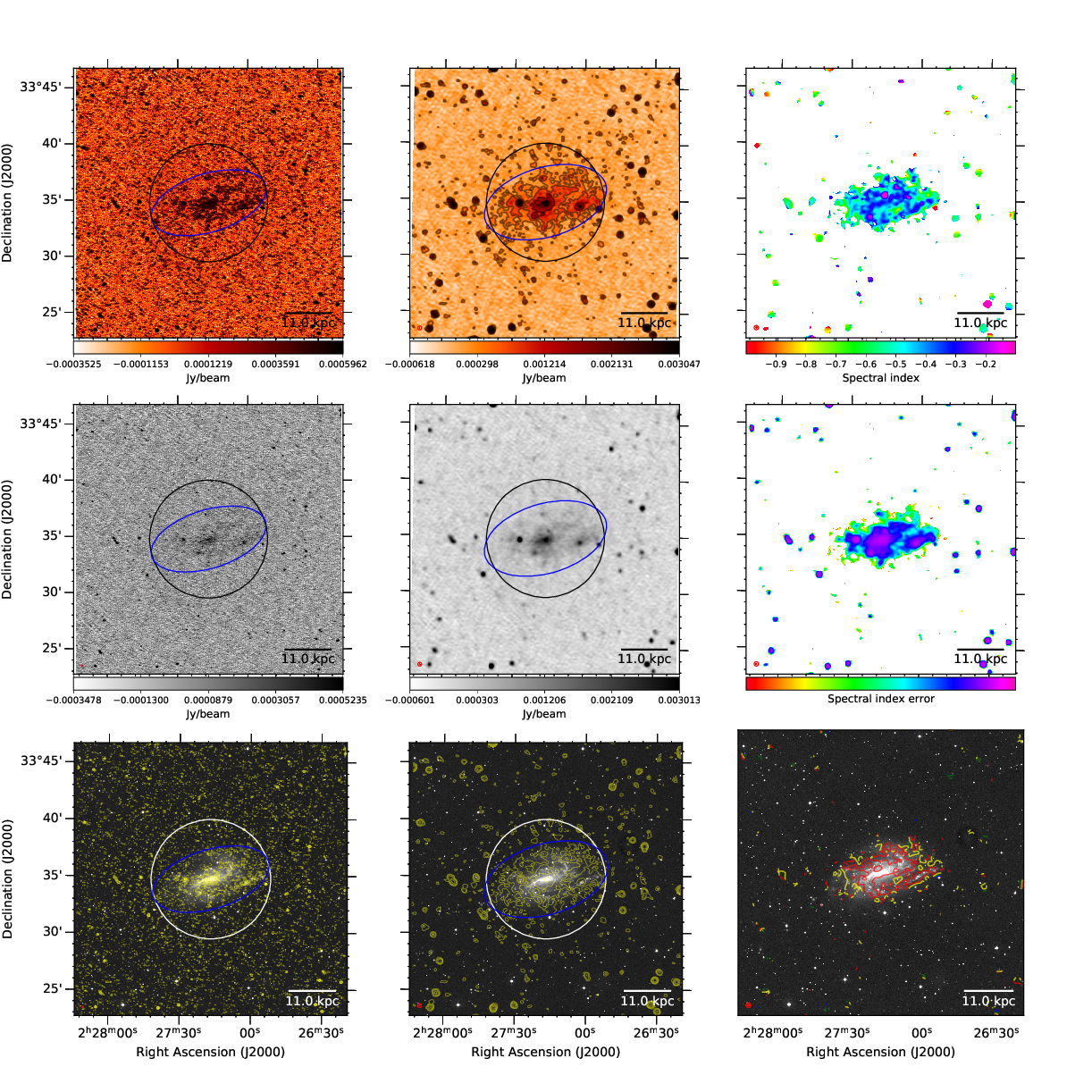}
    \caption{NGC~925. \spix{1365}{23}}
    \label{fig:n925}
\end{figure*}
\addcontentsline{toc}{subsection}{NGC 925}

\begin{figure*}
	\centering
	\includegraphics[width=\textwidth]{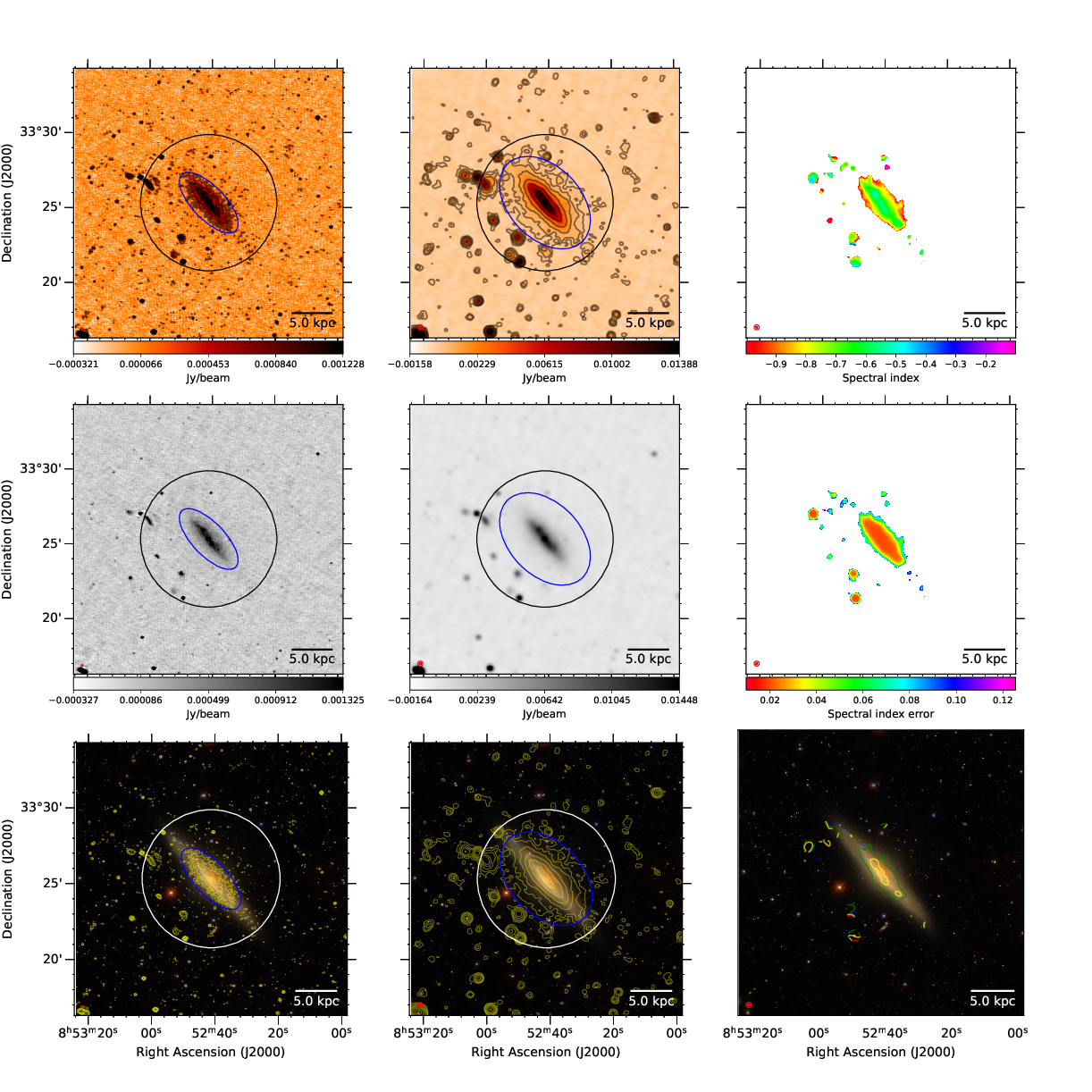}
    \caption{NGC~2683. \spix{6000}{20}.}
    \label{fig:n2683a}
\end{figure*}
\addcontentsline{toc}{subsection}{NGC 2683}

\begin{figure*}
	\centering
	\includegraphics[width=0.8\textwidth]{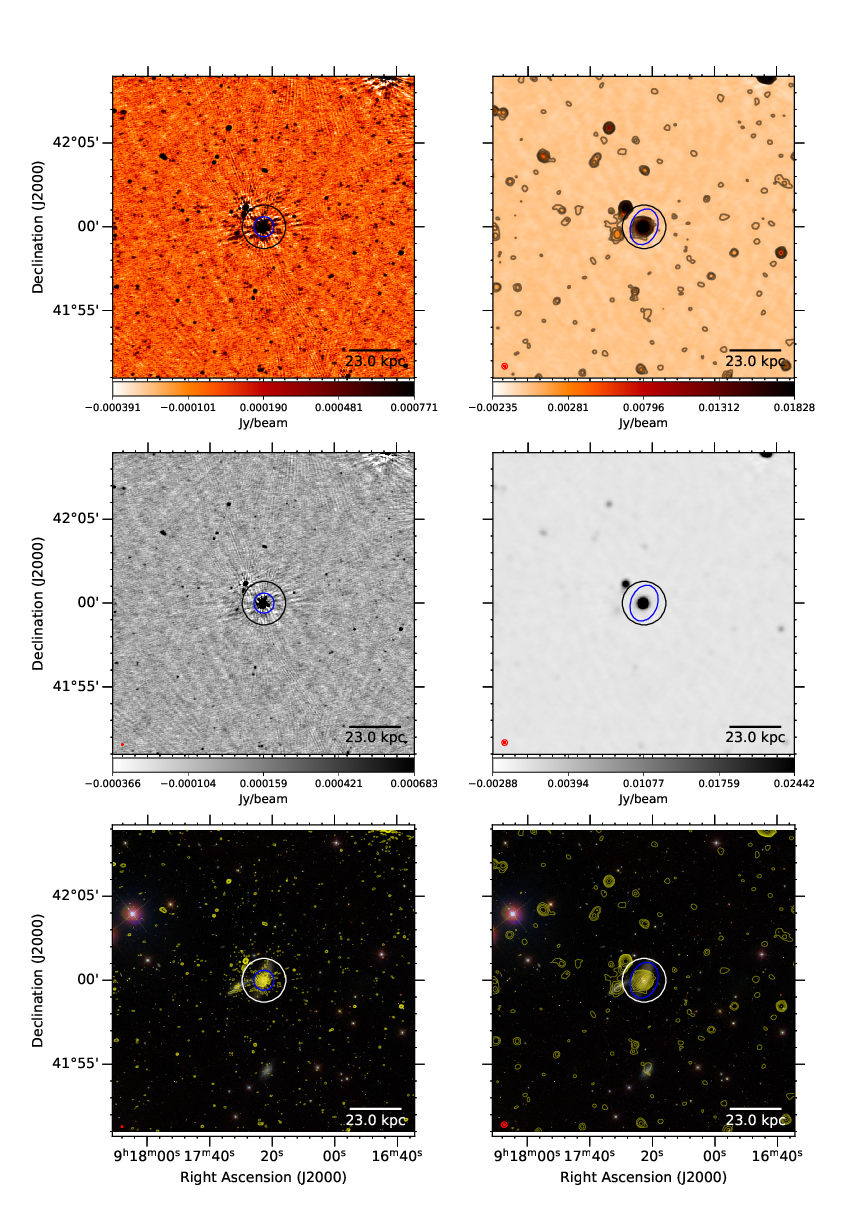}
    \caption{NGC~2798. \nospix}
    \label{fig:2798}
\end{figure*}
\addcontentsline{toc}{subsection}{NGC 2798}

\begin{figure*}
	\centering
	\includegraphics[width=\textwidth]{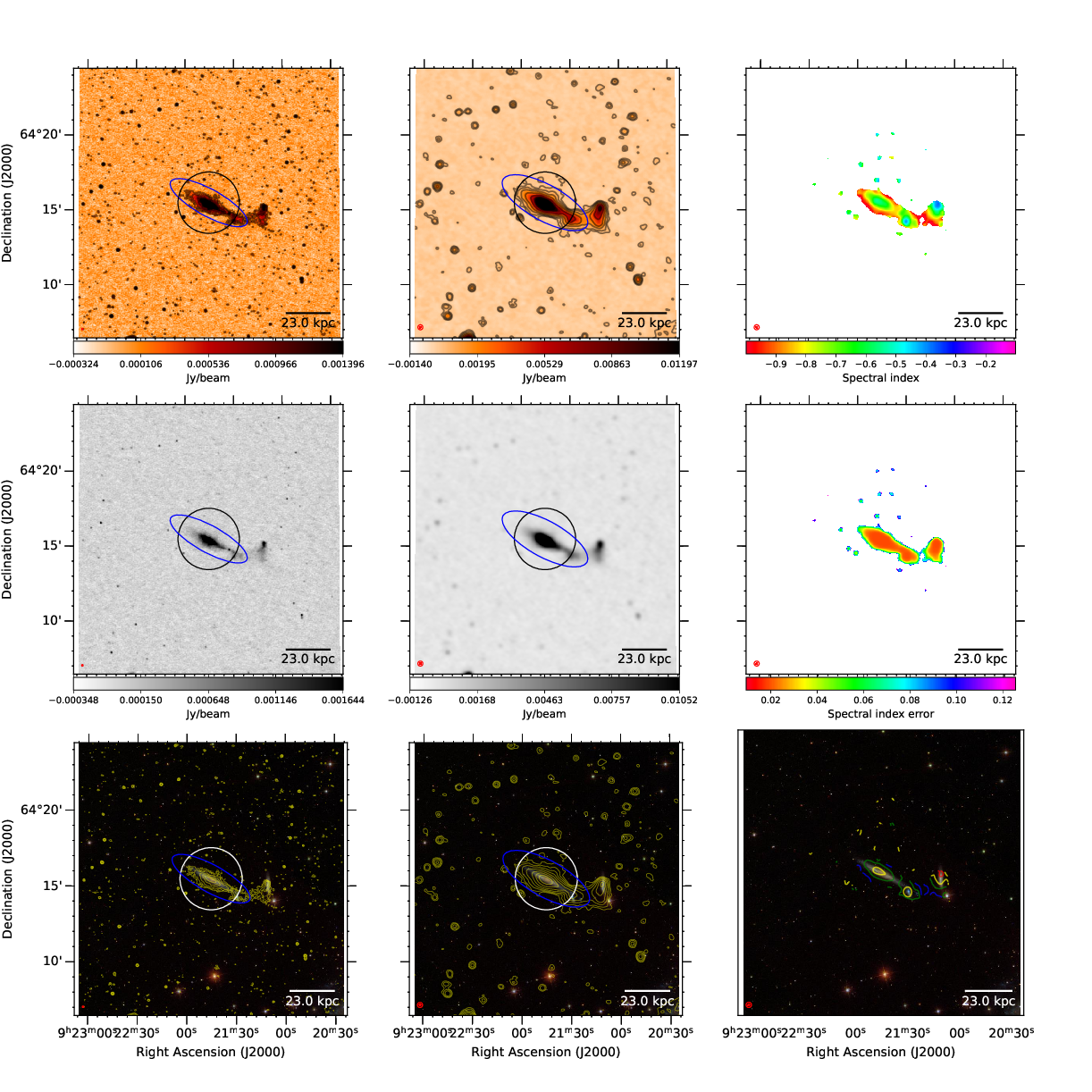}
    \caption{NGC~2820. \spix{6000}{20}}
    \label{fig:n2820}
\end{figure*}
\addcontentsline{toc}{subsection}{NGC 2820}

\begin{figure*}
	\centering
	\includegraphics[width=\textwidth]{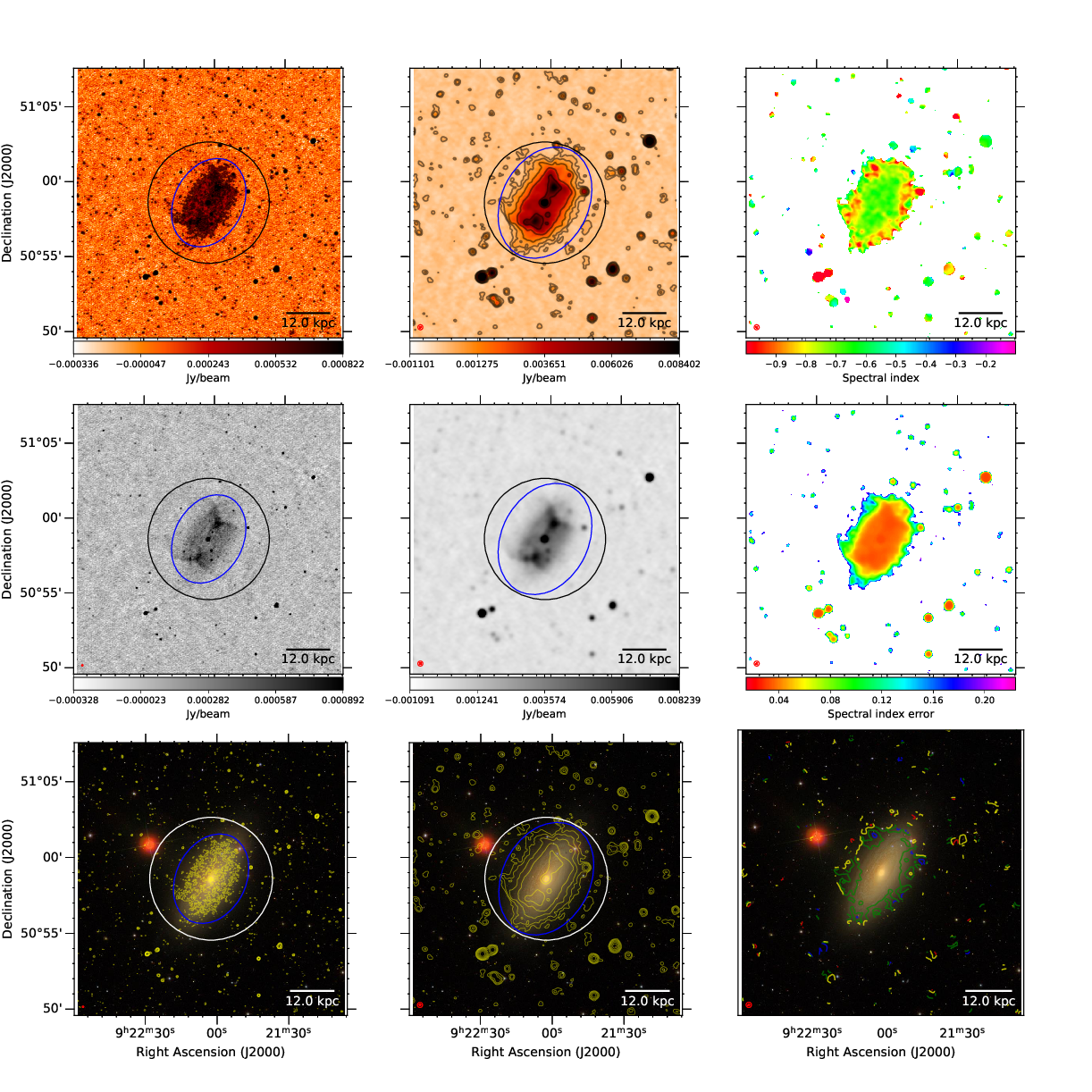}
    \caption{NGC~2841. \spix{1365}{20}}
    \label{fig:n2841}
\end{figure*}
\addcontentsline{toc}{subsection}{NGC 2841}

\begin{figure*}
	\centering
	\includegraphics[width=\textwidth]{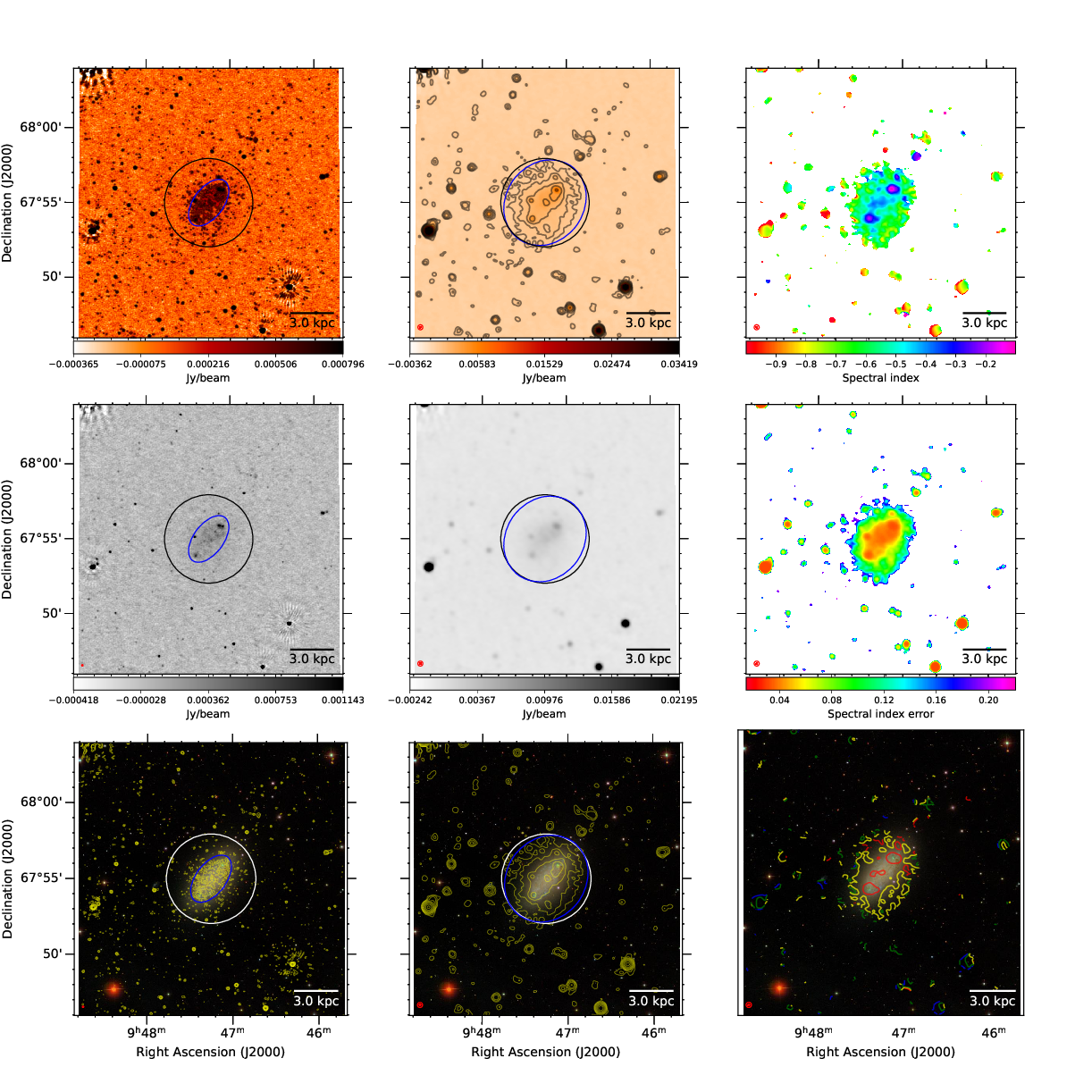}
    \caption{NGC~2976. \spix{1365}{20}}
    \label{fig:n2976}
\end{figure*}
\addcontentsline{toc}{subsection}{NGC 2976}

\begin{figure*}
	\centering
	\includegraphics[width=\textwidth]{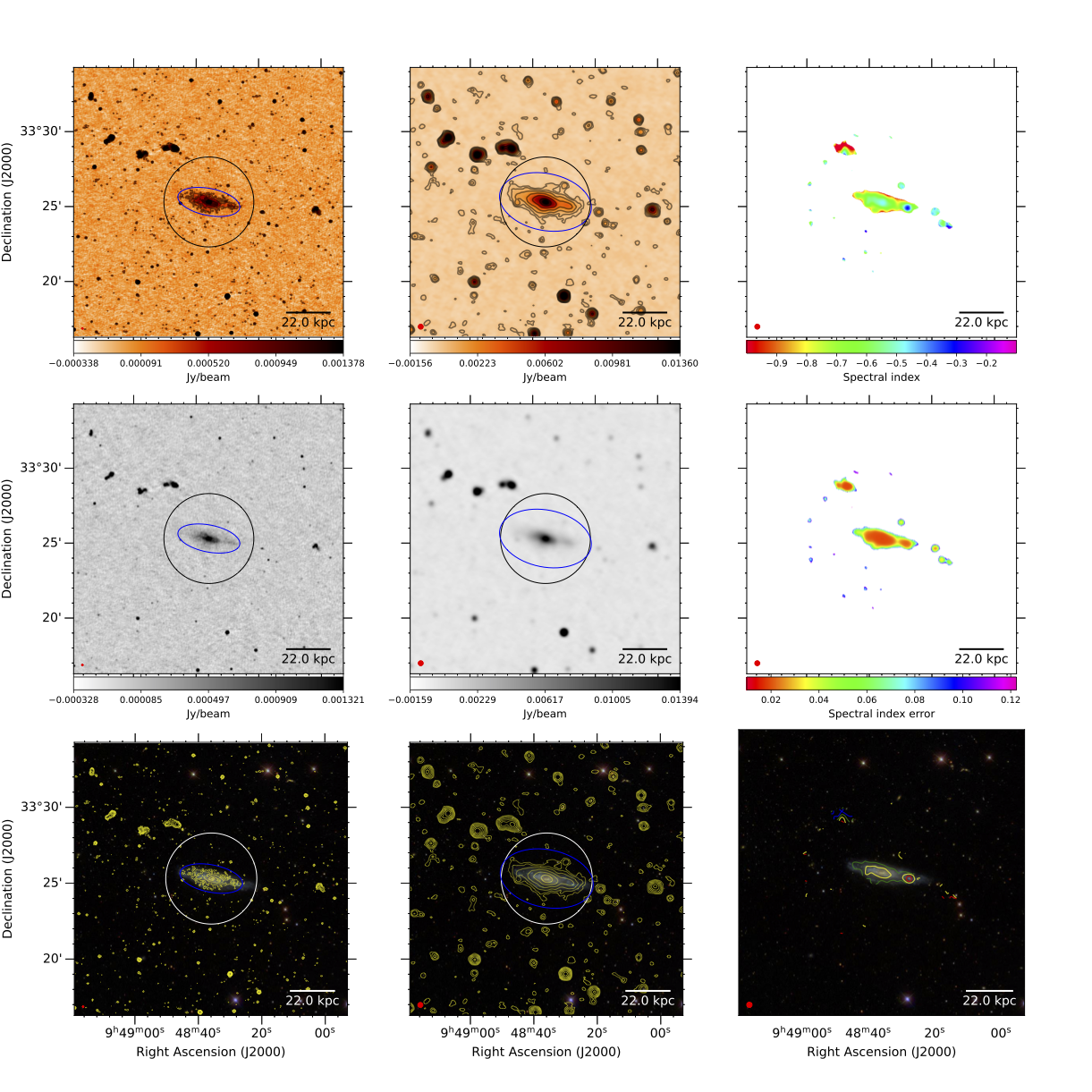}
    \caption{NGC~3003. \spix{6000}{20}}
    \label{fig:n3003}
\end{figure*}
\addcontentsline{toc}{subsection}{NGC 3003}

\begin{figure*}
	\centering
	\includegraphics[width=0.85\textwidth]{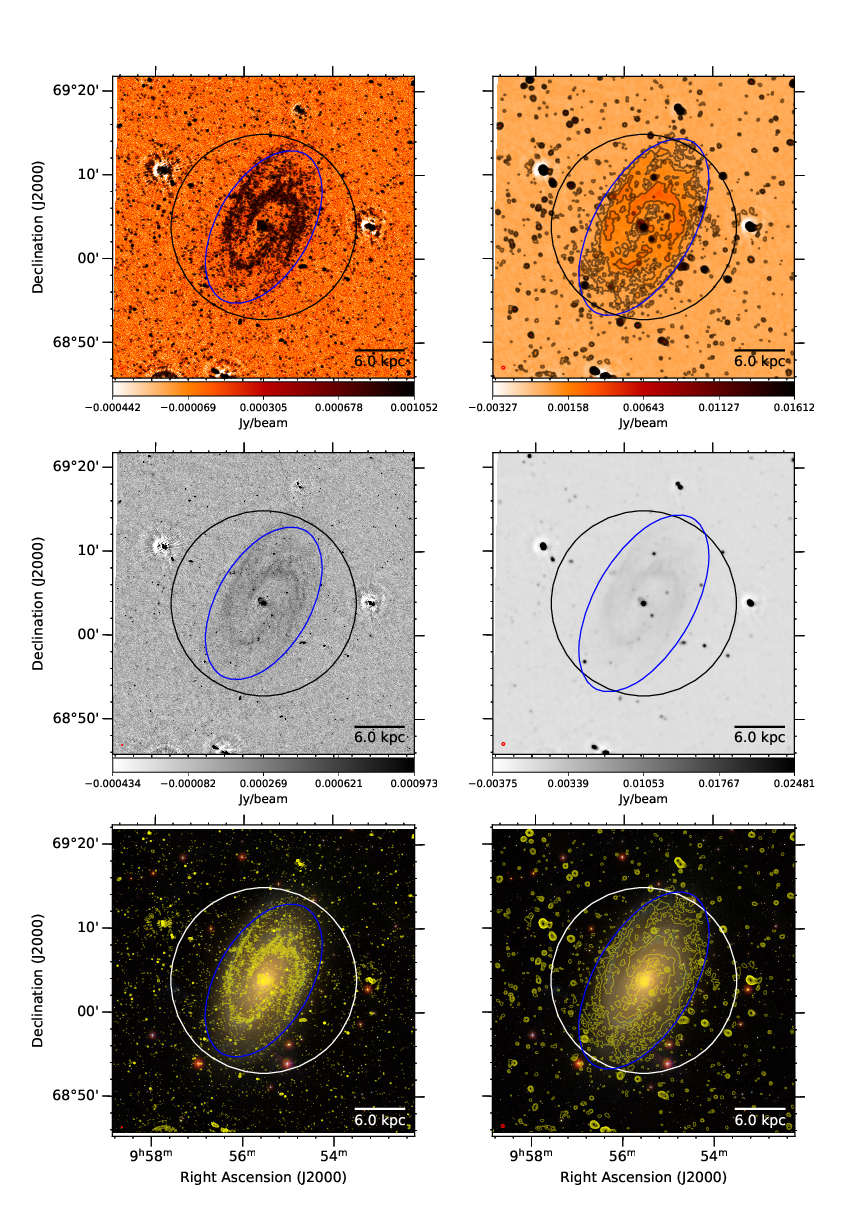}
    \caption{NGC~3031 (M~81). \nospix}
    \label{fig:n3031}
\end{figure*}
\addcontentsline{toc}{subsection}{NGC 3031}

\begin{figure*}
	\centering
	\includegraphics[width=\textwidth]{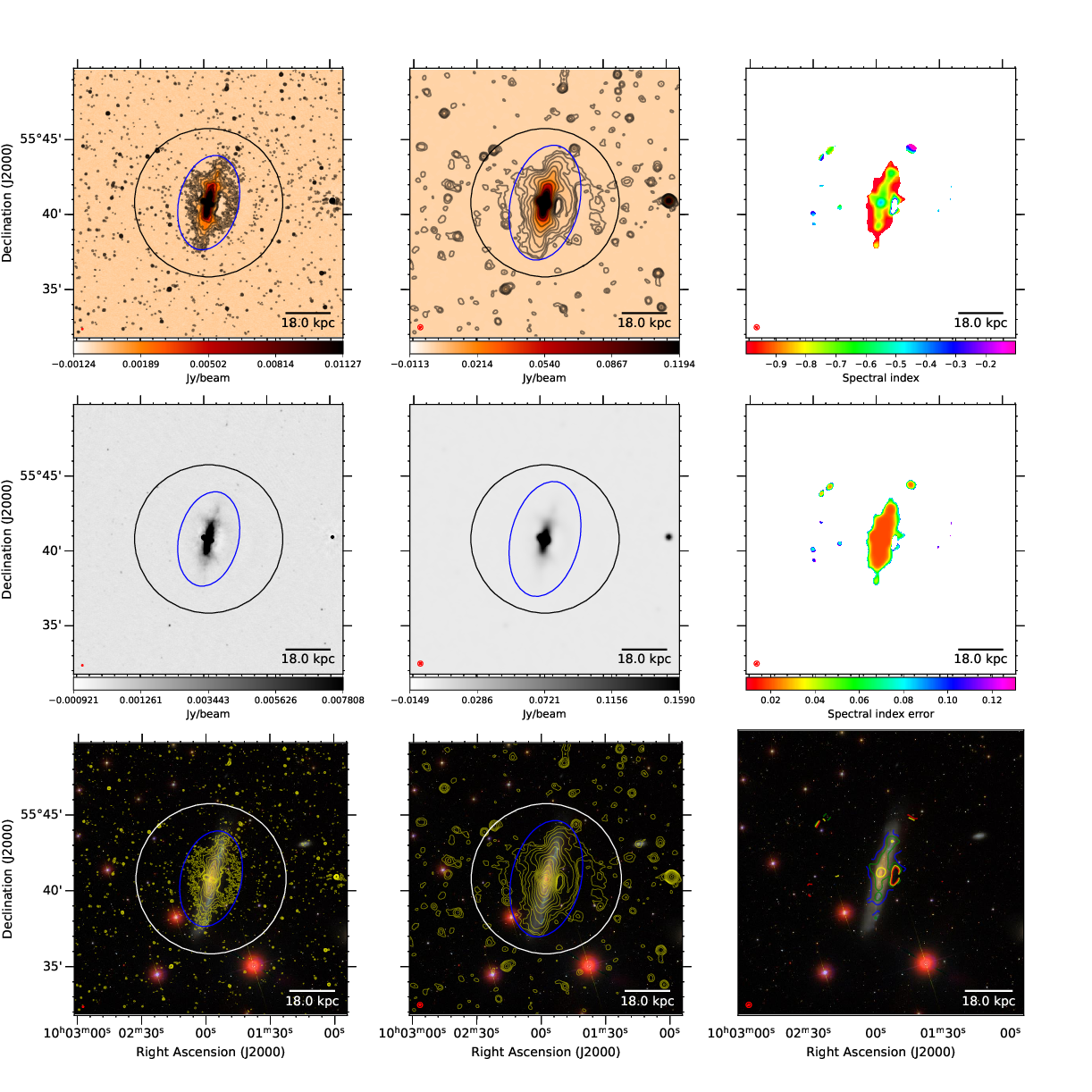}
    \caption{NGC~3079. \spix{6000}{20}}
    \label{fig:n3079}
\end{figure*}
\addcontentsline{toc}{subsection}{NGC 3079}

\begin{figure*}
	\centering
	\includegraphics[width=0.8\textwidth]{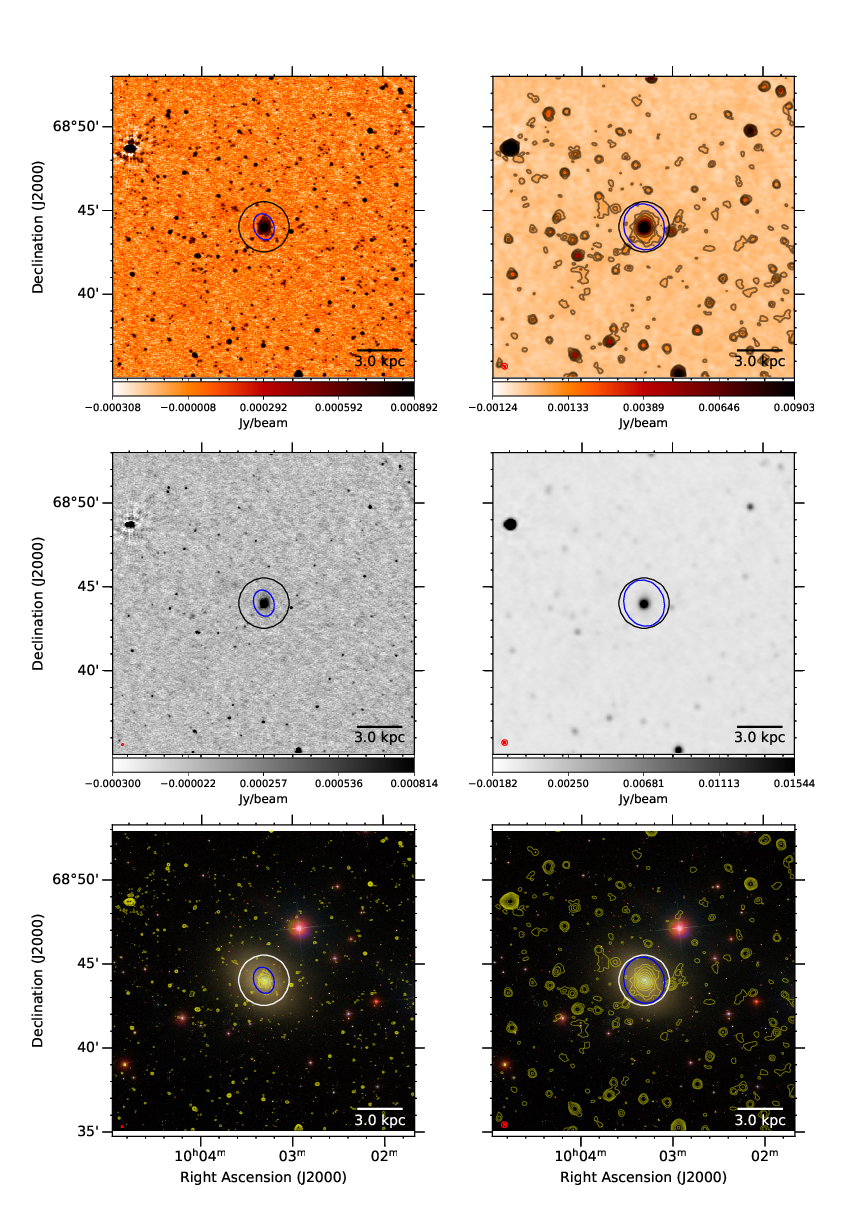}
    \caption{NGC~3077. \nospix}
    \label{fig:n3077}
\end{figure*}
\addcontentsline{toc}{subsection}{NGC 3077}

\begin{figure*}
	\centering
	\includegraphics[width=\textwidth]{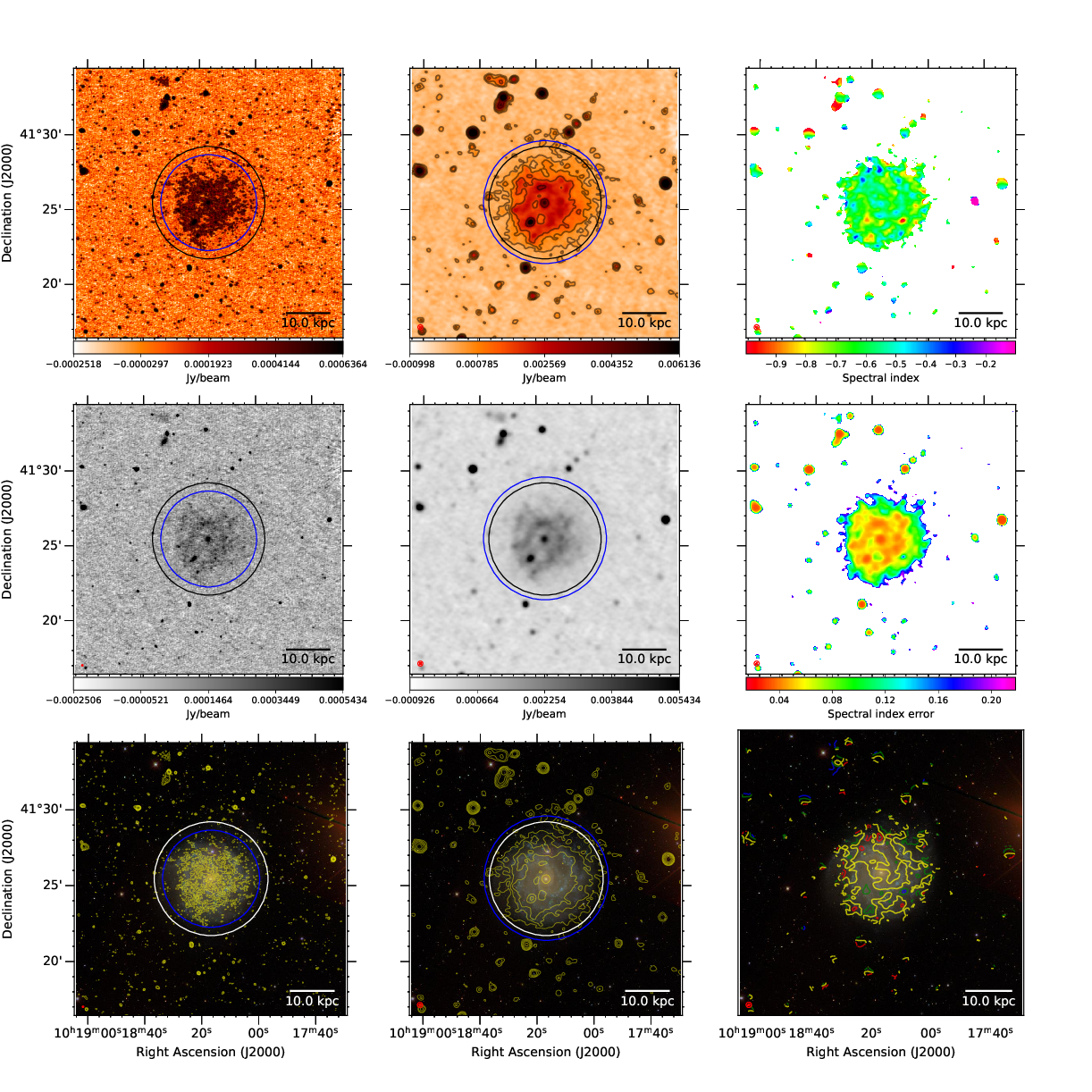}
    \caption{NGC~3184. \spix{1365}{20}}
    \label{fig:n3184}
\end{figure*}
\addcontentsline{toc}{subsection}{NGC 3184}

\begin{figure*}
	\centering
	\includegraphics[width=\textwidth]{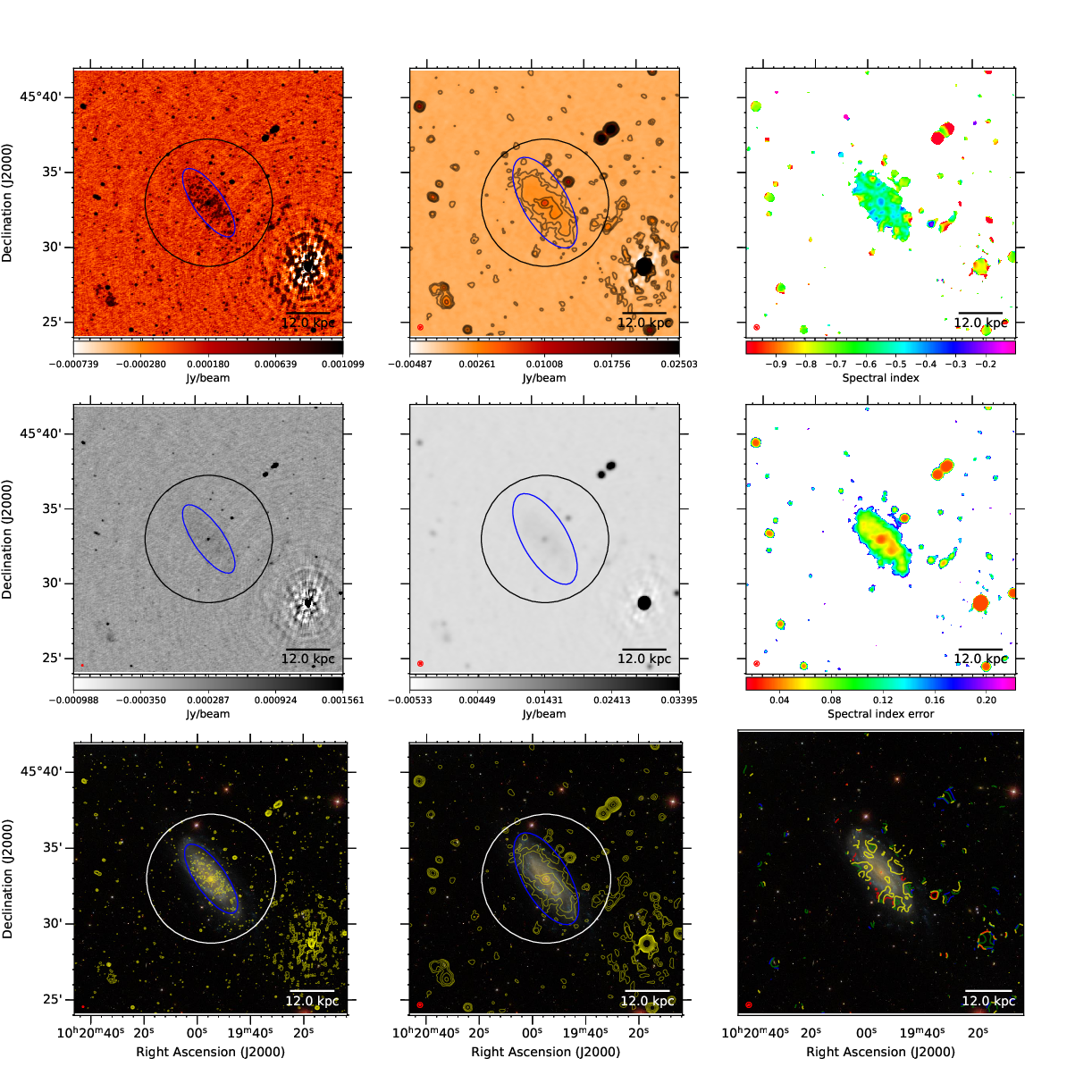}
    \caption{NGC~3198. \spix{1365}{20}}
    \label{fig:n3198}
\end{figure*}
\addcontentsline{toc}{subsection}{NGC 3198}

\begin{figure*}
	\centering
	\includegraphics[width=0.8\textwidth]{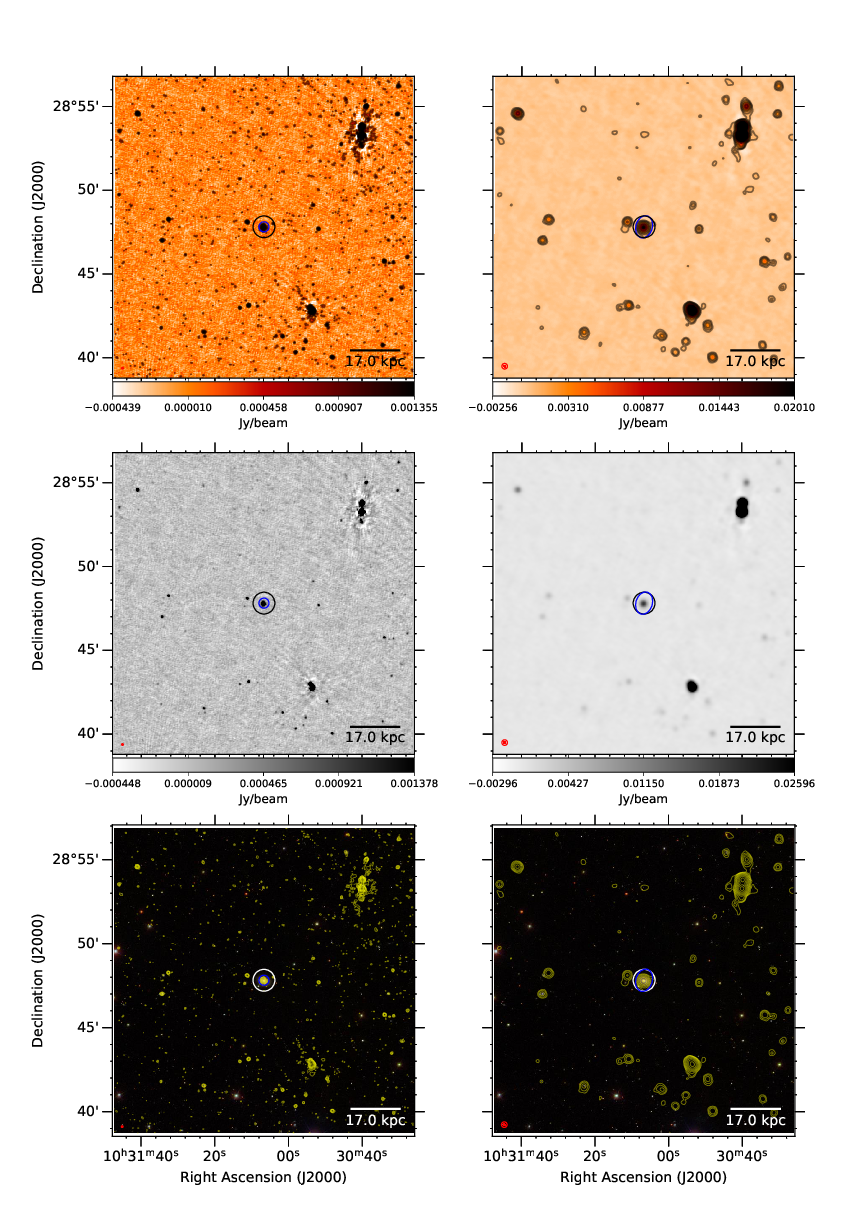}
    \caption{NGC~3265. \nospix}
    \label{fig:n3265}
\end{figure*}
\addcontentsline{toc}{subsection}{NGC 3265}

\begin{figure*}
	\centering
	\includegraphics[width=0.8\textwidth]{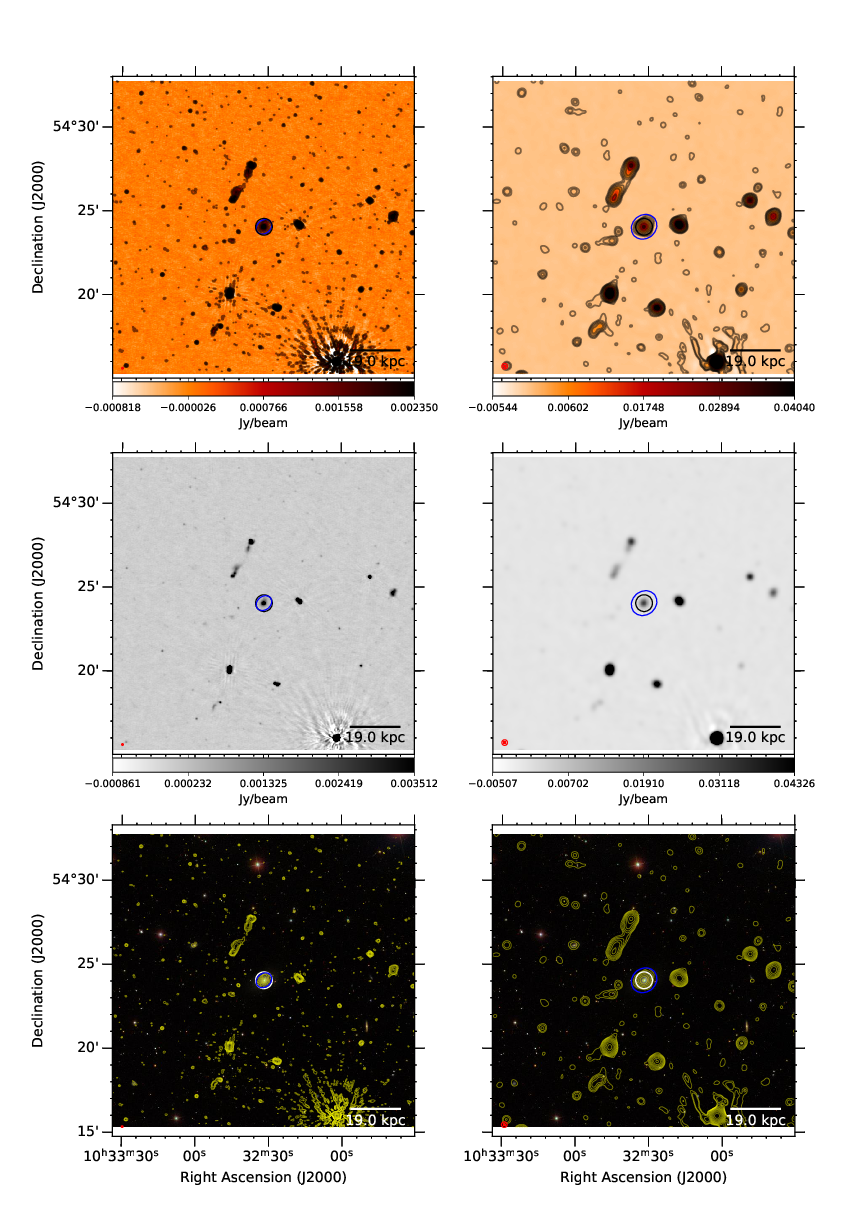}
    \caption{Mrk~33. \nospix}
    \label{fig:mrk33}
\end{figure*}
\addcontentsline{toc}{subsection}{Mrk 33}

\begin{figure*}
	\includegraphics[width=\textwidth]{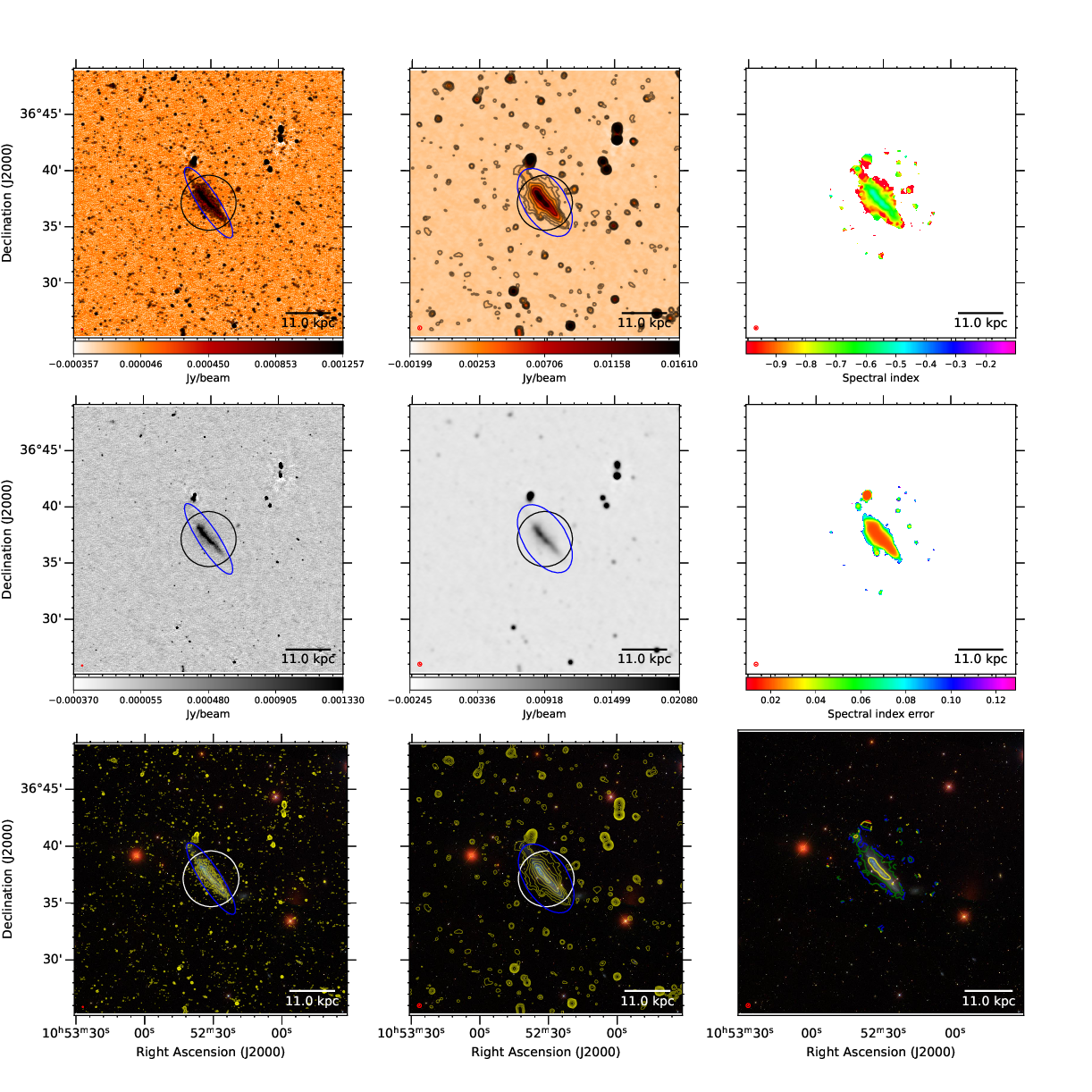}
    \caption{NGC~3432. \spix{6000}{20}}
    \label{fig:n3432}
\end{figure*}
\addcontentsline{toc}{subsection}{NGC 3432}

\begin{figure*}
	\includegraphics[width=\textwidth]{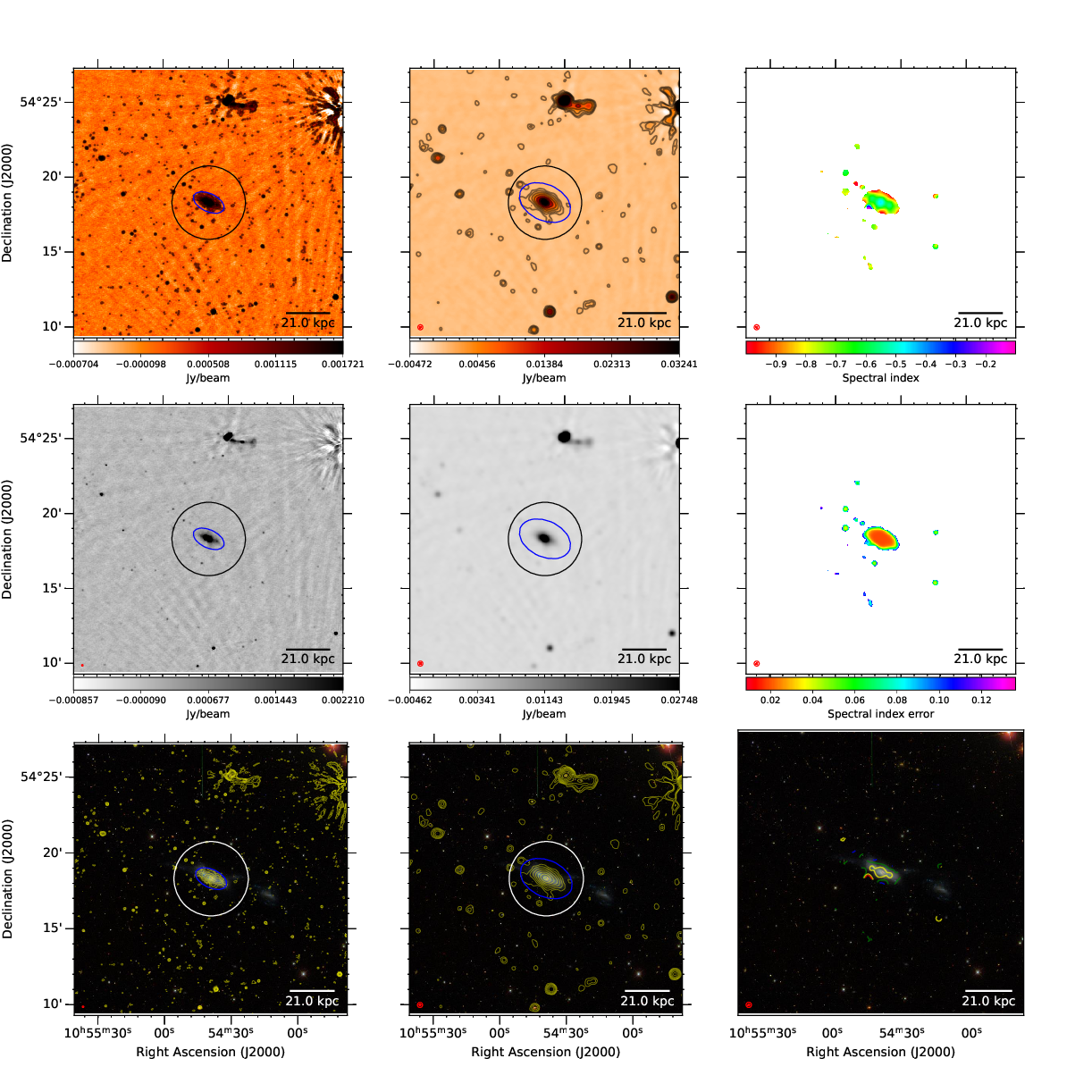}
    \caption{NGC~3448. \spix{6000}{20}}
    \label{fig:n3448}
\end{figure*}
\addcontentsline{toc}{subsection}{NGC 3448}

\begin{figure*}
	\includegraphics[width=\textwidth]{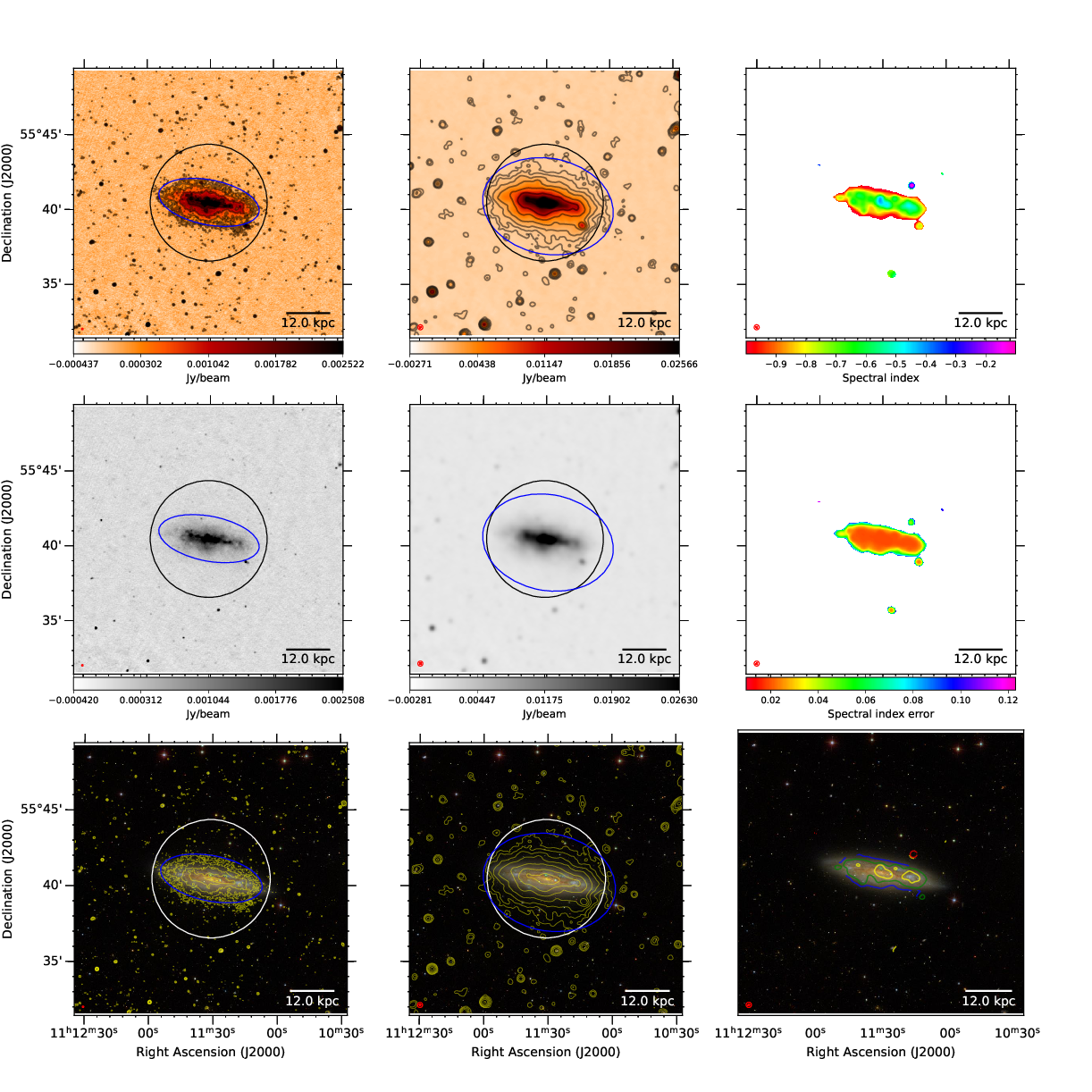}
	\centering
    \caption{NGC~3556. \spix{6000}{20}}
    \label{fig:n3556}
\end{figure*}
\addcontentsline{toc}{subsection}{NGC 3556}

\begin{figure*}
	\centering
	\includegraphics[width=\textwidth]{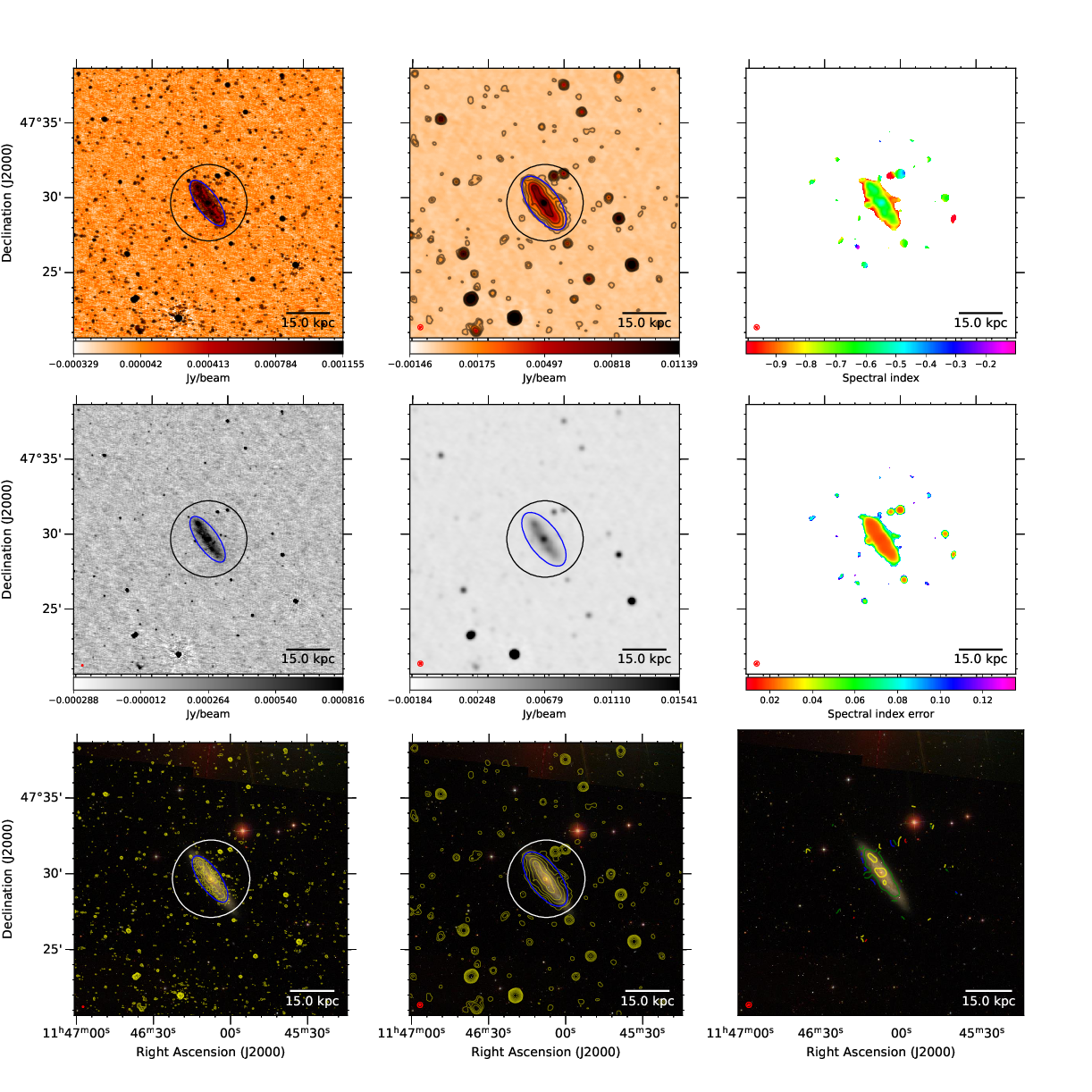}
    \caption{NGC~3877. \spix{6000}{20}}
    \label{fig:n3877}
\end{figure*}
\addcontentsline{toc}{subsection}{NGC 3877}

\begin{figure*}
	\centering
	\includegraphics[width=\textwidth]{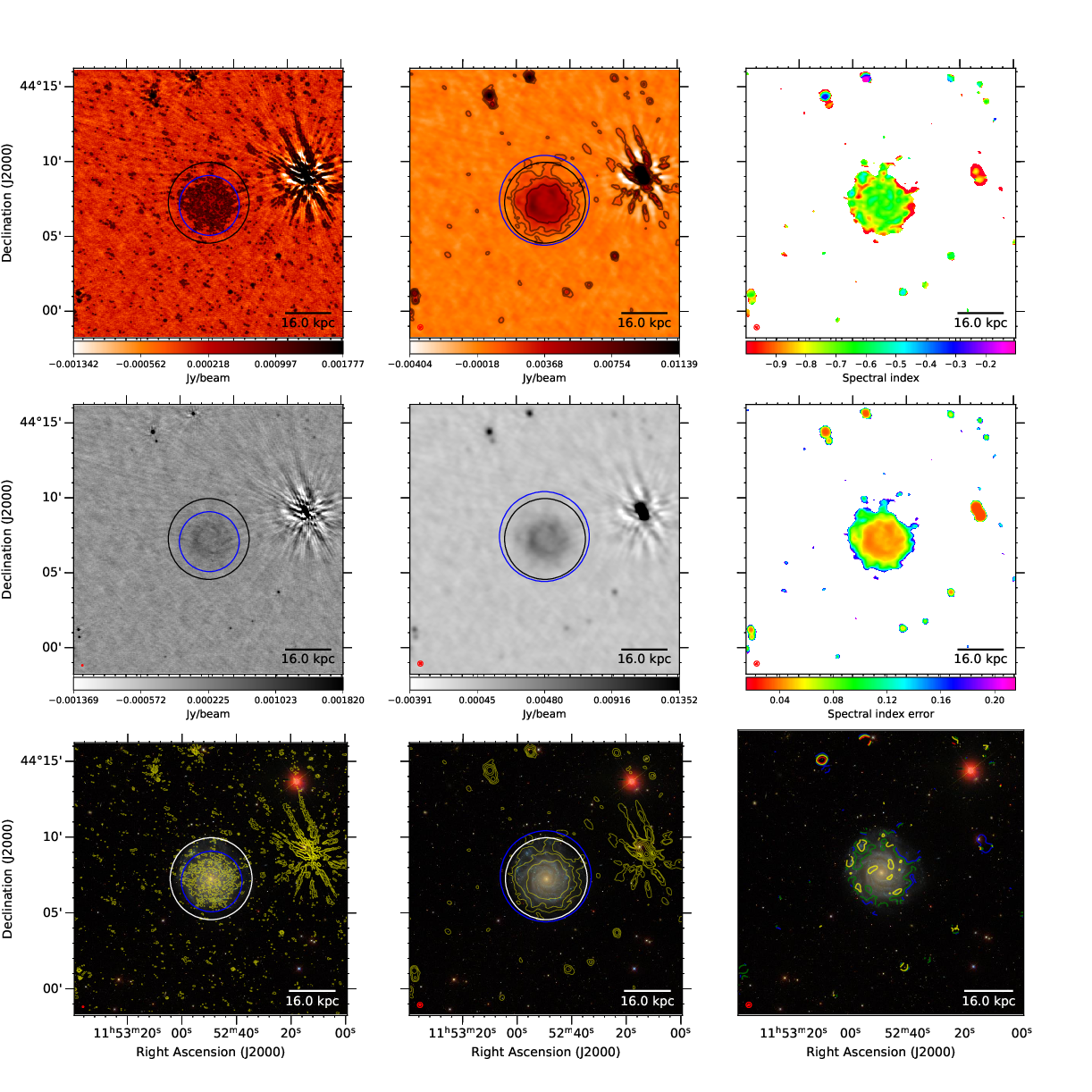}
    \caption{NGC~3938. \spix{1365}{20}}
    \label{fig:n3938}
\end{figure*}
\addcontentsline{toc}{subsection}{NGC 3938}

\begin{figure*}
	\centering
	\includegraphics[width=\textwidth]{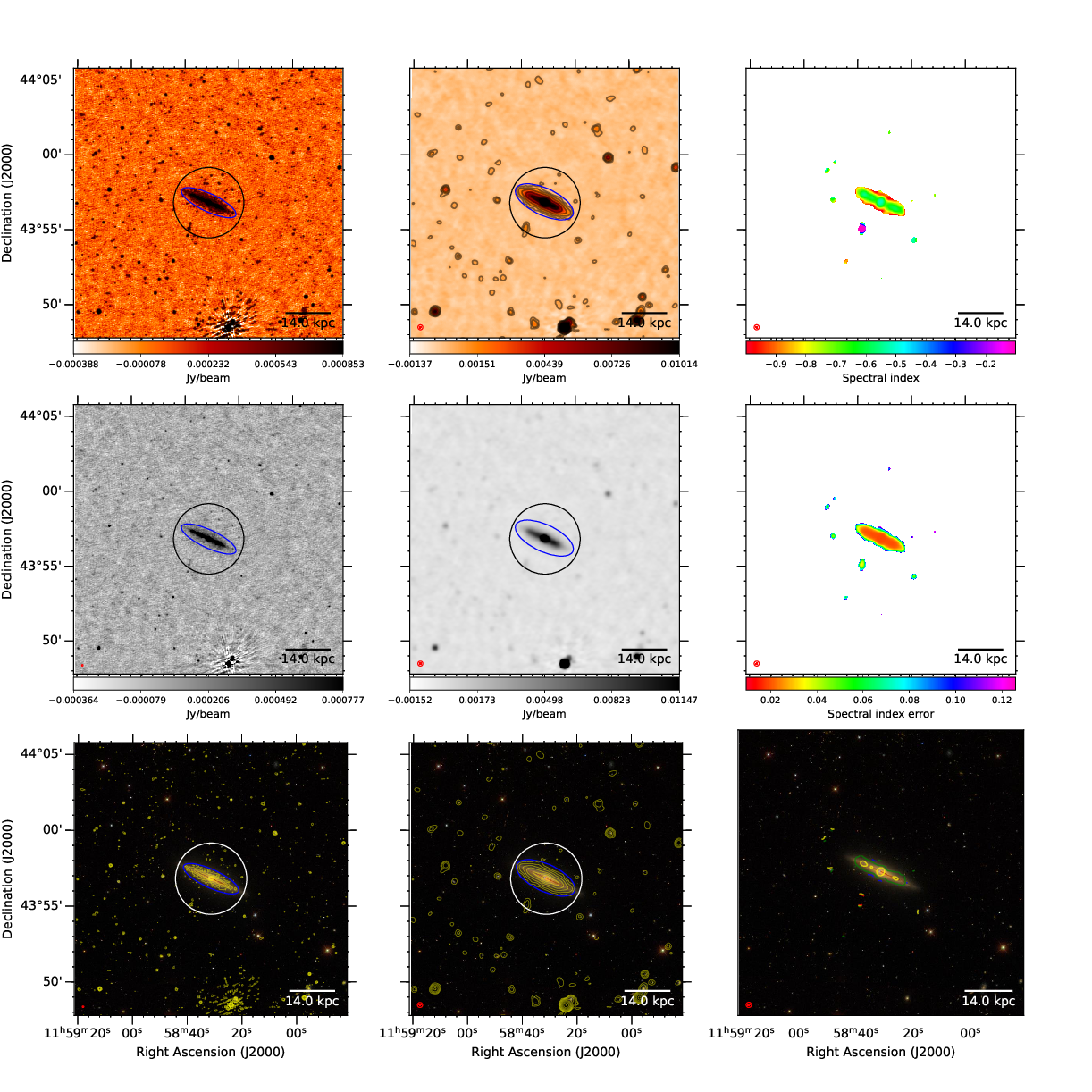}
    \caption{NGC~4013. \spix{6000}{20}}
    \label{fig:n4013}
\end{figure*}
\addcontentsline{toc}{subsection}{NGC 4013}

\begin{figure*}
	\centering
	\includegraphics[width=\textwidth]{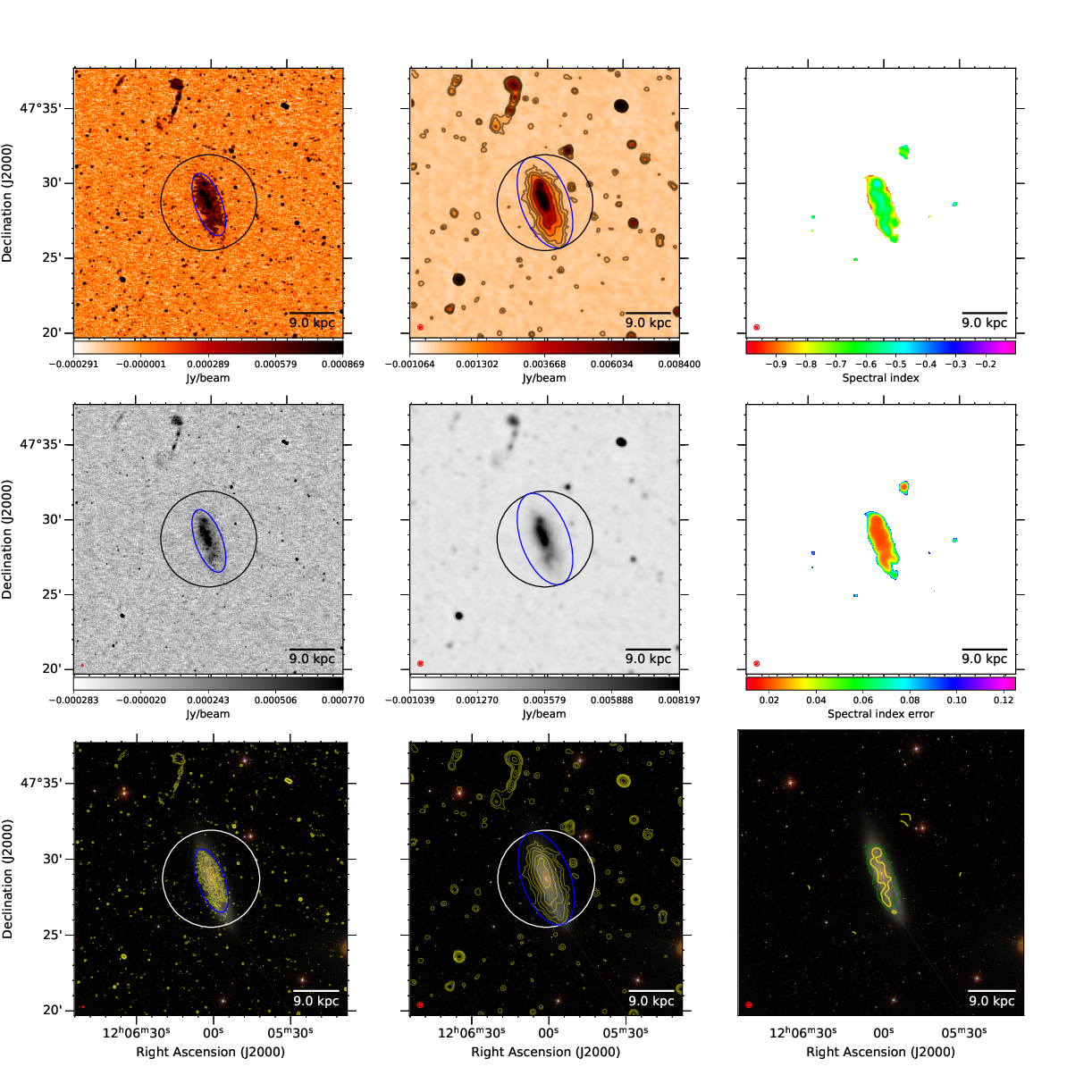}
    \caption{NGC~4096. \spix{6000}{20}}
    \label{fig:n4096}
\end{figure*}
\addcontentsline{toc}{subsection}{NGC 4096}

\begin{figure*}
	\centering
	\includegraphics[width=\textwidth]{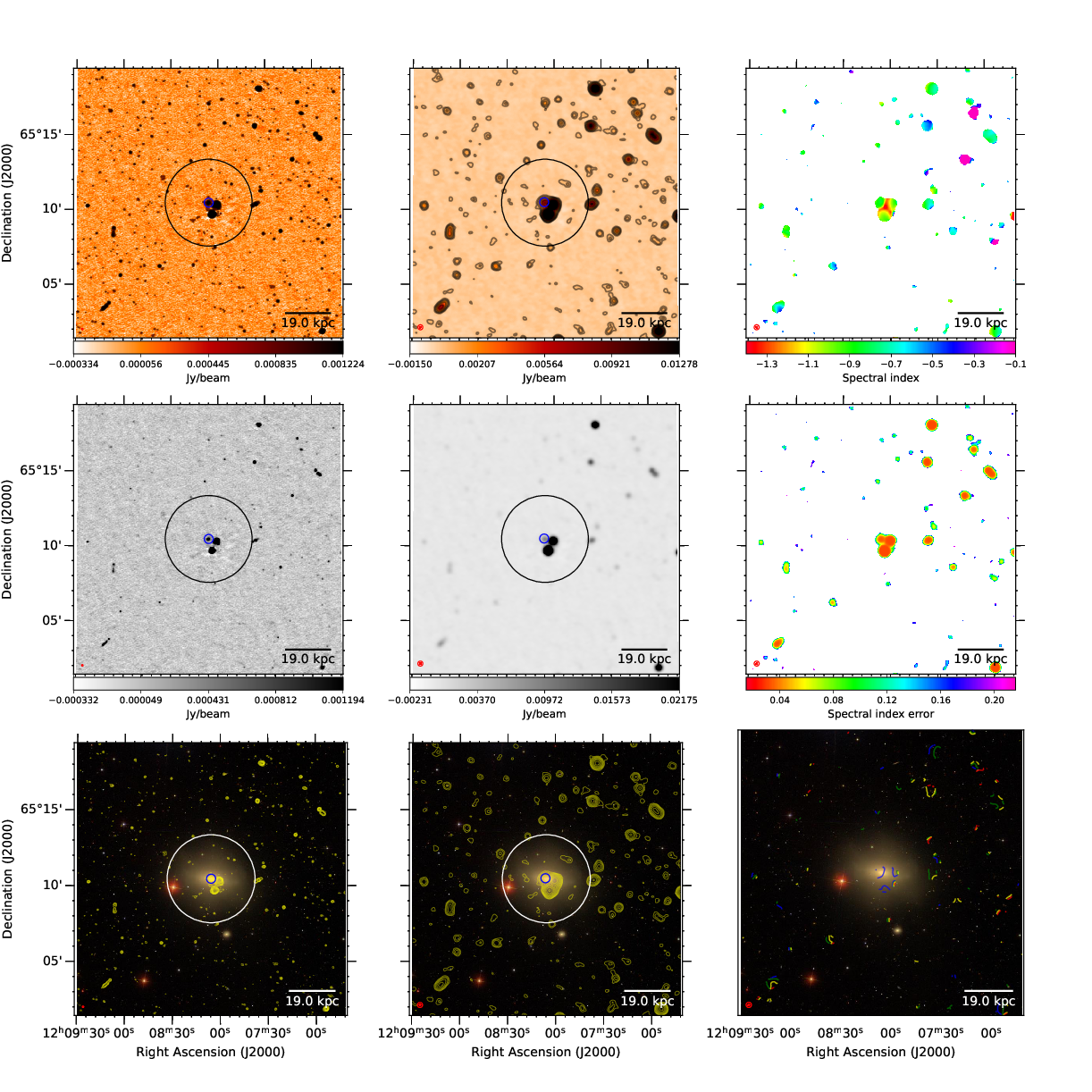}
    \caption{NGC~4125. \spix{1365}{20}}
    \label{fig:n4125}
\end{figure*}
\addcontentsline{toc}{subsection}{NGC 4125}

\begin{figure*}
	\centering
	\includegraphics[width=\textwidth]{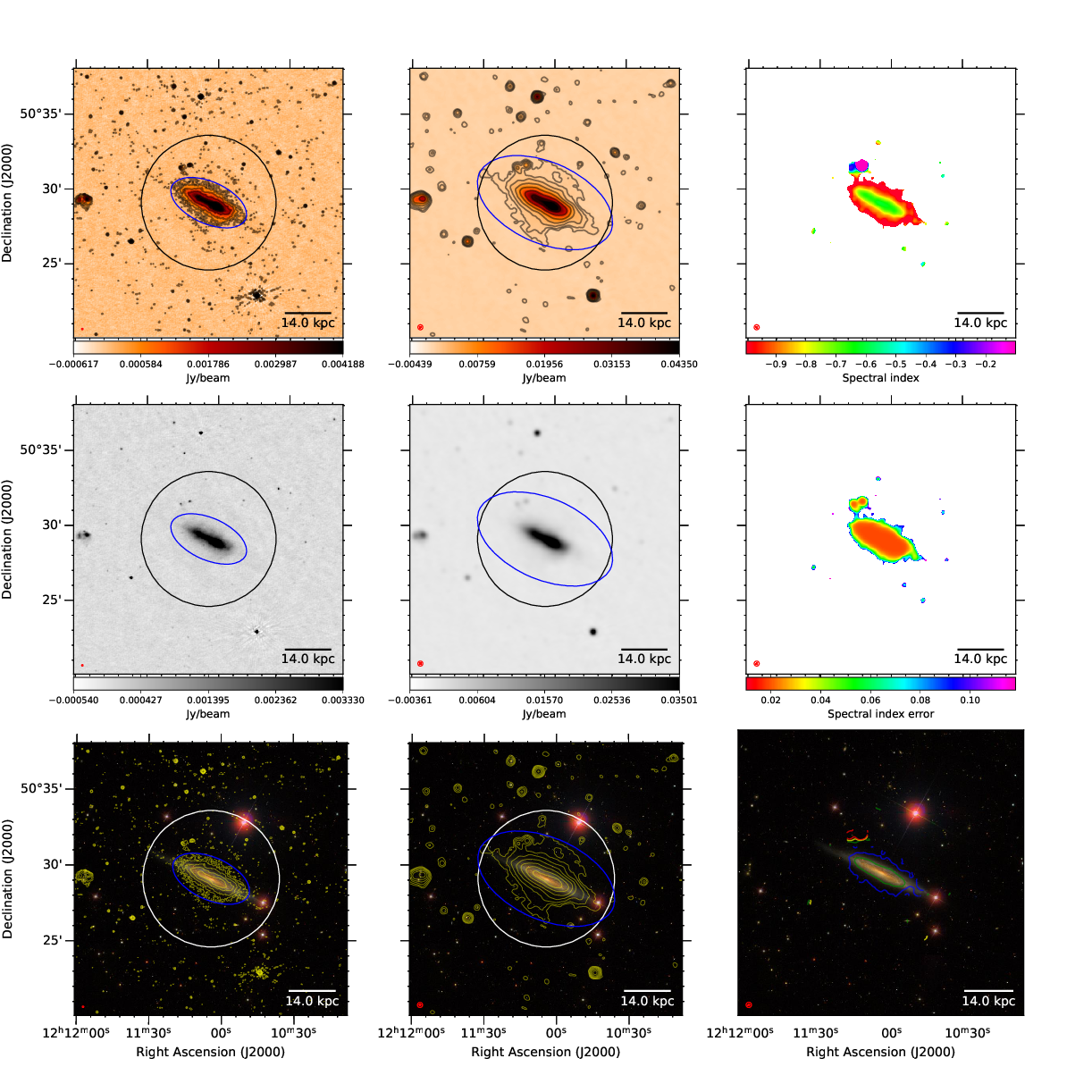}
    \caption{NGC~4157. \spix{6000}{20}}
    \label{fig:n4157}
\end{figure*}
\addcontentsline{toc}{subsection}{NGC 4157}

\begin{figure*}
	\centering
	\includegraphics[width=0.8\textwidth]{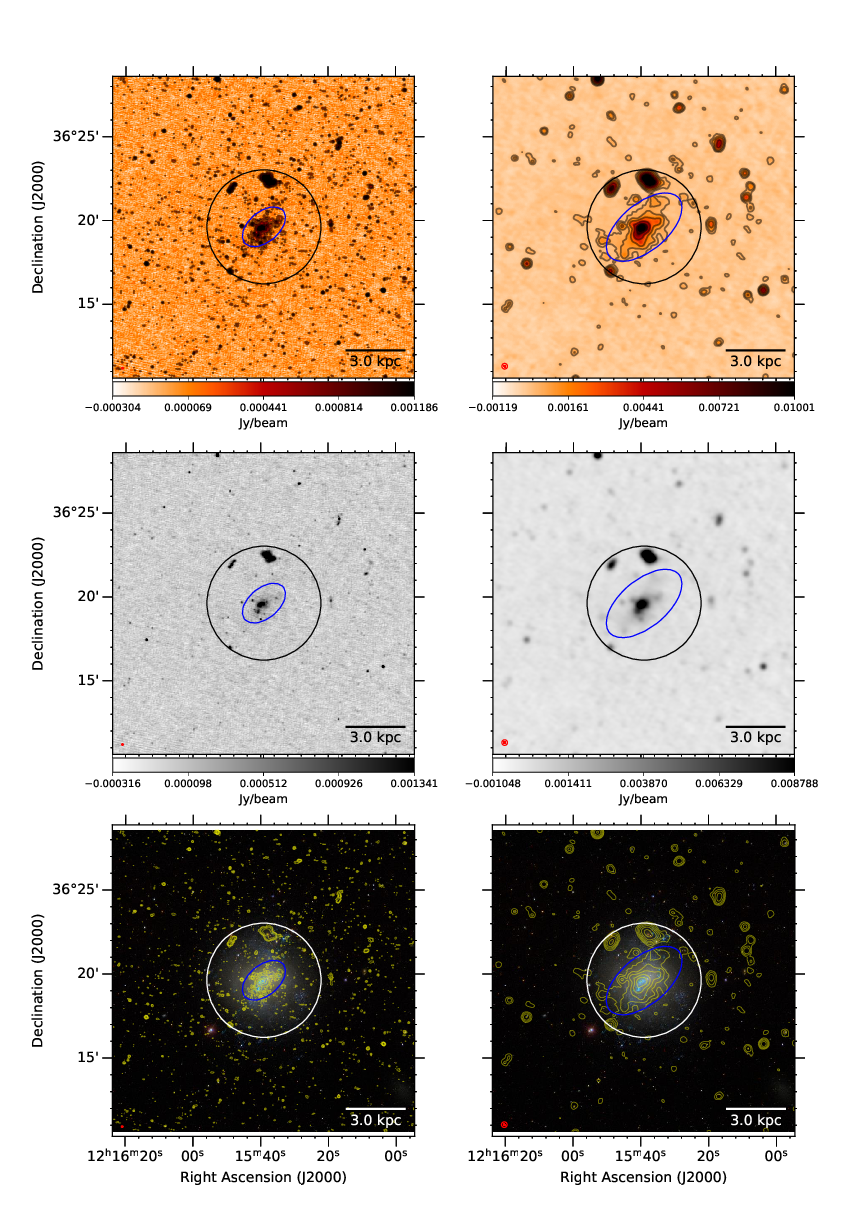}
    \caption{NGC~4214. \nospix}
    \label{fig:n4214}
\end{figure*}
\addcontentsline{toc}{subsection}{NGC 4214}

\begin{figure*}
	\centering
	\includegraphics[width=\textwidth]{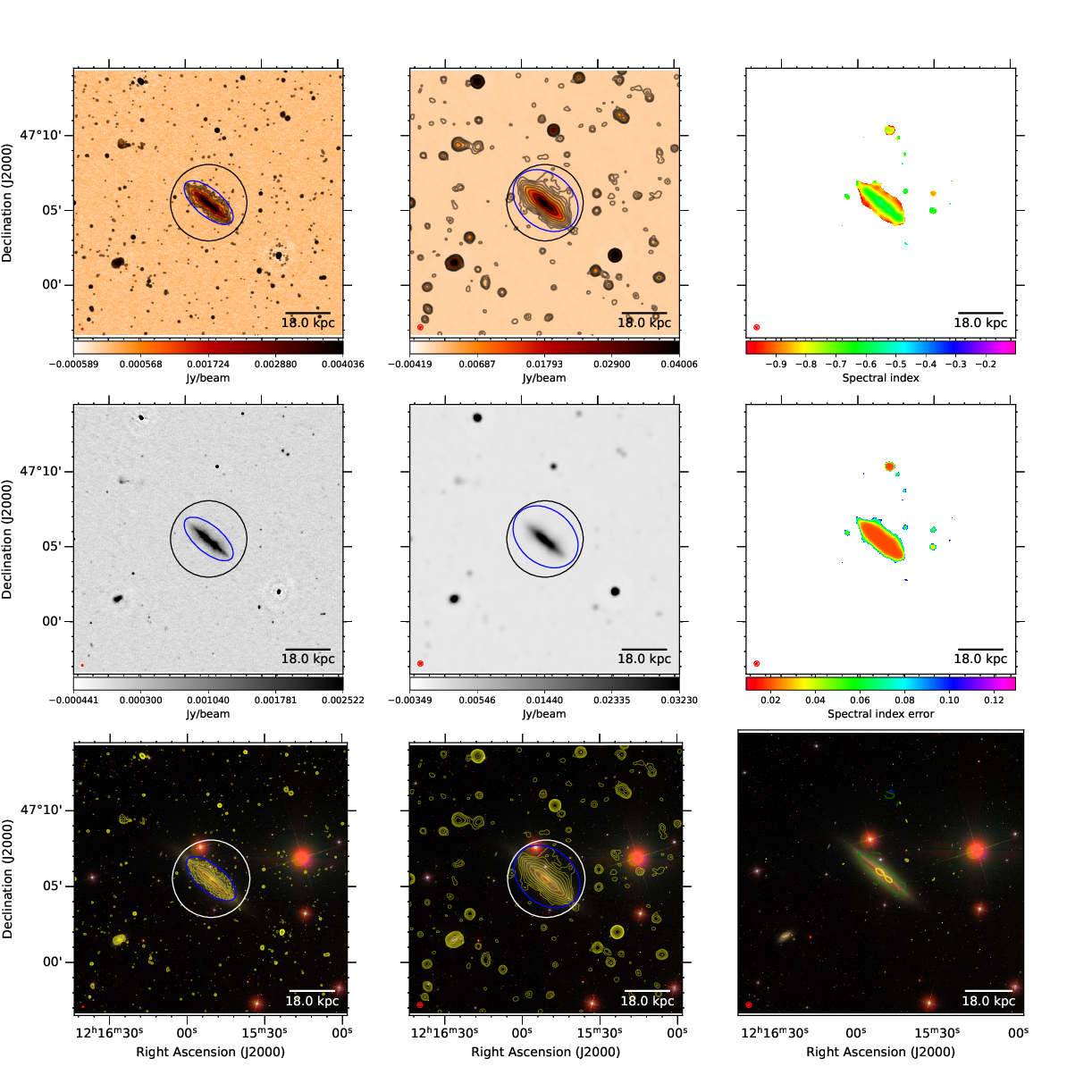}
    \caption{NGC~4217. \spix{6000}{20}}
    \label{fig:n4217}
\end{figure*}
\addcontentsline{toc}{subsection}{NGC 4217}

\begin{figure*}
	\centering
	\includegraphics[width=\textwidth]{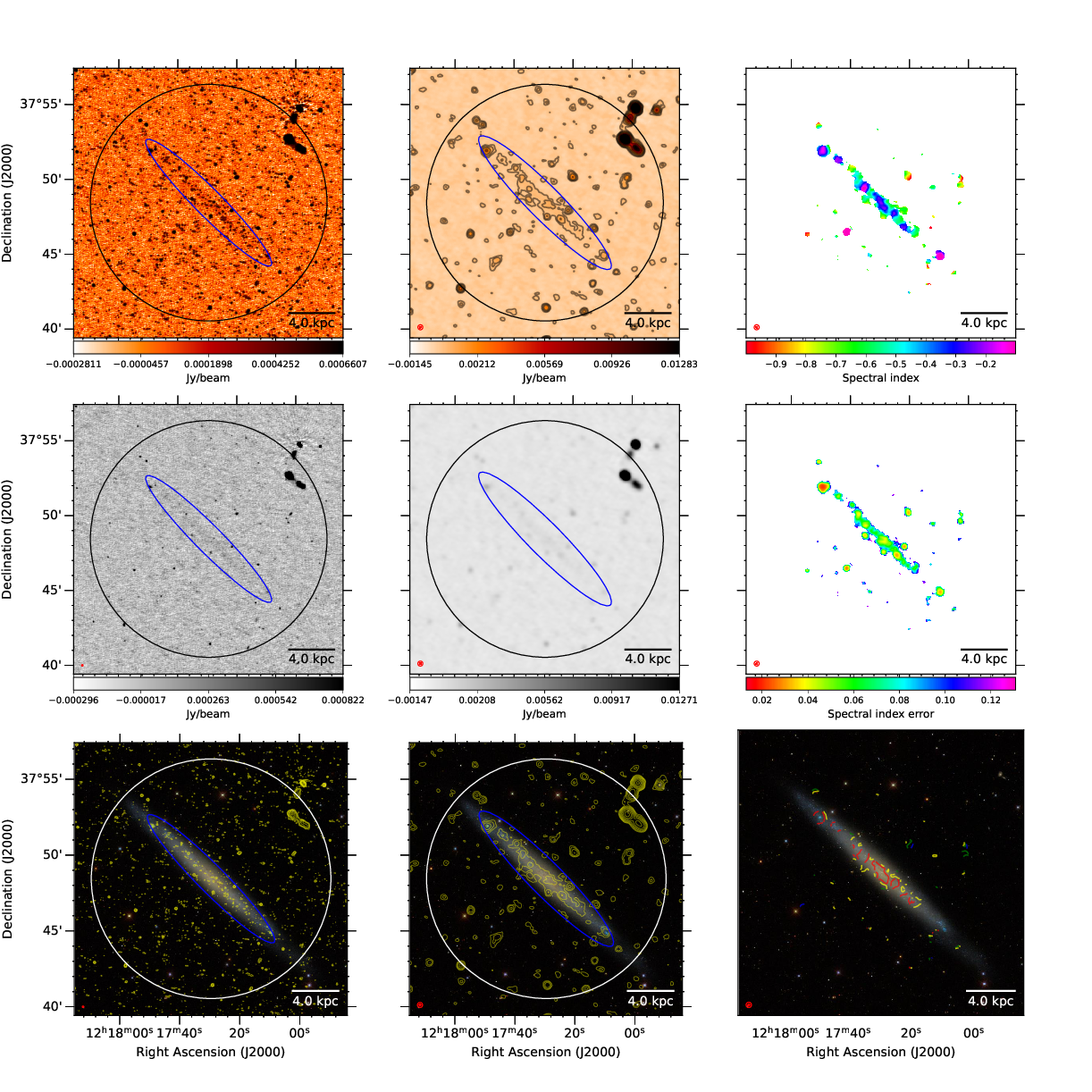}
    \caption{NGC~4244. \spix{6000}{20}}
    \label{fig:n4244}
\end{figure*}
\addcontentsline{toc}{subsection}{NGC 4244}

\begin{figure*}
	\centering
	\includegraphics[width=0.8\textwidth]{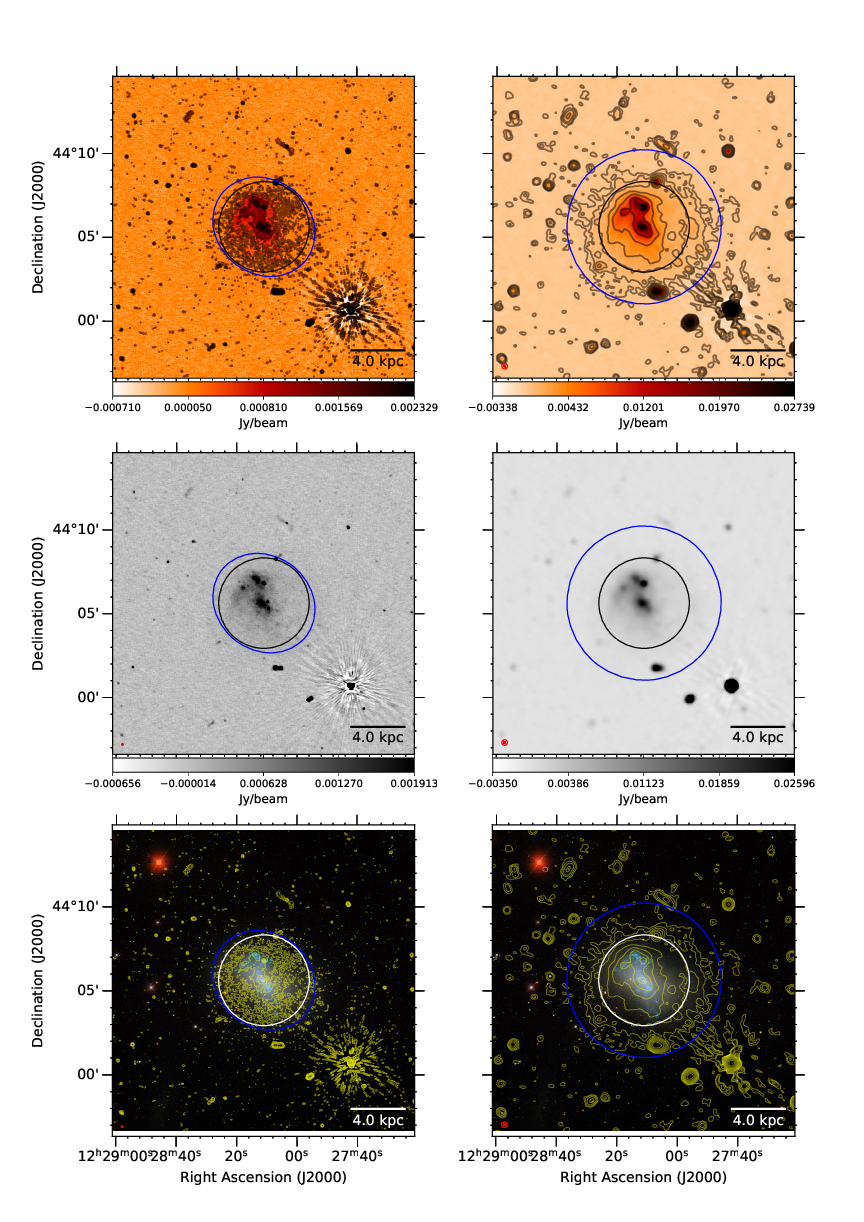}
    \caption{NGC~4449. \nospix}
    \label{fig:n4449}
\end{figure*}
\addcontentsline{toc}{subsection}{NGC 4449}

\begin{figure*}
	\centering
	\includegraphics[width=\textwidth]{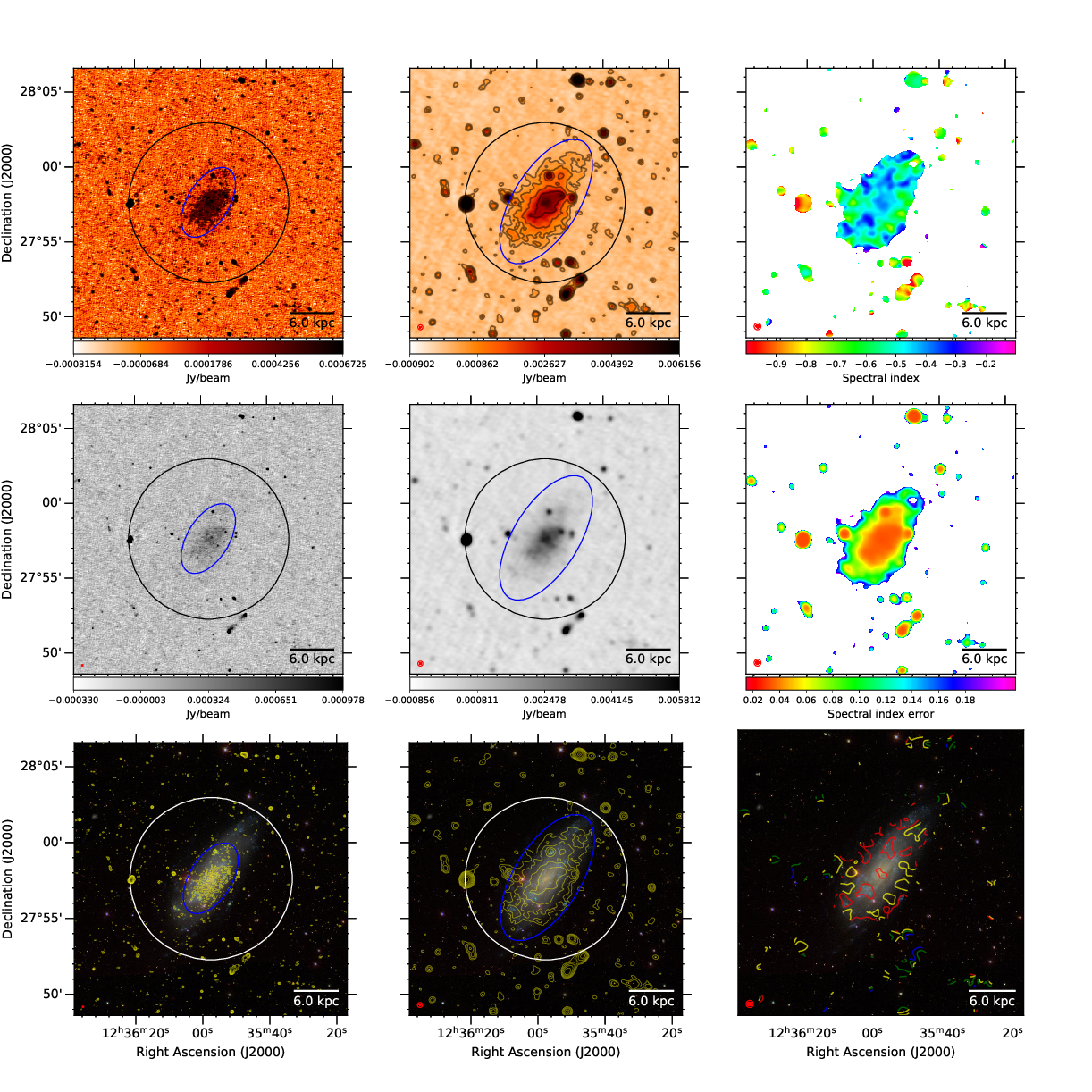}
    \caption{NGC~4559. \spix{1365}{27}}
    \label{fig:n4559}
\end{figure*}
\addcontentsline{toc}{subsection}{NGC 4559}

\begin{figure*}
	\centering
	\includegraphics[width=0.8\textwidth]{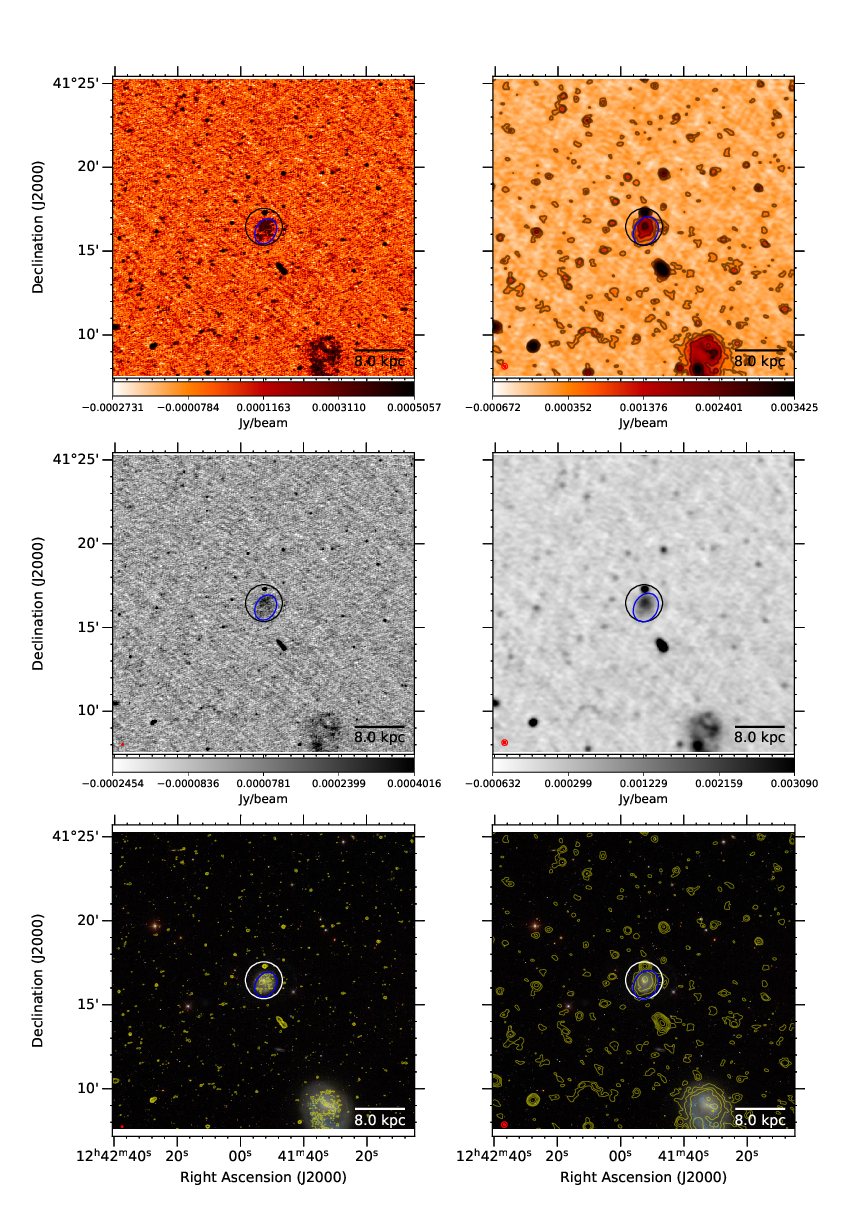}
    \caption{NGC~4625. \nospix}
    \label{fig:n4625}
\end{figure*}
\addcontentsline{toc}{subsection}{NGC 4625}

\begin{figure*}
	\centering
	\includegraphics[width=\textwidth]{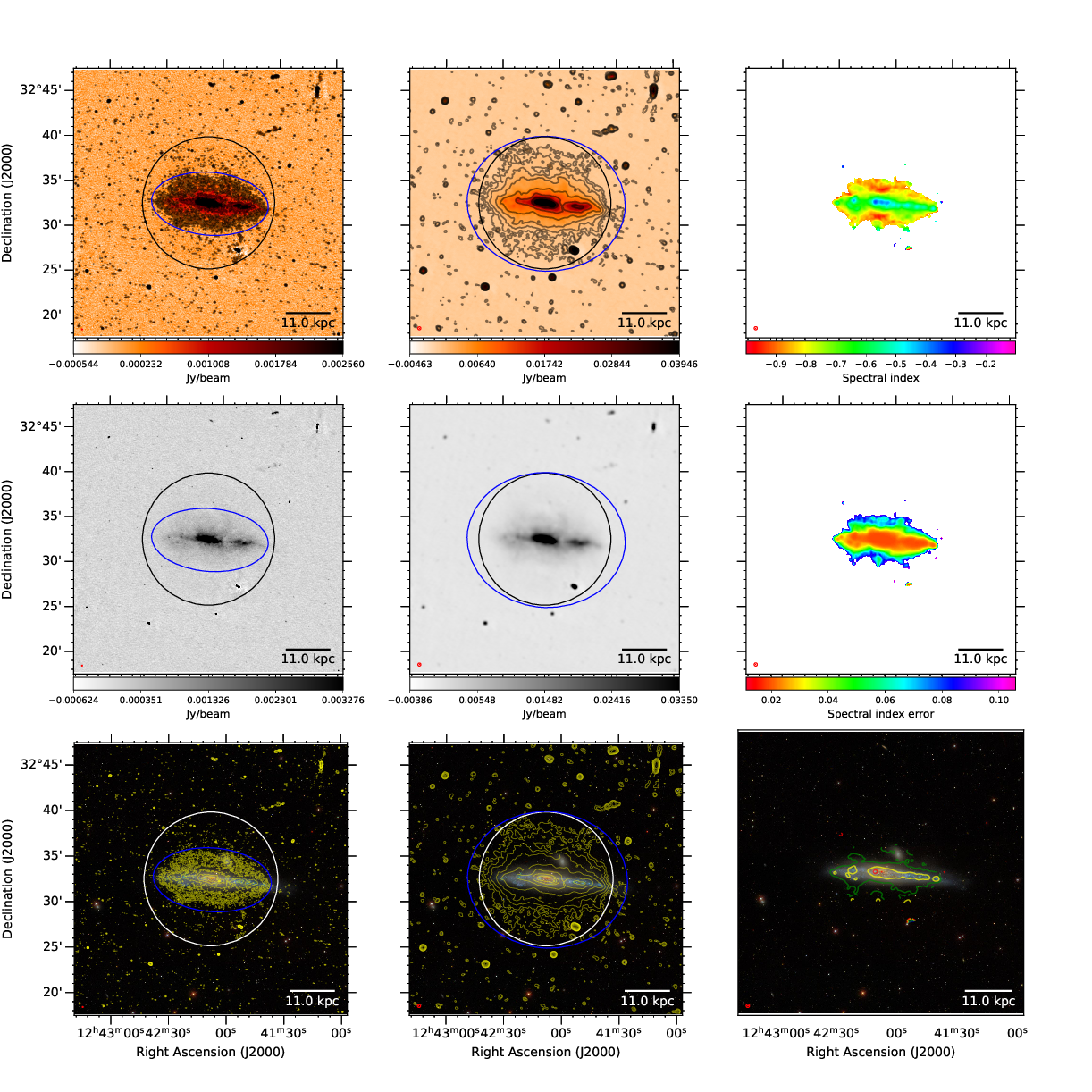}
    \caption{NGC~4631. \spix{6000}{20}}
    \label{fig:n4631}
\end{figure*}
\addcontentsline{toc}{subsection}{NGC 4631}

\begin{figure*}
	\centering
	\includegraphics[width=\textwidth]{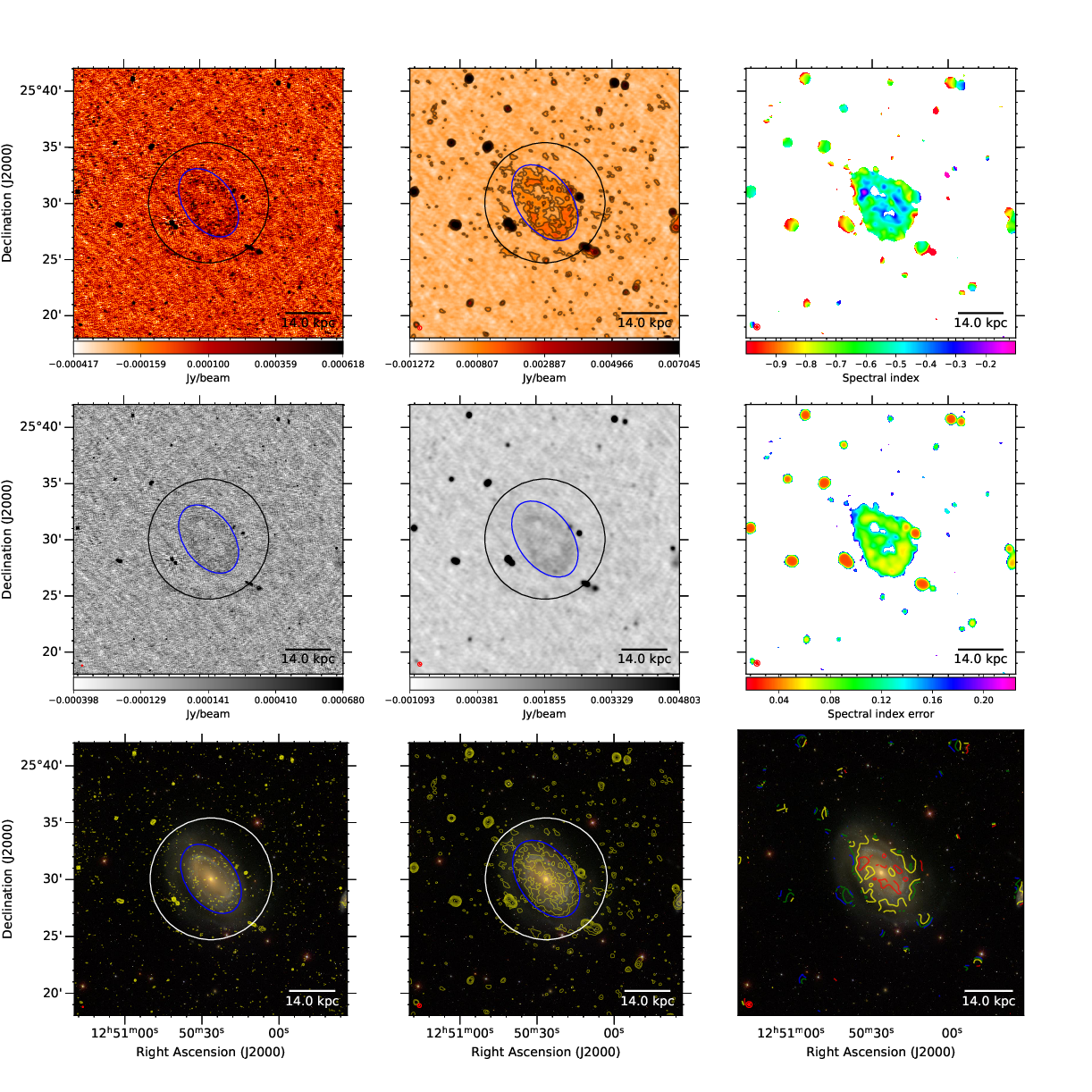}
    \caption{NGC~4725. \spix{1365}{30}}
    \label{fig:n4725}
\end{figure*}
\addcontentsline{toc}{subsection}{NGC 4725}

\begin{figure*}
	\centering
	\includegraphics[width=\textwidth]{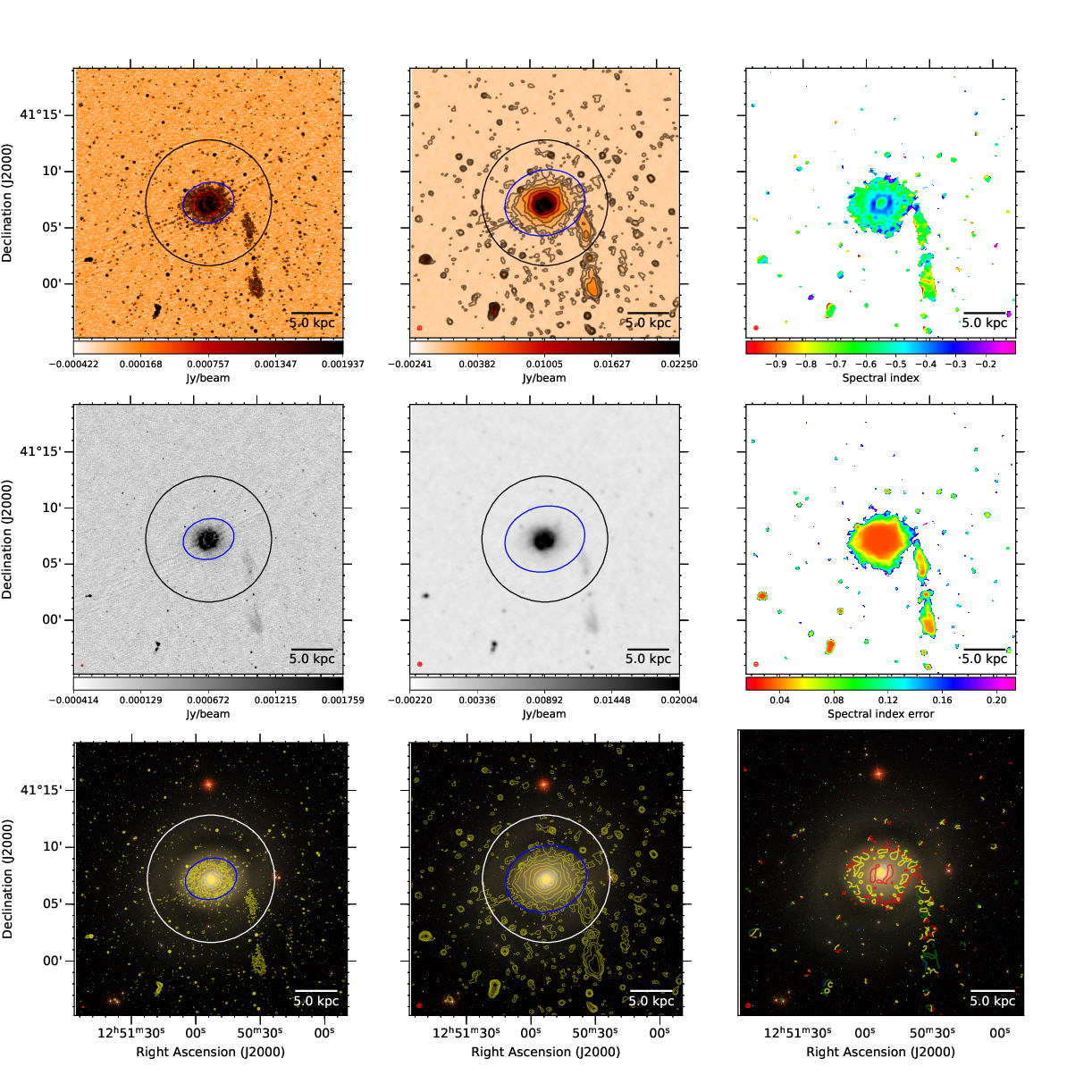}
    \caption{NGC~4736. \spix{1365}{20}}
    \label{fig:n4736}
\end{figure*}
\addcontentsline{toc}{subsection}{NGC 4736}

\begin{figure*}
	\centering
	\includegraphics[width=\textwidth]{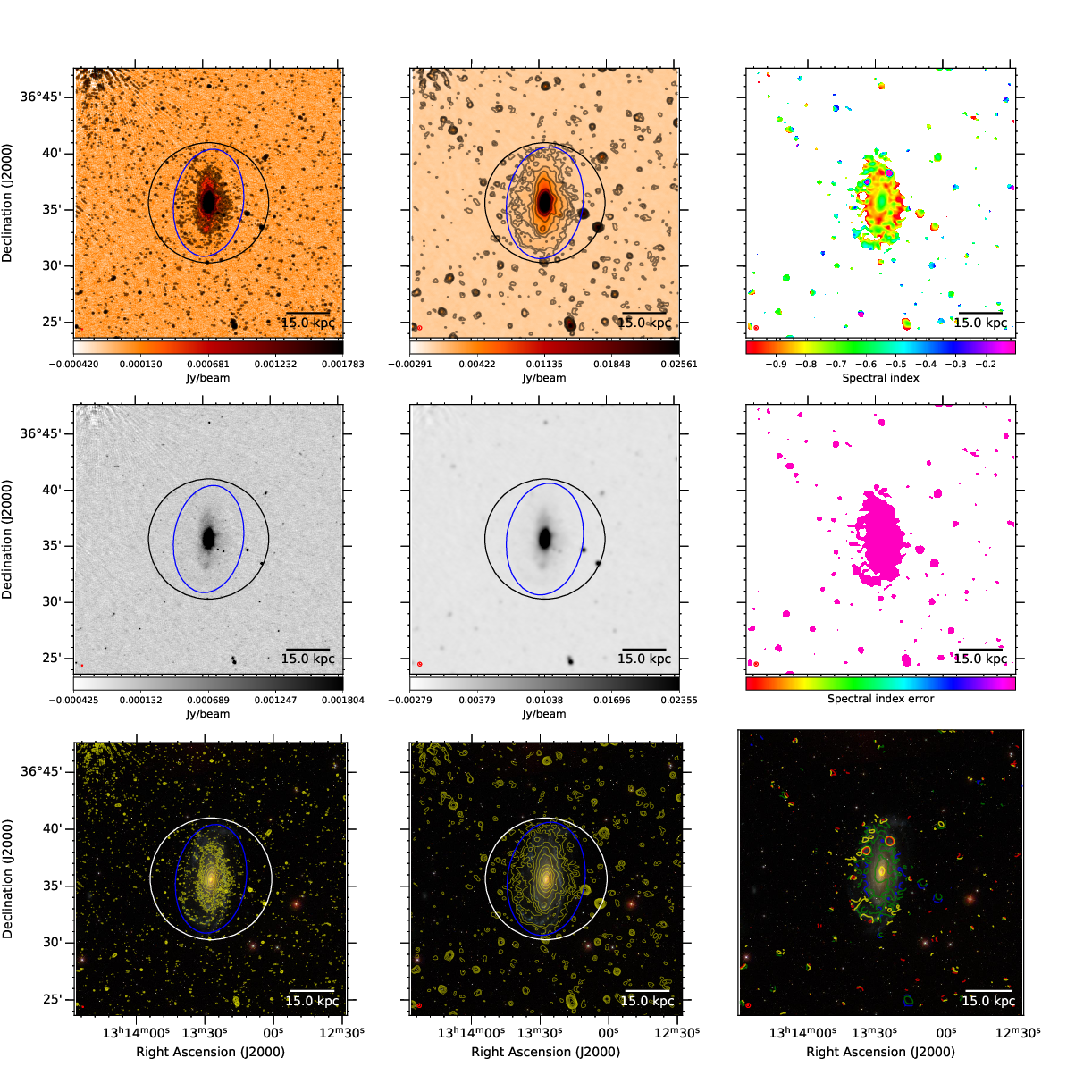}
    \caption{NGC~5033. \spix{1365}{21}}
    \label{fig:n5033}
\end{figure*}
\addcontentsline{toc}{subsection}{NGC 5033}

\begin{figure*}
	\centering
	\includegraphics[width=\textwidth]{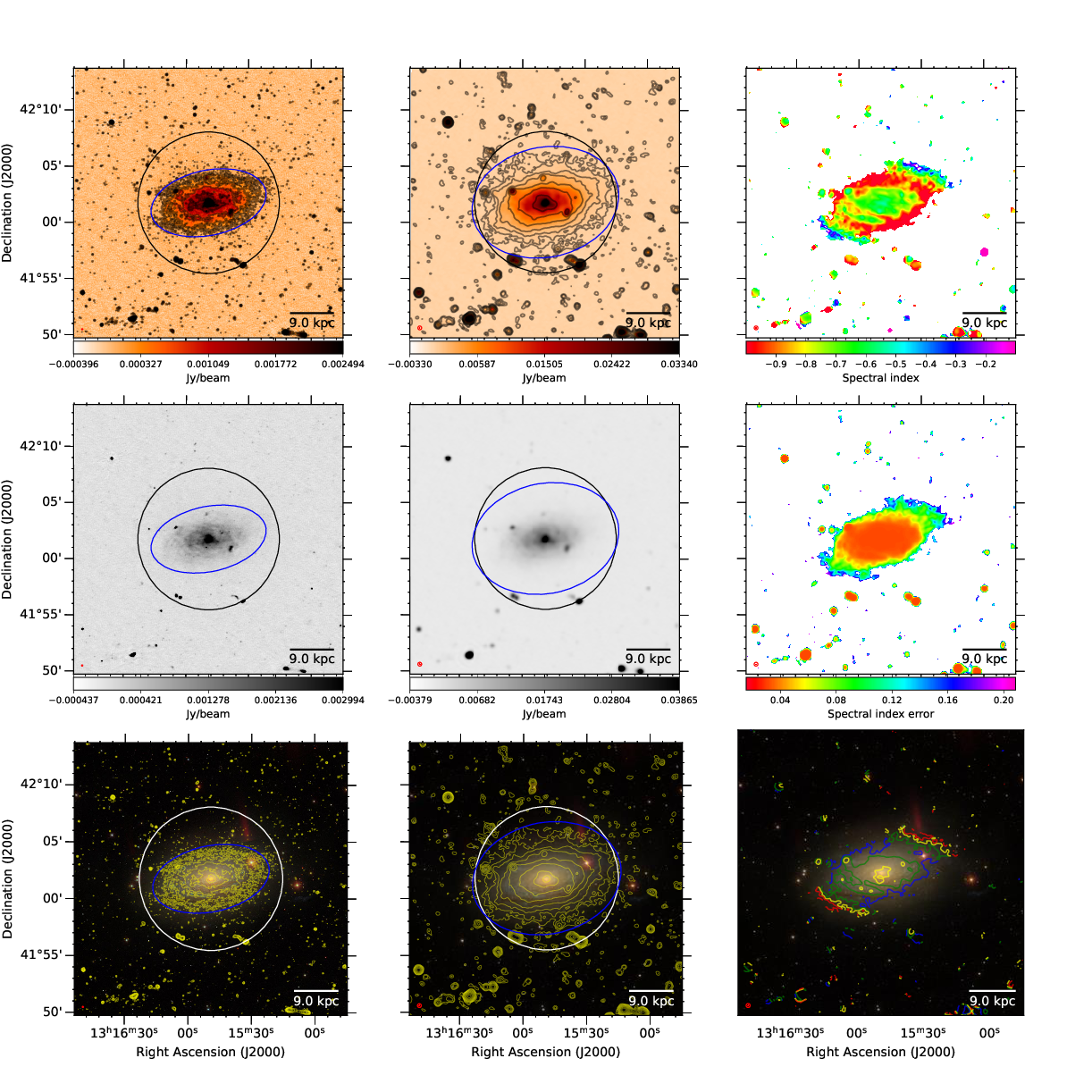}
	\centering
    \caption{NGC~5055. \spix{1365}{20}}
    \label{fig:n5055}
\end{figure*}
\addcontentsline{toc}{subsection}{NGC 5055}

\begin{figure*}
	\centering
	\includegraphics[width=\textwidth]{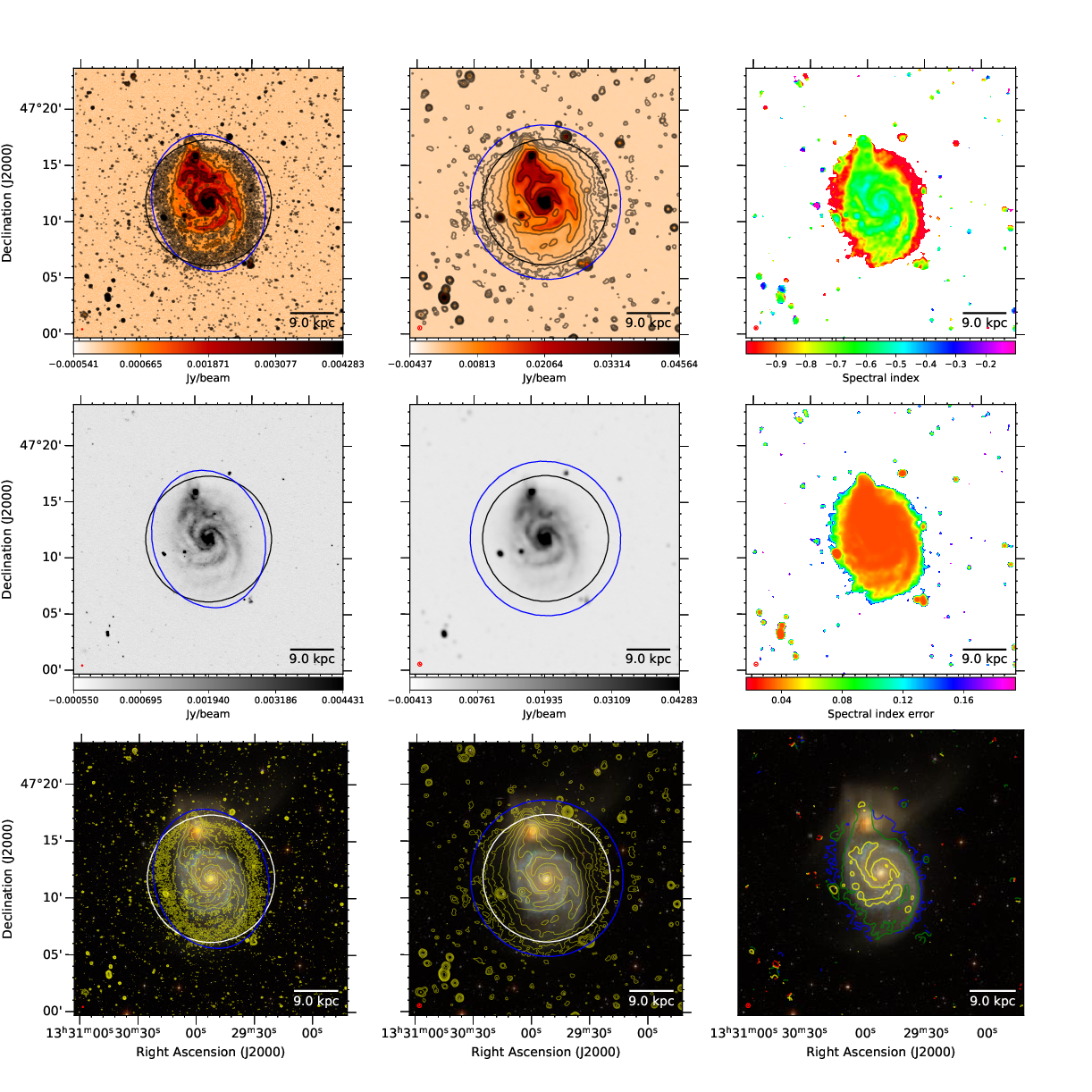}
    \caption{NGC~5194. \spix{1365}{20}}
    \label{fig:n5194_app}
\end{figure*}
\addcontentsline{toc}{subsection}{NGC 5194}

\begin{figure*}
	\centering
	\includegraphics[width=0.8\textwidth]{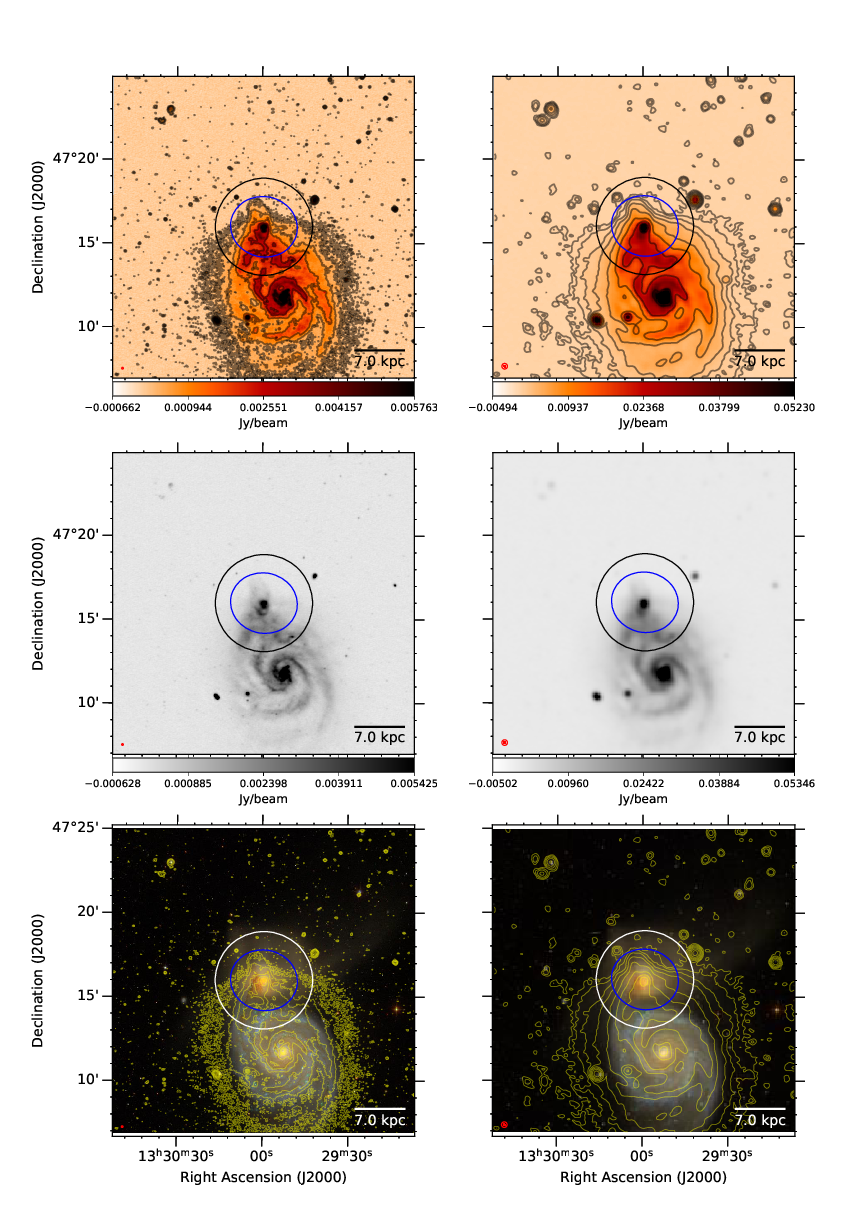}
    \caption{NGC~5195. \spix{1365}{20}}
    \label{fig:n5195}
\end{figure*}
\addcontentsline{toc}{subsection}{NGC 5195}

\begin{figure*}
	\centering
	\includegraphics[width=\textwidth]{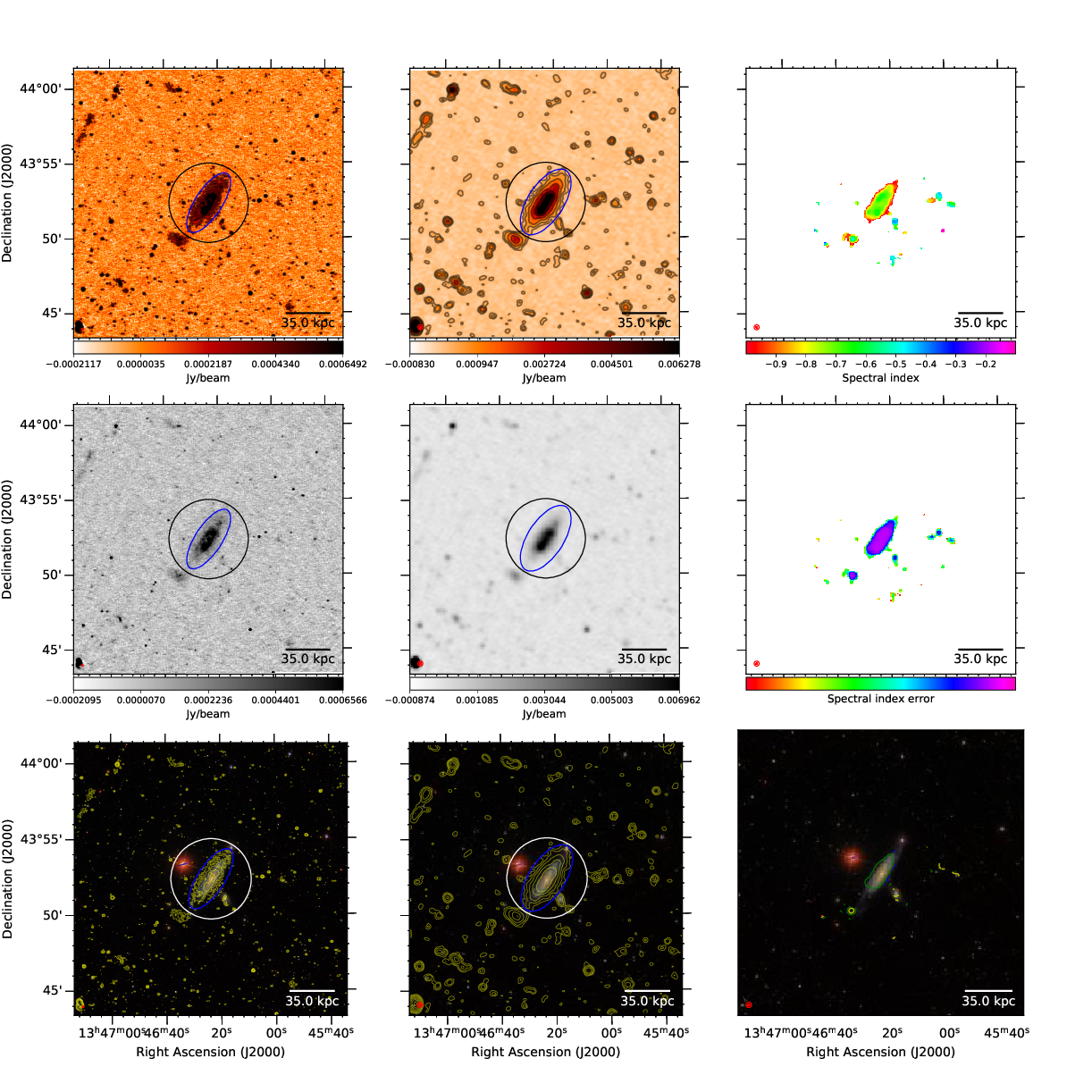}
    \caption{NGC~5297. \spix{1365}{20}}
    \label{fig:n5297}
\end{figure*}
\addcontentsline{toc}{subsection}{NGC 5297}

\begin{figure*}
	\centering
	\includegraphics[width=0.8\textwidth]{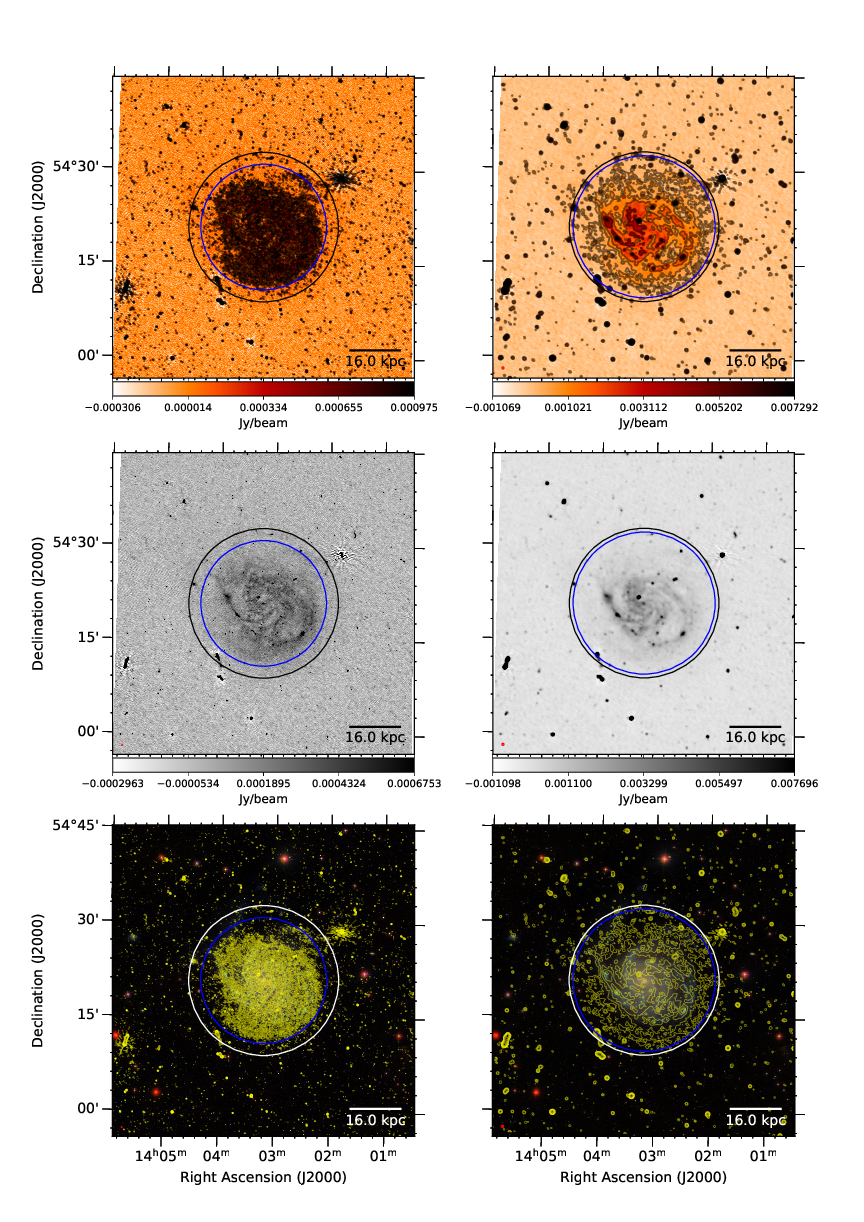}
    \caption{NGC~5457. \nospix}
    \label{fig:n5457}
\end{figure*}
\addcontentsline{toc}{subsection}{NGC 5457}

\begin{figure*}
	\centering
	\includegraphics[width=0.8\textwidth]{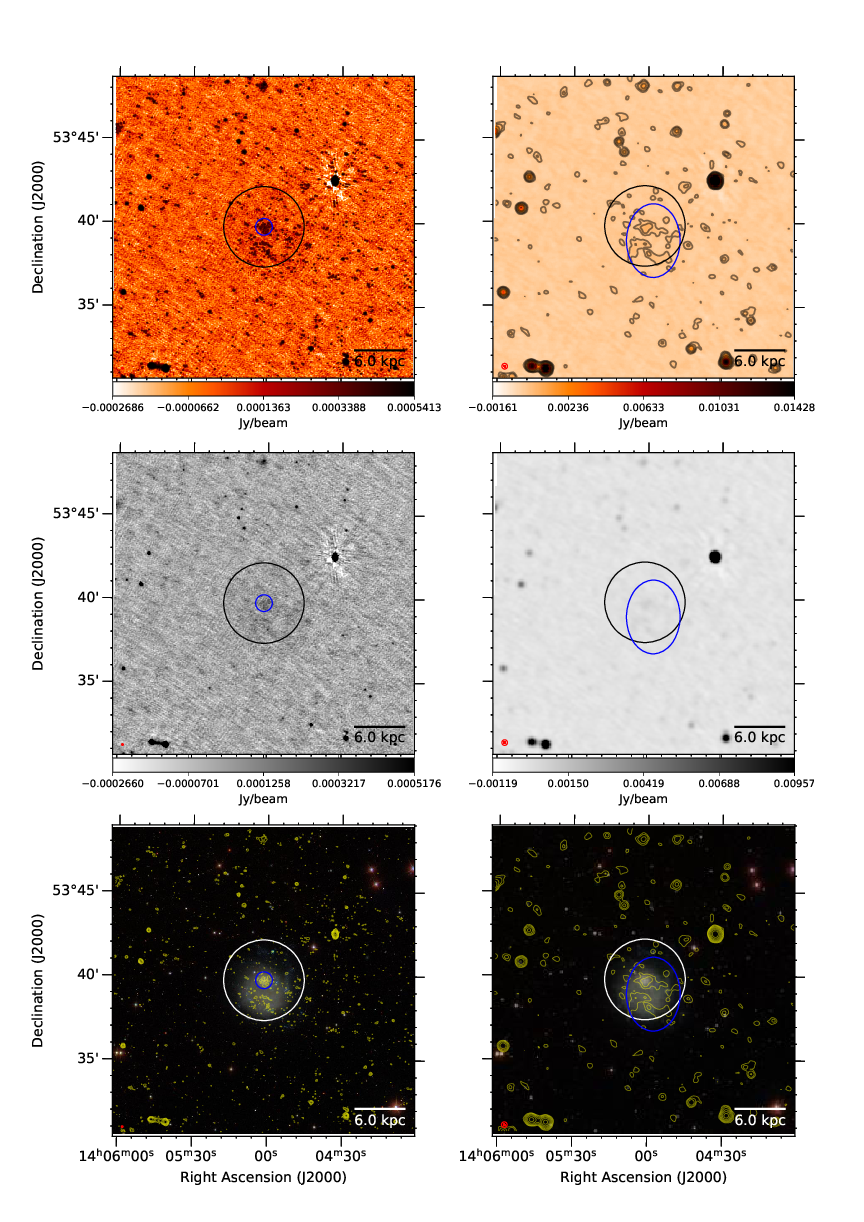}
    \caption{NGC~5474. \nospix}
    \label{fig:n5474}
\end{figure*}
\addcontentsline{toc}{subsection}{NGC 5474}

\begin{figure*}
	\centering
	\includegraphics[width=0.8\textwidth]{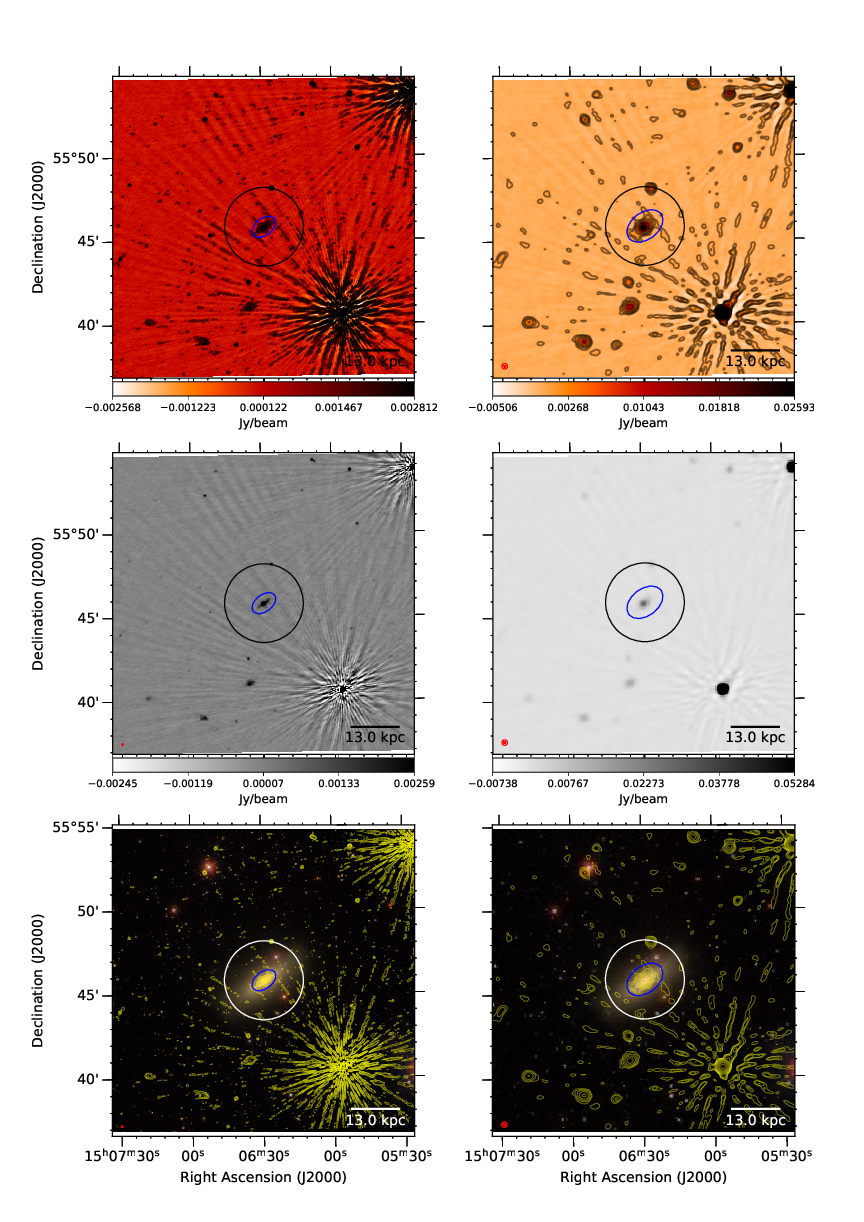}
    \caption{NGC~5866. \nospix}
    \label{fig:n5866}
\end{figure*}
\addcontentsline{toc}{subsection}{NGC 5866}

\begin{figure*}
	\centering
	\includegraphics[width=\textwidth]{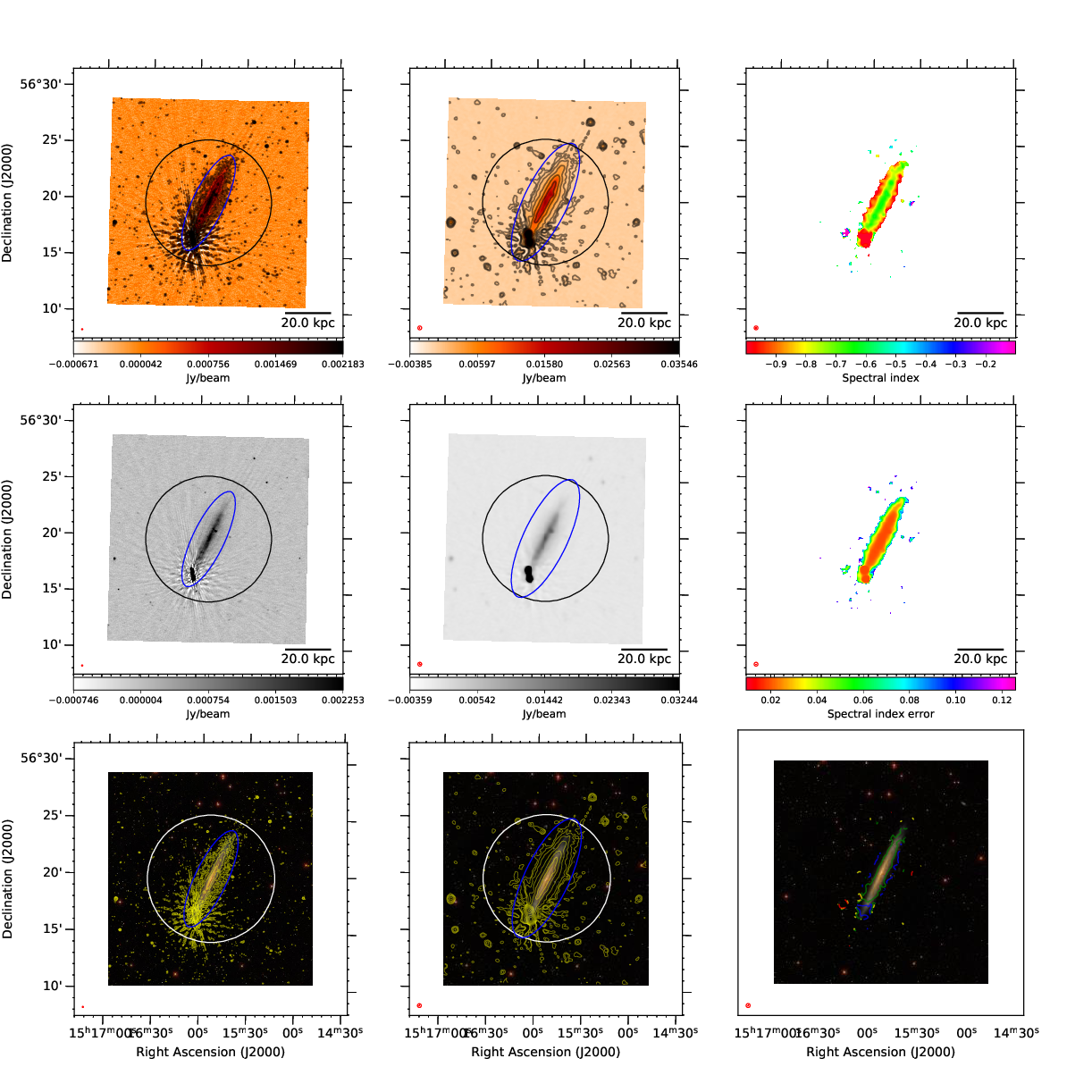}
    \caption{NGC~5907. \spix{6000}{20}}
    \label{fig:n5907}
\end{figure*}
\addcontentsline{toc}{subsection}{NGC 5907}

\begin{figure*}
	\includegraphics[width=\textwidth]{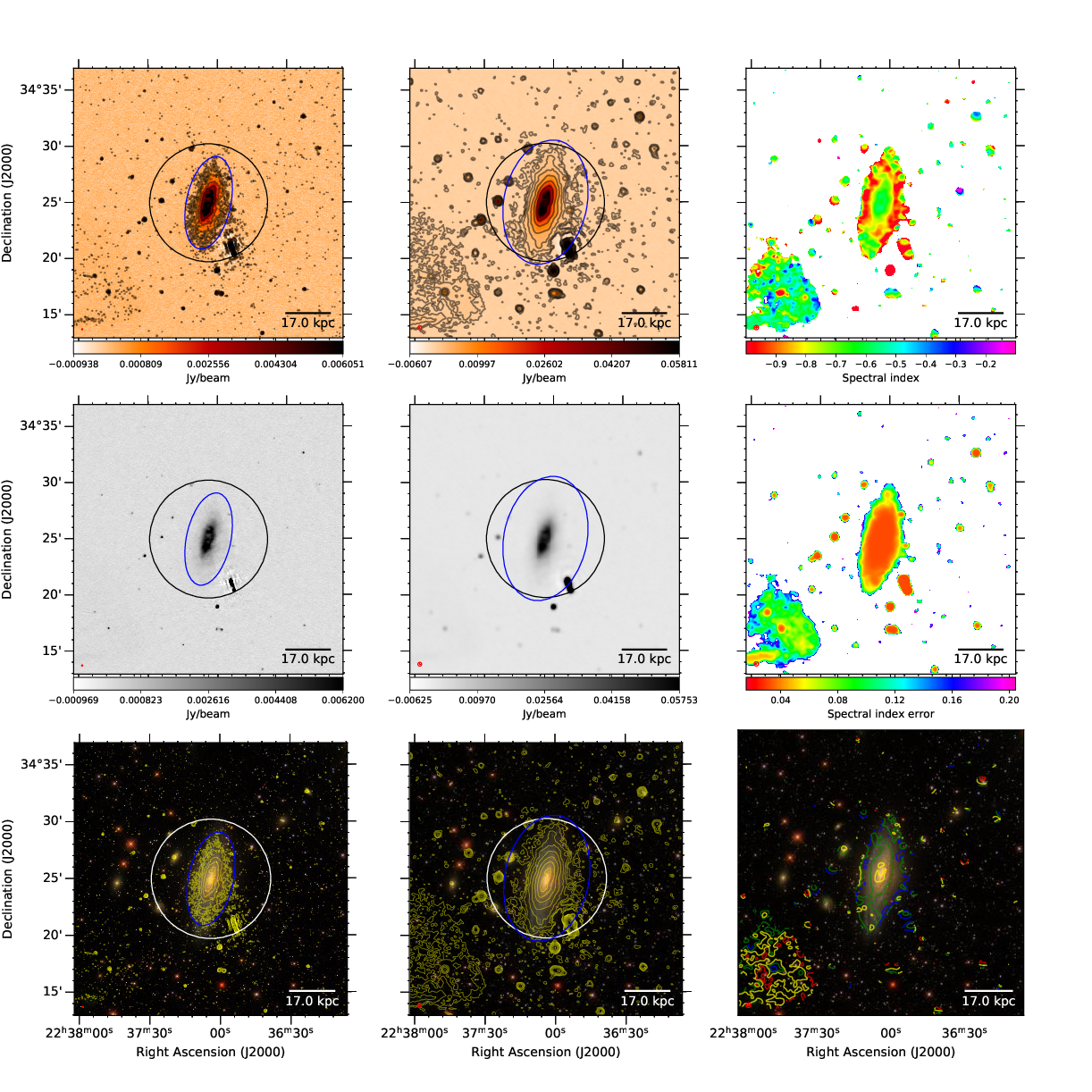}
    \caption{NGC~7331. \spix{6000}{23}}
    \label{fig:n7331}
\end{figure*}
\addcontentsline{toc}{subsection}{NGC 7331}

\end{document}